\documentclass[showpacs,aps,pre]{revtex4}
\usepackage{amsmath,amssymb}
\usepackage{graphicx}
\include{epsf}

\def\(({\left(}
\def\)){\right)}                       
\def\[[{\left[}
\def\]]{\right]}
\def\e{{\rm e}}
\def\POP{{\rm POP}}

\newcommand{\be}{\begin{equation}}
\newcommand{\ee}{\end{equation}}
\newcommand{\bea}{\begin{eqnarray}}
\newcommand{\eea}{\end{eqnarray}}

\begin{document}

\title{Phase Transitions in the Coloring of Random Graphs}

\author{Lenka Zdeborov\'a$^1$ and Florent Krz\c{a}ka{\l}a$^2$}

\affiliation{
  $^1$ LPTMS, UMR 8626 CNRS et Univ. Paris-Sud, 91405 Orsay CEDEX, France\\
  $^2$ PCT, UMR Gulliver 7083 CNRS-ESPCI, 10 rue Vauquelin, 75231 Paris, France \\
}

\begin{abstract}
  We consider the problem of coloring the vertices of a large sparse random
  graph with a given number of colors so that no adjacent vertices
  have the same color. Using the cavity method, we present a detailed and
  systematic analytical study of the space of proper colorings (solutions).
  
  We show that for a fixed number of colors and as the average vertex degree
  (number of constraints) increases, the set of solutions undergoes several
  phase transitions similar to those observed in the mean field theory of
  glasses.  First, at the clustering transition, the entropically dominant
  part of the phase space decomposes into an exponential number of pure
  states so that
  beyond this transition a uniform sampling of solutions becomes
  hard.  Afterward, the space of solutions condenses over a finite number
  of the largest states and consequently
  the total entropy of solutions becomes smaller than the annealed one.
  Another transition takes place when
  in all the entropically dominant states a finite fraction of nodes freezes 
  so that each of these nodes is allowed a single color in all the solutions 
  inside the state.  Eventually, above the coloring threshold, no more solutions
  are available.  We compute all the critical connectivities for
  Erd\H{o}s-R\'enyi and regular random graphs 
  and determine their asymptotic values for large number of colors.
  
  Finally, we discuss the algorithmic consequences of our findings. We argue
  that the onset of computational hardness is not associated
  with the clustering transition and we suggest instead 
  that the freezing transition might be the relevant phenomenon.
  We also discuss the performance of a simple local Walk-COL algorithm 
  and of the belief propagation algorithm in the light of our results.
\end{abstract}

\pacs{89.20.Ff, 75.10.Nr, 05.70.Fh, 02.70.-c} 


\maketitle

\section{Introduction}
\label{sec:Intro}

Graph coloring is a famous yet basic problem in combinatorics. Given a graph
and $q$ colors, the problem consists in coloring the vertices in such a way
that no connected vertices have the same color~\cite{GaJo}. The celebrated
four-colors theorem assures that this is always possible for planar graphs
using only four colors~\cite{4colors}. For general graphs, however, the
problem can be extremely hard to solve and is known to be
NP-complete~\cite{NPcol}, so that it is widely believed that no algorithm can
decide in a polynomial time (with respect to the size of the graph) if a given
arbitrary instance is colorable or not. Indeed, the problem is often taken as a
benchmark for the evaluation of the performance of algorithms in computer
science. 
It has also important practical application as timetabling, scheduling, register
allocation in compilers or frequency assignment in mobile radios.

In this paper, we study colorings of sparse random graphs~\cite{Bo,Janson}.
Random graphs are one of the most fundamental source of challenging problems
in graph theory since the seminal work of Erd\H{o}s and R\'enyi~\cite{Erdos}
in 1959.  Concerning the coloring problem, a crucial observation was made by
focusing on typical instances drawn from the ensemble of random graphs with a
given average vertex connectivity $c$, as $c$ increases a threshold phenomenon
is observed. Bellow a critical value $c_s$ a proper coloring of the graph
with $q$ colors exists with a probability going to one in the large size
limit, while beyond $c_s$ it does not exist in the same sense. This sharp
transition also appears in other Constraint Satisfaction Problems (CSPs) such
as the satisfiability of Boolean formulae~\cite{GaJo}. 
The existence of the sharp COLorable/UNCOLorable (COL/UNCOL) 
transition was partially {\footnote{More precisely~\cite{Friedgut} proves the
    existence of a sharp threshold for a possibly size dependent sequence of
    thresholds, whose convergence is not proven.}} proven in~\cite{Friedgut}, and computing rigorously
its precise location is a major open problem in graph theory. Many upper and
lower bounds were established
\cite{Lu,Nature,DuboisMandler,AchlioptasMoore03,AchlioptasRegular,Wormlad1,Wormlad2,Diaz}
for Erd\H{o}s-R\'enyi and regular random graphs.

It was also observed empirically~\cite{CKT91,Selman} that deciding
colorability becomes on average much harder near to the coloring threshold
$c_s$ than far away from it. This onset of computational hardness cannot be
explained only by the simple fact that near to the colorable threshold the
number of proper colorings is small \cite{Hardness}. Some progress in the
theoretical understanding has been done by the analysis of search algorithms
\cite{Hogg,Cocco}.  For the coloring problem, it was proven~\cite{AcMo} that a
simple algorithm $q$-colors almost surely in linear time random graphs of
average connectivity $c \le q \log{q} - 3q/2$ for all $q\ge 3$
(see~\cite{AcMo} for references on previous works).  For 3-coloring the best
algorithmic lower bound is $c=4.03$~\cite{AchlioptasMoore03}.  An important
and interesting open question~\cite{FrMcDi} is the existence of an
$\epsilon>0$ and of a polynomial algorithm which $q$-colors almost surely a
random graph of connectivity $c=(1+\varepsilon) q\log q$ for arbitrary large
$q$.

\begin{figure}[h]
\begin{center}
  \epsfxsize=17.5cm\epsfbox{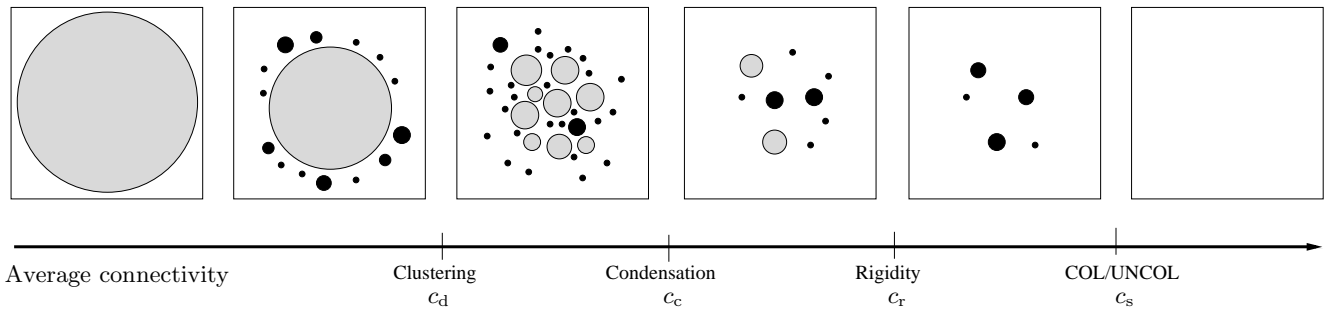} 
  \put(-340,-7){$c_{\rm d}$}
  \put(-251,-7){$c_{\rm c}$}
  \put(-166,-7){$c_{\rm r}$}
  \put(-80,-7){$c_{\rm s}$}
  \put(-500,1){Average connectivity}
\end{center}
\caption{\label{fig:zero_T_3} A sketch of the set of solutions when the
  average connectivity (degree) is increased. At low connectivities (on the left), all solutions are in a
  single cluster. For larger $c$, clusters of solutions appear but the single
  giant cluster still exists and dominates the measure.  At the
  dynamic/clustering transition $c_d$, the phase space slits in an exponential
  number of clusters. At the condensation/Kauzmann transition $c_c$, the measure
  condenses over the largest of them. Finally, no solutions exist above the
  COL/UNCOL transition $c_s$. The rigidity/freezing transition $c_r$ (which might come before or after the condensation transition) takes place when the
  dominating clusters start to contain frozen variables (dominating clusters
  is a minimal set of clusters such that it covers almost all proper
  colorings).  The clusters containing frozen variables are colored in black
  and those which do not are colored in grey.}
\end{figure}

The sharp coloring threshold and the onset of hardness in its vicinity has
also triggered a lot of interest within the statistical physics community
following the discovery of a close relation between constraint satisfaction
problems and spin glasses~\cite{NatureMonasson,Science}. In physical terms,
solving a CSP consists in finding groundstates of zero energy.  The limit of
infinitely large graphs corresponds to the thermodynamic limit.  In the case
of the $q-$coloring problem, for instance, one studies the zero temperature
behavior of the anti-ferromagnetic Potts model~\cite{WU}.  Using this
correspondence, a powerful heuristic tool called the {\it cavity method} has
been developed~\cite{Science,MP99,MP03,MeZe}; it allows an exact analytical
study of the CSPs on sparse random graphs. Unfortunately, some pieces are
still missing to make the cavity method a rigorous tool although many of its
results were rigorously proven.  The cavity method is equivalent to the
famous replica method~\cite{MPVkniha} (and unfortunately for clarity, it has
also inherited some of its notations, as we shall see).

The cavity method and the statistical physics approach have been used to study
the $q$-coloring of random graphs in~\cite{Coloring,ColoringFlo,ColoringSaad}.
The coloring threshold $c_s$ was calculated~\cite{Coloring}, the
self-consistency of the solution checked~\cite{ColoringFlo} and the large $q$
asymptotics of the coloring threshold computed~\cite{ColoringFlo}.  These
results are believed to be {\it exact} although proving their validity
rigorously remains a major subject in the field. Nevertheless ---as the
results obtained for the random satisfiability
problem~\cite{MeZe,MeZeII,MPR03}--- they agree perfectly with rigorous
mathematical
bounds~\cite{Lu,Nature,DuboisMandler,AchlioptasMoore03,AchlioptasRegular,Wormlad1,Wormlad2},
and with numerical simulations. The coloring threshold is thus fairly well
understood, at least at the level of cavity method.

A maybe more interesting outcome of the statistical physics analysis of the
CSPs was the identification of a new transition which concerns the structure
of the set of solutions, and that appears before the coloring
threshold~\cite{GiulioRemiMartin,Science,MeZe}: while at low connectivities
all solutions are in a single pure state (cluster)
\footnote{In this paper we use the words "cluster" and "state" as synonyms.}, 
the set of solutions 
splits in an exponential number of different states (clusters) at 
a connectivity strictly smaller than $c_s$. 
Roughly speaking, clusters are groups of near-by solutions that are 
in some sense disconnected from each other.
Recently,
the existence of the clustered phase was proven rigorously in some cases for
the satisfiability problem~\cite{Mora,Fede}.  A major step was made
by applying the cavity equations on a single instance: this led to the
development of a very efficient message-passing algorithm called {\it Survey
  Propagation} (SP) that was originally used for the satisfaction problem
in~\cite{Science} and later adapted for the coloring problem
in~\cite{Coloring}. Survey propagation allows one to find solutions of large
random instances even in the clustered phase and very near to the coloring
threshold.

Despite all this success, the cavity description of the clustered phase was
not complete in many aspects, and a first improvement
has been made with the introduction of a refined zero temperature cavity
formalism that allows a more detailed description of the geometrical
properties of the clusters~\cite{ColoringMarc,OlivierThesis}. We pursue in
this direction and give for the first time a detailed characterization of the
structure of the set of solutions. We observed in particular that the
clustering threshold was not correctly computed, that other important
transitions were overlooked and the global picture was mixed up.
The corrected picture that we describe in this paper is the following: 
when the connectivity is increased, the set of solutions undergoes several
phase transitions similar to those observed in mean field structural glasses
(we sketch these successive transitions in fig.~\ref{fig:zero_T_3}). First,
the phase space of solutions decomposes into an exponential number of states
which are entropically negligible with respect to one large cluster.  Then, at
the clustering threshold $c_d$, even this large state decomposes into an
exponential number of smaller states.  Subsequently, above the condensation
threshold $c_c$, most of the solutions are found in a finite number of the
largest states.  Eventually, the connectivity $c_s$ is reached beyond which no
more solutions exist.  Another important transition, that we refer to as {\it
  rigidity}, takes place at $c_r$ when a finite fraction of frozen variables
appears inside the dominant pure states (those containing almost all the
solutions).
All those transitions are sharp and we computed the values of the corresponding critical connectivities.

A nontrivial ergodicity breaking takes place at the clustering transition, 
in consequence uniform sampling of solutions becomes hard. 
On the other hand, clustering itself is not responsible for the hardness 
of finding a solution.  
Moreover, until the condensation transition many results obtained by
neglecting the clustering effect are correct. In particular for all $c<c_c$:
i) the number of solutions is correctly given by the annealed entropy (and,
for general CSP, by the replica symmetric entropy), and ii) simple message
passing algorithm such as Belief Propagation (BP)~\cite{Yedidia,BP-Pearl}
converges to a set of exact marginals ({\it i.e.} the probability that a given
node takes a given color). In consequence we can use BP plus a decimation-like
strategy to find proper colorings on a given graph.
Finally we give some arguments to explain why the rigidity transition 
is a better candidate for the onset of computational hardness in finding solutions.

Our results are obtained within the one-step replica symmetry breaking
approach, and we believe (and argue partially in section \ref{sec:cavityRSB}),
that our results would not be modified by 
considering further steps of RSB (as opposed to previous conclusions \cite{ColoringFlo}).

A shorter and partial version of our results, together with a study of similar
issues in the satisfiability problem, was already published in~\cite{US}. We
refer to~\cite{THEM} for a detailed discussion of the satisfiability problem.
The paper is organized as follows: In section~\ref{sec:Model}, we present the
model. In section~\ref{sec:cavityRS}, we introduce the cavity formalism at the
so-called replica symmetric level, and discuss in detail why and where this
approach fails.  In section~\ref{sec:cavityRSB} we take into account the
existence of clusters of solutions and employ the so called one-step
replica-symmetry breaking formalism to describe the properties of clusters.
The results for several ensembles of random graphs are then presented in
section~\ref{sec:result}.  We finally discuss the algorithmic implications of
our findings in section~\ref{sec:algo} and conclude by a general discussion.
Some appendixes with detailed computations complete the paper.

\section{The Model}
\label{sec:Model}

\subsection{Definition of the model}
For the statistical physics analysis of the $q-$coloring
problem~\cite{ColoringSaad,Kanter,Coloring} we consider a Potts~\cite{WU} spin
model with anti-ferromagnetic interactions where each variable $s_i$ (spin,
node, vertex) is in one of the $q$ different states (colors) $s=1,\dots,q$.
Consider a graph $G = ({\cal V,E})$ defined by its vertices ${\cal
  V}=\{1,\dots,N\}$ and edges $(i,j)\in {\cal E}$ which connect pairs of
vertices $i,j\in {\cal V}$; we write the Hamiltonian as
\be
{\cal H}(\{s\}) = \sum_{(i,j) \in {\cal E}} \delta(s_i,s_j)\, .
\label{Ham}
\ee
With this choice there is no energy contribution for neighbors with different
colors, but a positive contribution otherwise. The ground state energy is thus
zero if and only if the graph is $q$-colorable. 
The Hamiltonian leads to a Gibbs measure~\cite{GIBBS} over configurations
(where $\beta$ is the inverse temperature) :
\be
    \mu(\{s\}) = \frac{1}{Z_0} \e^{-\beta {\cal H}(\{s\}) }\, ,
    \label{Gibbs}
\ee
In the zero temperature limit, where $\beta \to \infty$, this measure becomes
uniform over all the proper colorings of the graph.

\subsection{Ensembles of Random Graphs}
We will consider ensembles of graphs that 
are given by a degree distribution ${\cal Q}(k)$. The required property of 
${\cal Q}(k)$
is that its parameters should not depend on the size of the graph. 
All the analytical results will concern only very large sparse graphs
($N\to\infty$). Provided the second and higher moments of ${\cal Q}(k)$ do not
diverge, such graphs are locally tree-like in this limit \cite{Bo,Janson}. More
precisely, call a $d$-neighborhood of a node $i$ the set of nodes which are at
distance at most $d$ from $i$. For $d$ arbitrary but finite the
$d$-neighborhood is almost surely a tree graph when $N\to\infty$. This
property is connected with the fact that the length $\ell$ of the shortest
loops (up to a finite number of them) scales with the graph size as $\ell \sim
\log(N)$.
We will consider the two following canonical degree distribution functions:
\begin{enumerate}
\item[(i)] Uniform degree distribution, ${\cal Q}(k)=\delta(k-c)$,
  corresponding to the {\it $c$-regular random graphs}, where every vertex has
  exactly $c$ neighbors.
\item[(ii)] Poissonian degree distribution, ${\cal Q}(k) = \e^{-c} c^k/ k!$,
  corresponding to the Erd\H{o}s and R\'enyi (ER) random graphs~\cite{Erdos}.
  A simple way to generate graphs with $N$ vertices from this ensemble is to
  consider that each link is present with probability $c/(N-1)$. The binomial
  degree distribution converges to the Poissonian in the large size limit.
\end{enumerate}
It will turn out that the cavity technics simplify considerably for the
regular graphs. However, 
ER graphs have the advantage that their average connectivity is a real number
that can be continuously tuned, which is obviously very convenient when one
wants to study phase transitions. It is thus useful to introduce a third
ensemble, which still has the computational advantage of regular graphs, but
that at the same time gives more freedom to vary the connectivity:
\begin{enumerate}
\item[(iii)] bi-regular random graphs, where nodes with connectivity $c_1$ are
  all connected to nodes with connectivity $c_2$, and vice-versa. There are
  thus two sets of nodes with degree distributions ${\cal
    Q}(k)=\delta(k-c_1)$ and ${\cal Q}(k)=\delta(k-c_2)$.
\end{enumerate}
Notice that these graphs are bipartite by definition and therefore have
always a trivial $2$-coloring which we will have to dismiss in the
following.  This can be easily done within the cavity formalism (it is
equivalent to neglecting the ordered ``crystal'' phase in glass
models~\cite{GLASSFERRO}).

In the first two cases, the parameter $c$ plays the role of the average
connectivity, $c=\sum_k k {\cal Q}(k)$.  In the cavity approach, a very
important quantity is the {\it excess degree} distribution ${\cal Q}_1(k)$,
{\it i.e.} the distribution of the number of neighbors, different from $j$, of a
vertex $i$ adjacent to a random edge $(ij)$:
\begin{equation}
  \label{eq:neighbors}
  {\cal Q}_1(k) = \frac {(k+1) {\cal Q}(k+1)} c \ .
\end{equation}
This distribution remains Poissonian for Erd\H{o}s-R\'enyi graphs, whereas the
excess degree is equal to $c-1$ in the case of regular graphs. In the
bi-regular case, there are two sets of nodes with excess degrees $c1-1$ and
$c2-1$.

\section{The cavity formalism at the replica symmetric level}
\label{sec:cavityRS}

We start by reviewing the {\it replica symmetric} (RS) version of the cavity
method~\cite{MP99, MP03} and its implications for the coloring problem. 
In the last part of the section we show when, and why, 
the RS approach fails.
 

\subsection{The replica symmetric cavity equations}
\label{sec:RSeq}

The coloring problem on a tree is solved exactly by an iterative method called
the belief propagation algorithm~\cite{BP-Pearl} (note some boundary
conditions have to be imposed, otherwise a tree is always 2-colorable) that is 
equivalent to the replica symmetric cavity method~\cite{Yedidia}. At this level, 
the method is in fact a classical tool in statistical physics to deal with that tree
structure that dates back to the original ideas of Bethe, Peierls and
Onsager~\cite{BetheOnsager}. It allows one to compute the marginal
probabilities that a given node takes a given color
and, in the language of statistical physics,
observables like energy, entropy, average magnetization, etc.  The applicability
of the method goes however beyond tree graphs and we will discuss when it is correct
for random locally tree-like graphs in section~\ref{RS_stab}.

\begin{figure}[!ht]
\begin{center}
  \epsfxsize=10.cm\epsfbox{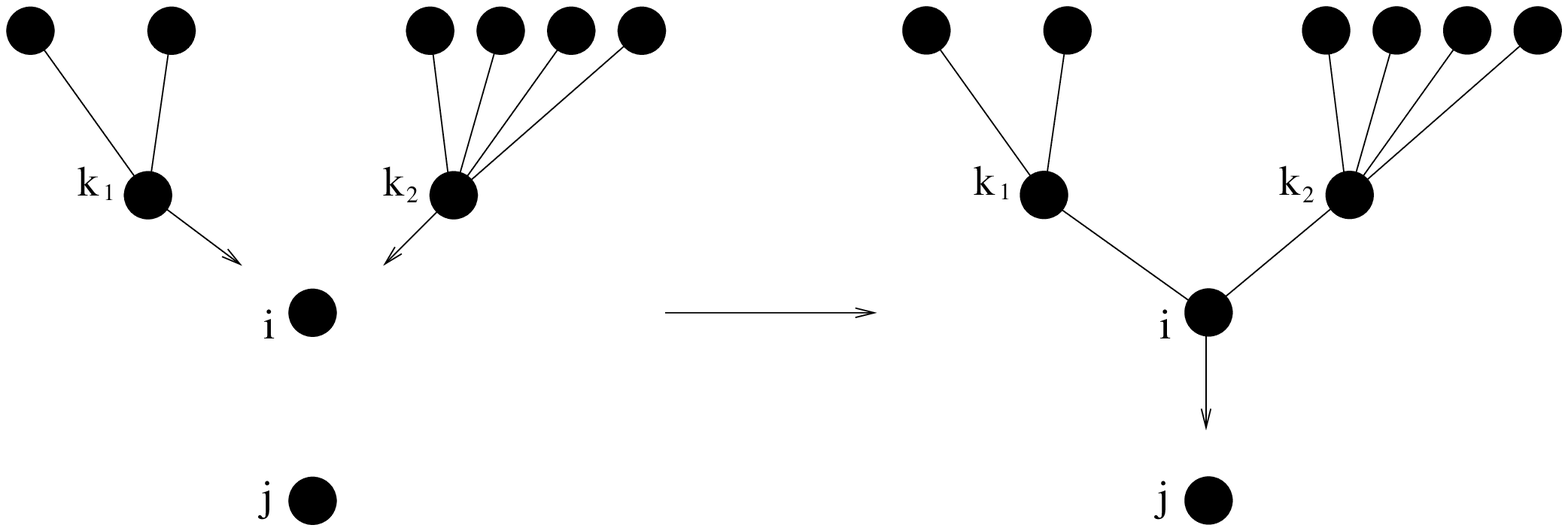}
\end{center}
\caption{\label{fig:iter} Iterative  construction of a tree by adding a new Potts spin.}
\end{figure}

Let us now describe the RS cavity 
equations. Denote $\psi^{i\to j}_{s_i}$ the probability that the spin $i$ has
color $s_i$ when the edge $(ij)$ is not present and consider the iterative
construction of a tree in fig.~\ref{fig:iter}. The probability follows
a recursion:
\begin{eqnarray}
     \psi_{s_i}^{i\to j} = \frac{1}{Z_0^{i\to j}} \prod_{k\in i - j}
     \sum_{s_k} \e^{-\beta \delta_{s_i s_k}} \psi_{s_k}^{k\to i} 
  = \frac{1}{Z_0^{i\to j}} \prod_{k\in i - j} \left(1- (1-\e^{-\beta}) \psi_{s_i}^{k\to i} \right). \label{update}
\end{eqnarray}
where $Z_0^{i\to j}$ is a normalization constant (the cavity partition sum)
and $\beta$ the inverse temperature. The notation $ k\in i - j$ means the
set of neighbors of $i$ except $j$.  The normalization $Z_0^{i\to j}$ is
related to the free energy shift after the addition of the node $i$ and the
edges around it, except $(ij)$, as
\begin{equation}
      Z_0^{i\to j} = \e^{-\beta \Delta F^{i\to j} }.       \label{reweighting}
\end{equation}
In the same way, the free energy shift after the addition of the node $i$ and all
the edges around it is
\begin{equation}
      \e^{-\beta \Delta F^{i} }=Z_0^i =\sum_s \prod_{k\in i } \left(1- (1-\e^{-\beta}) \psi_{s}^{k\to i} \right). \label{free_i}
\end{equation}
The $Z_0^i$ is at the same time the normalization of the total probability
that a node $i$ takes color $s_i$ (the marginal of $i$):
\begin{eqnarray}
     \psi_{s_i}^{i} = \frac{1}{Z_0^{i}} \prod_{k\in i} \left(1- (1-\e^{-\beta}) \psi_{s_i}^{k\to i} \right). \label{update_full}
\end{eqnarray}
The free energy shift after the addition of an edge $(ij)$ is
\begin{equation}
     \e^{-\beta \Delta F^{ij} }= Z_0^{ij} = 1- (1-\e^{-\beta})\sum_s \psi_{s}^{j\to i}\psi_{s}^{i\to j} .
  \label{free_ij}
\end{equation}
The free energy density in the thermodynamic limit is then given by (see for instance~\cite{MP99})
\begin{equation}
  f(\beta) = \frac 1N \(( \sum_i \Delta F^{i} - \sum_{ij} \Delta F^{ij}  \)).
  \label{free_en}
\end{equation}
Note that this relation for the free energy is variational, {\it i.e.} that if one
differentiates with respect to $\beta$, then only the {\it explicit
  dependence} needs to be considered (see~\cite{MP99} for details).  The
energy (the number of contradictions) and the corresponding entropy (the
logarithm of the number of solutions with a given energy) densities can be
then computed from the Legendre transform
\begin{equation}
 -\beta f(\beta) = -\beta e + s(e)\, ,
  \label{Legrenge_RS}
\end{equation}
were $f=F/N$, $e=E/B$ and $s=S/N$ are intensive variables.

The learned reader will notice that in some previous works using the cavity
method~\cite{Coloring,ColoringFlo}, these equations were often written for a
different object than the probabilities $\psi$.  Instead, the so-called cavity
magnetic fields $h$ and biases $u$ where considered. The two approaches are
related via
\begin{equation}
        \e^{-\beta h_{s_i}^{i\to j}} \equiv Z_0^{i\to j} \psi_{s_i}^{i\to j} \equiv
        \prod_{k\in i -j} \e^{-\beta u_{s_i}^{k\to i}}. \label{uh}
\end{equation}
Strictly speaking the $\psi$ are ``cavity probabilities'' while $h$ and $u$
are ``cavity fields'', however $\psi$ are sometimes also referred to as cavity
fields (or messages) in the literature, and the reader will thus forgive us if we do so. 

Note that each of the two notations is suitable for performing the zero 
temperature limit in a different way: in the first one we fix the energy 
zero and we obtain the zero temperature BP recursion which gives the marginal 
probabilities for each variable, while in
the second case, we obtain a simpler recursion called {\it Warning 
Propagation} (\ref{WP}) that deals with the energetic 
contributions but neglects the entropic ones \cite{MeZe,Coloring}. This is the origin of the
discrepancy between the RS results of refs.~\cite{ColoringSaad} and
\cite{Coloring}. As we shall see, the differences between these two limits will
be an important point in the paper.

\subsection{Average over the ensemble of graphs and the RS solution}

To compute (quenched) averages of observables over the considered ensemble of
random graphs given by the degree distribution ${\cal Q}(k)$ we need
to solve the self-consistent cavity functional equation
\begin{equation}
     {\cal P}(\psi) = \sum_k {\cal Q}_1(k) \prod_{i=1}^{k-1} \int 
    d\psi^i {\cal P}(\psi^i) \delta({\psi}-{\cal F}(\{\psi^i\})),  \label{RS_pop}
\end{equation}
where ${\cal Q}_1(k)$ is the distribution of the number of neighbors given
that there is already one neighbor, and the function ${\cal F}(\{\psi^i\})$ is
given by eq.(\ref{update}). Beware that $\psi$ here is a $q$-components
vector while we omit the vector notation to lighten the reading. This equation
is quite complicated since the order parameter is nontrivial, but we can solve
it numerically using the population dynamics method described
in~\cite{MP99,MP03}.

Throughout this paper we will search only for the color symmetric functions
${\cal P}(\psi)$, {\it i.e.} invariant under permutation of colors.  Clearly with
this assumption we might miss some solutions of~(\ref{RS_pop}). Consider for
example $q>2$ colors and the ensemble of random bi-regular graphs. Since every
bipartite graph is 2-colorable there are $q(q-1)/2$ corresponding color
asymmetric solutions for ${\cal P}(\psi)$.  
For the ensemble of random graphs considered here, we later argue that this assumption is 
however justified.

Another important observation is that for regular graphs the
equation~(\ref{RS_pop}) crucially simplifies: the solution {\it
  factorizes}~\cite{Factorizes} in the sense that the order parameter $\psi$
is the same for each edge. This is due to the fact that, locally, every edge
in such a regular graph has the same environment. All edges are therefore
equivalent and thus the distribution ${\cal P}(\psi)$ has to be a delta
function. For the bi-regular graphs, the solution of~(\ref{RS_pop}) also
factorizes, but the two sets of nodes of connectivity~$c_1$ and $c_2$ (each of
them being connected to the other) have to be considered separately.

It is immediate to observe that ${\cal P}(\psi)=\delta(\psi-1/q)$ ({\it i.e.} each
of the $q$ components of each cavity field $\psi$ equals $1/q$), is always a
solution of~(\ref{RS_pop}). By analogy with magnetic systems we shall call
this solution {\it paramagnetic}. Numerically, we do not find any other
solution in the colorable phase.  For regular random graphs the paramagnetic
solution is actually the only factorized one.
The number of proper colorings predicted by the RS approach is thus easy to
compute. Since all messages are of the type ${\cal P}(\psi)=\delta(\psi-1/q)$,
the free energy density simply reads
\be -\beta f_{\rm RS} =
\log{q} + \frac c2 \log\((1-\frac{1-e^{-\beta}}{q}\)). 
\label{RS_free}
\ee 
The entropy density at zero temperature thus follows
\be s_{\rm RS} = \log{q} + \frac {c}{2} \log{\((1-\frac{1}{q}\))}\, .\label{S_RS} \ee 
It coincides precisely with the annealed (first moment) entropy. We will see
in the following that, surprisingly, the validity of this formula goes well
beyond the RS phase (actually until the so-called condensation transition).

\subsection{Validity conditions of the replica symmetric solution}
\label{RS_stab}

We used the main assumption of the replica symmetric approach when we wrote
eq.~(\ref{update}): we supposed that the cavity probabilities
$\psi_{s_i}^{k\to i}$ for the neighbors $k$ of the node $i$ are ``sufficiently''
independent in absence of the node $i$, because only then the joint
probability factorizes. This assumption would be true if the lattice 
were a tree with
non-correlated boundary conditions, but loops, or correlations in the
boundaries, may create correlations between the neighbors of node $i$ 
(in absence of $i$) and the RS
cavity assumption might thus cease to be valid in a general graph.
The aim here is to precise and quantify this statement both from a rigorous 
and heuristic point of view.

\subsubsection{The Gibbs measure uniqueness condition}
Proving rigorously the correctness of the RS
cavity assumption for random graphs 
is a crucial step that has not yet been successfully overcome in
most cases. The only success so far was obtained by proving 
a far too strong condition: {\it the Gibbs measure uniqueness}~\cite{Aldous01,BG05,Jon,MSh06}. 
Roughly speaking: the Gibbs measure~(\ref{Gibbs}) is unique if the behavior of a spin $i$ 
is totally independent 
from the boundary conditions ({\it i.e.} very distant spins) for  {\it any} possible boundary
conditions. 
More precisely, let us define $\{s_l\}$ colors of all the spins at distance at least $l$
from the spin $i$. The Gibbs measure $\mu$ is unique if and only if the
following condition holds for every $i$ (and in the limit $N \to
\infty$)
\begin{equation}
     \mathbb{E} \[[ \sup_{\{s_l\},\{s'_l\}} \sum_{s_i=1}^q
     |\mu(s_i|\{s_l\})-\mu(s_i|\{s'_l\})|\]] \mathop{\rightarrow}_{l \to
     \infty} 0\, ,
    \label{Gibbs_uni}
\end{equation}
where the average is over the ensemble of graphs.  In~\cite{BG05,Jon}, it was
proven that the Gibbs measure in the coloring problem on random regular graphs
is unique only for graphs of degree $c<q$. To the best of our
knowledge, this has not been computed for Erd\H{o}s-R\'enyi graphs (later in
this section, we argue on the basis of a physical argument that it should be
$c<q-1$ in this case).

\subsubsection{The Gibbs measure extremality condition}
In many cases, the RS approach is observed to be correct beyond the uniqueness
threshold. It was thus suggested in~\cite{MM05} (see also~\cite{US}) that the
Gibbs measure extremality provides a proper criterion for the correctness of
the replica symmetric assumption. Roughly speaking, the difference between
uniqueness and extremality of a Gibbs measure is that although there may
exist boundary conditions for which the spin $i$ is behaving differently
than for others, such boundary conditions have a
null measure if the extremality condition is fulfilled. Formally (and keeping the notations from the previous section),
the extremality corresponds to
\begin{equation}
  \mathbb{E}\[[ \sum_{\{s_l\}} \mu(\{s_l\}) \sum_{s_i=1}^q |\mu(s_i|\{s_l\})-\mu(s_i)|\]] \mathop{\rightarrow}_{l \to \infty}  0 \, .
  \label{extremality}
\end{equation}
In mathematics the ``extremal Gibbs measure'' is often used as a
synonym for a ``pure state''. Recently, the authors of~\cite{MM05} provided
rigorous bounds for the Gibbs measure extremality of the coloring problem on
trees. 

There exist two heuristic equivalent approaches to check the extremality
condition.  In the first one, one studies the divergence of the so called
``point-to-set'' correlation length~\cite{BiroliBouchaud,MS05}. The second
one is directly related to the cavity formalism: one should check
for the existence of a nontrivial solution of the one-step replica symmetry 
breaking equations (1RSB) at $m=1$ (see section~\ref{sec:result}). 
Both these analogies were remarked
in~\cite{MM05} and exploited in~\cite{US}. We will show in
section~\ref{sec:result} that the extremality condition ceases to be valid at
the clustering threshold $c_d$, beyond which the 1RSB formalism will be
needed.

\subsubsection{The local stability: a simple self-consistency check}
\label{sec:stab}
A necessary, simple to compute but not sufficient, validity condition for the 
RS assumption is the non-divergence of the spin glass
susceptibility (see for instance~\cite{Rivoire}). If it diverges, a spin glass
transition occurs, and the replica symmetry has to be
broken~\cite{MPVkniha}. 
The local stability analysis thus gives an upper bound to the
Gibbs extremality condition, which remarkably coincides with the rigorous upper
bound of \cite{KS66}. This susceptibility is defined as
\be  \chi_{\rm SG} = \frac{1}{N} \sum_{i,j} \langle s_i s_j\rangle^2_c \, . \ee
The connectivities above which it diverges at zero
temperature can be computed exactly within the cavity formalism (we refer to appendix
\ref{app:stab} for the derivation). It follows for regular and Erd\H{o}s-R\'enyi
graphs:
\be
c^{\rm reg}_{\rm RS}= q^2 -2q+2\, \text{,~~~~~~~}
c^{\rm ER}_{\rm RS}= q^2 -2q+1\, ,
\ee
while the stability of the bi-regular graphs of connectivities $c_1,c_2$
is equivalent to regular graphs with $c=1+\sqrt{(c_1-1)(c_2-1)}$.

Note that for regular and ER graphs the RS instability threshold is in the
colorable phase only for $q=3$. Indeed the 5-regular graphs are 3-colorable~\cite{ColoringFlo} (and rigorous results in~\cite{Diaz, Wormlad2}) and exactly critical since $c^{\rm reg}_{\rm RS}(3)=5$, and for ER graphs the
COL/UNCOL transition appears at $c_s \approx 4.69$~\cite{Coloring} while
$c^{\rm ER}_{\rm RS}(3)=4$.  This means that the replica symmetry breaking
transition appears continuously at the point $c_{\rm RS}$ so that above it the RS approach
is not valid anymore. For all $q\ge 4$, however, the local stability
point is found beyond the best upper bound on the coloring threshold for both
regular and ER graphs. In this case, the extremality condition will not be
violated by the continuous mechanism, but we will see that, instead, a discontinuous
phase transition, as happens in mean field structural glasses, will take
place.

Interestingly enough, a similar computation can be made for the ferromagnetic
susceptibility $\chi_{F} = \frac{1}{N} \sum_{i,j} \langle s_i s_j\rangle_c$
(see again appendix~\ref{app:stab}).  It diverges at $c=q$ for regular graph
and $c=q-1$ for Erd\H{o}s-R\'enyi graphs.  This divergence (called {\it
  modulation} instability in~\cite{Rivoire}) would announce the transition
towards an anti-ferromagnetic ordering on a tree, which is however
incompatible with the frustrating loops in a random graph (although such order
might exist on the bi-regular graphs). This is precisely the solution we
dismiss when considering only the color symmetric solution of (\ref{RS_pop}).
Note however that the presence of this instability shows that the problem
ceases to a have a unique Gibbs state (although it is still extremal) as for
some specific (and well-chosen) boundary conditions, an anti-ferromagnetic
solution may appear.  Indeed it coincides perfectly with the rigorous
uniqueness condition $c=q$ for regular graph, and suggests strongly that the
uniqueness threshold (or at least an upper bound) for Erd\H{o}s-R\'enyi graphs
is $c=q-1$.

\section{One-step replica symmetry breaking framework}
\label{sec:cavityRSB}

So far we described the RS cavity method for coloring random graphs and
explained that the extremality of the Gibbs measure gives a validity
criterion. We now describe the one-step replica symmetry breaking 
cavity solution~\cite{MP99,MP03}.  In this approach, the
non-extremality of the Gibbs measure is cured, by decomposing it into several parts
(pure states, clusters) in such a way that within each of the states the
Gibbs measure becomes again extremal. 

This decomposition has many elements/states, not just a finite numbers like
the $q$-states of the usual ferromagnetic Potts models. It is actually found
that the number of pure states is growing exponentially with the size of the
system. Let us define the state-entropy function $\Sigma(f)$ ---called the
{\it complexity}--- which is just the logarithm of the number of states with
internal free energy density $f$, {\it i.e.} ${\cal{N}}(f)= \exp[N\Sigma(f)]$.
In the glass transition formalism, this complexity is usually referred to as
the {\it configurational entropy}.  Dealing with exponentially many pure
states is obviously a nightmare for all known rigorous approaches to the
thermodynamic limit. The heuristic cavity method overcomes this problem
elegantly, as was shown originally in the seminal work of~\cite{MP99,MP03}.

Another very useful intuition about the 1RSB cavity method comes from the
identification of states $\alpha$ with the fixed points $\{\psi\}$ of the
belief propagation equations (\ref{update}). The goal is thus to compute the
statistical properties of these fixed points. Each of the states is weighted
by the corresponding free energy (\ref{free_en}) to the power $m$, where $m$
is just a parameter analogous to the inverse temperature $\beta$ (in the
literature $m$ is often called the Parisi replica symmetry breaking
parameter~\cite{Parisi80, MPVkniha}).  The probability measure over states
$\{\psi\}$ is then
\begin{equation}  
    \tilde\mu(\{\psi\}) = \frac{Z_0(\{\psi\})^m}{Z_1} =\frac{1}{Z_1} e^{-\beta m N f(\{\psi\})}\, , \label{Gibbs_1RSB}
\end{equation}
where $Z_1$ is just the normalization constant. To write the analog of the
belief propagation equations we need to define the probability (distribution)
$P^{i\to j}(\psi^{i\to j})$ of the fields $\psi^{i \to j}$. This can be
obtained from those of incoming fields as
\begin{eqnarray}
      P^{i\to j}(\psi_{s_i}^{i\to j}) &=& \frac{1}{Z_1^{i\to j}} 
      \prod_{k\in i-j} \int d \psi_{s_i}^{k\to i}  P^{k\to i}(\psi_{s_i}^{k\to i}) \delta(\psi_{s_i}^{i\to j}- {\cal F}(\{\psi_{s_i}^{k\to i}\}) ) 
      \left( Z_0^{i\to j}\right)^m, \nonumber \\  
      &\equiv& \frac{1}{Z_1^{i\to j}} \int_{\POP} \delta(\psi - {\cal F})
      \left( Z_0^{i\to j}\right)^m \, .  \label{1RSB}
\end{eqnarray}
The function ${\cal F}$ is given by eq.~(\ref{update}) and the delta function
ensures that the set of fields $\psi^{i \to j}$ is a fixed point of the belief
propagation (\ref{update}). The re-weighting term $\left( Z_0^{i\to
    j}\right)^m$ takes into account the change of the free energy of a state
after the addition of a cavity spin $i$ and its adjacent edges except $(ij)$,
as defined in eq. (\ref{reweighting}). This term appears for the same reason as
a Boltzmann factor $e^{-\beta \delta_{s_i,s_k}}$ in eq.~(\ref{update}): it
ensures that the state $\alpha$ is weighted by $(Z_\alpha)^m$ in the same way
a configuration $\{s\}$ is weighted by $e^{-\beta {\cal H}(\{s\})}$ in
(\ref{Gibbs}).  Finally $Z_1^{i\to j}$ is a normalization constant. 
In the second line of (\ref{1RSB}) we introduced an abbreviation that will be
used from now on to make the equations more easily readable. The notation
\POP\, comes from ``population dynamics'' which refers to the numerical
method we use to solve eq.~(\ref{1RSB}). The probability distribution
$P(\psi)$ can be represented numerically 
by a set of fields taken from $P(\psi)$, and then the
probability measure $ P(\psi){\rm d}\psi$ becomes uniform sampling from this
set, for more details see appendix~\ref{app:num}.

We define the ``replicated free energy'' and compute it in analogy with
eq.~(\ref{free_en}) as
\begin{equation}
\Phi(\beta, m) \equiv  - \frac{1}{\beta m N} \log(Z_1) = \frac 1N \(( \sum_i \Delta \Phi^{i} - \sum_{ij} \Delta
\Phi^{ij}\))\, ,  \label{Psi}
\end{equation}
where
\begin{equation}
      \e^{-\beta m \Delta \Phi^i} = \int_{\POP}   \e^{-\beta m \Delta F^i}\, , \text{~~~~}
      \e^{-\beta m \Delta \Phi^{ij}} = \int_{\POP}   \e^{-\beta m \Delta F^{ij}}.
\end{equation}
Putting together (\ref{Gibbs_1RSB}) and (\ref{Psi}) we have
\begin{equation}
Z_1 = \e^{-\beta m N \Phi(\beta, m)} = \sum_{\{\psi \}} e^{-\beta m N f(\{\psi\})} = \int_{f} {\rm d}f\,  \e^{ - N\left[ \beta m f (\beta) - \Sigma(f)\right]},
\label{saddle}
\end{equation}
where the sum over $\{ \psi \}$ is over all states (or BP fixed
points). In the interpretation of~\cite{Remi} $m$ is the number
of replicas of the system, thus the name ``replicated free energy'' for
$\Phi(\beta,m)$. Note that we are using the word ``replica'' only to refer
to the established terminology as no replicas are needed within the cavity formalism.
From the saddle point method, it follows that the Legendre transform of
complexity function $\Sigma(f)$ gives the replicated free energy $\Phi(m)$
\begin{equation}
-\beta m \Phi(\beta, m) = -\beta m f(\beta) + \Sigma(f). \label{comp}
\end{equation}
Notice that this equation is correct only in the highest order in the system
size $N$, {\it i.e}. in densities and at the thermodynamic limit.  From the
properties of the Legendre transform we have
\begin{eqnarray}
\Sigma = \beta m^2 \partial_m \Phi(\beta,m)\; , \qquad
f = \partial_m[m\, \Phi(\beta,m)]\;  , \qquad 
\beta m = \partial_f \Sigma(f)  .\label{Legendre}
\end{eqnarray} 
Thus, from eq.(\ref{Psi}), the free energy reads
\begin{eqnarray}
      f(\beta) = 
   \sum_i \frac{\int_{\POP}  \Delta F^i \e^{-\beta m \Delta F^i}}{ \int_{\POP}   \e^{-\beta m \Delta F^i}} - \sum_{ij} \frac{\int_{\POP}  \Delta F^{ij} \e^{-\beta m \Delta F^{ij}}}{\int_{\POP}   \e^{-\beta m \Delta F^{ij}}}\, . \label{free_energy}
\end{eqnarray}

When the parameter $m$ is equal to one (the number of replicas is actually one in the approach of~\cite{Remi}), then $-\beta \Phi(\beta,1)=-\beta
f(\beta) +\Sigma(f)$ reduces to the usual free energy function considered in
the RS approximation
\be \Phi(\beta,1)= e - \frac{\Sigma+s}{\beta}=e-T s_{\rm tot} \, ,
\label{phi_m1}
\ee 
where $s$ in the internal entropy density of the corresponding clusters and
$s_{\rm tot}$ the total entropy density of the system.

\subsection{Analyzing the 1RSB equations}

Combining (\ref{Psi}) and 
(\ref{free_energy}) we can compute $\Sigma$ and $f$ for each value $m$, that gives
us implicitly $\Sigma(f)$. To compute the thermodynamic observables in the
model we have to minimize the total free energy $f_{\rm tot}= f(\beta) -
\Sigma /\beta$ over such values of $f$ where the complexity $\Sigma(f)$ is
non-negative (so that the states exist in the thermodynamic limit). The
minimum of the total free energy corresponds to a value of parameter $m=m^*$
and states with the free energy $f^*$ dominate the thermodynamics.
Three different cases are then observed:
\begin{enumerate}
\item[a)] If there is only the trivial (replica symmetric) solution at $m=1$, 
  then the Gibbs measure (\ref{Gibbs}) is extremal and the replica symmetric 
  approach is correct. If at the same time a nontrivial solution exists 
  for some $m\neq 1$, then the clusters corresponding to this solution have
  no influence on the thermodynamics.
\item[b)] If there is a nontrivial solution at $m=1$ with a positive
  complexity, then $m^*=1$ minimizes the total free energy. 
   The system is in a {\it clustered phase} with an exponential number 
   of dominating states.
\item[c)] If however the complexity is negative at $m=1$, then the
  corresponding states are absent with probability one in the thermodynamic
  limit. Instead the total entropy is dominated by clusters corresponding to 
  $m^*$ such that $\Sigma(m^*)=0$: the system is in 
  a {\it condensed phase}. Note that the condition $\Sigma(m^*)=0$ corresponds
  to the maximum of the replicated free energy (\ref{Psi}).
\end{enumerate}

The transitions between these cases are well known in structural glass
phenomenology where they appear when the temperature is
lowered~\cite{REM,GLASSTHEORY}. The transition from the paramagnetic phase to
the clustered one is usually referred to as the {\it dynamical transition}
\cite{MS06} or the {\it clustering transition}.  It is not a true
thermodynamic transition as the total free energy of the system at $m^*=1$ is
still equal to the replica symmetric one (\ref{free_en}) (see appendix
\ref{app:m1}) and thus is an analytical function of connectivity. 
However, the phase space is broken into exponentially many
components and, as a consequence, the dynamics fall out-of-equilibrium beyond
this transition.

The second transition from the clustered to the condensed phase is, however, a
genuine thermodynamic transition (the free energy has a discontinuity 
in the second derivative at $c_c$) and is called the {\it replica symmetry
  breaking transition}, or the static {\it glass transition}. At this point
the measure condenses into few clusters, and we shall call it the {\it
  condensation transition}. In structural glasses, it corresponds to the well
known {\it Kauzmann transition} \cite{Kauzmann}. The sizes of the clusters in
the condensed phase follow the so called Poisson-Dirichlet process which is
discussed shortly in appendix~\ref{app:PD}.

The procedure to compute the replicated free-energy (\ref{Psi}) and the
related observables was described above for a single large random graph. To
compute the averages over the ensemble of random graphs, we need to solve an
equation analogous to eq.~(\ref{RS_pop})
\begin{equation}
     {\cal P}[P(\psi)] = \sum_k {\cal Q}_1(k) \prod_{i=1}^{k-1} \int 
    d P^i(\psi^i) \,  {\cal P}[P^i(\psi^i)] \, \delta(P({\psi})-{\cal F}_2(\{P^i(\psi^i)\})),  \label{1RSB_pop}
\end{equation}
where the functional ${\cal F}_2$ is given by eq.~(\ref{1RSB}). Solving this
equation for a general ensemble of random graphs and a general parameter $m$ is a
numerically quite tedious problem. In the population dynamics
algorithm~\cite{MP99,MP03} we need to deal with a population of populations of
q-components fields.  It is much more convenient to look at the ensemble of
random regular graphs where a factorized solution ${\cal
  P}[P(\psi)]=\delta(P(\psi)-P_0(\psi))$ must exists. Then we are left with
only one functional equation (\ref{1RSB}).

Before discussing the zero temperature limit, we would like to point out that
there exists another very important case in which eq.~(\ref{1RSB_pop})
simplifies. For $m=1$, as first remarked and proved in~\cite{MM05},
when dealing with the problem of reconstruction on trees, the equations can be written (and numerically
solved) in a much simpler way. We again refer to the appendix~\ref{app:m1} for
details.  Especially for the Poissonian random graphs this simplification is
very useful.

\subsection{Zero temperature limit}
\label{twoT}

We now consider the zero temperature limit $\beta \to \infty$ of the 1RSB
equations (\ref{comp})-(\ref{1RSB_pop}) to study
the coloring problem. In most of the previous works~\cite{MP03, MeZe, Coloring}
the {\it energetic zero temperature limit} was employed. The $\beta \to
\infty$ limit of eq. (\ref{comp}) was taken in such a way that $m \beta=y$
remains constant.  The replicated free energy (\ref{comp}) then becomes
\be 
    -y \Phi_e(y) = -y e + \Sigma(e) \, . \label{energetic}
\ee 
It is within this approach that the survey propagation (SP) algorithm was
derived. The connectivity at which the complexity function $\Sigma(e=0)$
becomes negative is the coloring threshold. Above this connectivity
$\Sigma(e)$ was used to compute the minimal number of violated constraints
(the ground state energy). The reweighting in eq. (\ref{1RSB})
becomes $\e^{-y\Delta E^{i\to j}}$, and when $y\to \infty$ all the
configurations with positive energy are forbidden.

In this paper we adopt the {\it entropic zero temperature limit}, suggested
originally in~\cite{ColoringMarc,OlivierThesis}. The difference in the two approaches was
already underlined in sec.~\ref{sec:RSeq}. We want to study the structure
of proper colorings, {\it i.e.} the configurations of zero energy, and we thus fix
the energy to zero. Then we obtain the entropy by considering $-\beta f = s$ and
introduce a {\it free entropy} ---or in replica term a ``replicated
entropy''--- as $\Phi_s(m)=-\beta m \Phi(\beta,m)|_{\beta \to \infty}$. Eq.
(\ref{comp}) then becomes
\begin{equation}
  \Phi_s(m) = m s + \Sigma(s). \label{Sigma_ent}
\end{equation}
The belief propagation update (\ref{update}) becomes
\begin{eqnarray}
  \psi_{s_i}^{i\to j} = \frac{1}{Z_0^{i\to j}} \prod_{k\in i - j} \left(1- \psi_{s_i}^{k\to i} \right), \label{update_zero}
\end{eqnarray}
while the 1RSB equation (\ref{1RSB}) keeps the same expression (and thus the
same computational complexity).

The partition sum $Z_0$ in (\ref{Gibbs}) becomes in this limit the number of proper
colorings or solutions. The clusters are now sets of such solutions, and
are weighted by their size to the power $m$. The free entropy $\Phi_s(m)$ is
then computed as
\bea 
\Phi_s(m) = \frac 1N \(( \sum_i \Delta \Phi_s^{i} - \sum_{ij}
\Delta \Phi_s^{ij}\)) = \frac 1N \(( 
\sum_i \log { \int_{\POP}  \((\Delta Z^i\))^m} - \sum_{ij} \log {\int_{\POP}
  \((\Delta Z^{ij}\))^m} \))\, , \label{free_entropy}
\eea
where $\Delta Z^i$ and $\Delta Z^{ij}$ are given by eqs.~(\ref{reweighting})
and (\ref{free_i}).  The analysis from the previous section is valid also for
the entropic zero temperature limit. The information extracted from the number
of clusters of a given size, $\Sigma(s)$, is one of the main results of this paper and
will be discussed and interpreted in section \ref{sec:result}.

\subsection{The role of frozen variables}
\label{frozen}

In this section we discuss the presence and the role of the frozen variables and
explain the connection between the energetic and the entropic zero temperature
limits. This allows us to revisit (and extend) the survey propagation
equations.  Remember that the components of the cavity field $\psi^{i\to
  j}_{s_i}$ are the probabilities that the node $i$ takes the color $s_i$ when
the constraint on the edge $ij$ is not present. In the zero temperature limit
we can classify them in two categories:
\begin{enumerate}
\item[(i)] A {\it hard field} corresponds to the case when all components of
  $\psi^{i\to j}$ are zero except one, $s$.  Then only that color is allowed
  for the spin $i$, in absence of edge $(ij)$. 
\item[(ii)] A {\it soft field} corresponds to the case when more than one
  component of $\psi^{i\to j}_{s_i}$ is nonzero. The variable $i$ is thus not
  frozen in absence of edge $(ij)$, and the colors of all the nonzero
  components are allowed.
\end{enumerate}

This distinction is also meaningful for the full probabilities $\psi^i_{s_i}$
(\ref{update}), if $\psi^i_{s_i}$ is a hard field then the variable $i$ is
frozen.  In the colorable region there cannot exist a finite fraction of
frozen variables (even if we consider properly the permutational symmetry of
colors) since by adding a link the connectivity changes by $1/N$ but the
probability of becoming uncolorable would be finite.  On the contrary, in the 1RSB
picture, we observe that 
a finite fraction of variables can be frozen within a single cluster. In
other words, in all the solutions that belong to this given cluster a finite fraction of
variables can take one color only. By adding a link into the graph, the
connectivity grows by $1/N$, and there is a finite probability that a cluster
with frozen variables disappears.  The distinction between hard and soft
fields is useful not only for the intuition about clusters, but also for the
analysis of the cavity equations and it also leads to the survey propagation
algorithm.

The distribution of fields over states $P^{i\to j}(\psi^{i\to j})$ 
(\ref{1RSB}) can be decomposed into the hard- and soft-field 
parts
\begin{equation}
    P^{i\to j}(\psi^{i\to j}) = \sum_{s=1}^q \eta^{i\to j}_s 
    \delta(\psi^{i\to j}-r_s) + \eta^{i\to j}_0 
    \tilde P^{i\to j}(\psi^{i\to j})\, ,
\end{equation}
where $\tilde P^{i\to j}$ is the distribution of the soft fields and 
the normalization is $\sum_{s=0}^q \eta^{i\to j}_s =1$.

Interestingly, the presence of frozen variables in the entropically dominating
clusters is connected to the divergence of the size of average minimal
rearrangement~\cite{MS05,Guilhem_rear}. Precisely, choose a random proper
coloring $\{s\}$ and a random node $i$ in the graph. The average minimal rearrangement is 
 the Hamming distance to the nearest solution in which node $i$ has a color different from
$s_i$ averaged over the nodes $i$, the proper colorings, and graphs in the ensemble.

Another interesting role of the frozen variables arises within the {\it
  whitening procedure}, introduced in~\cite{ParisiWhit} and studied, between
others, for the satisfiability problem in~\cite{Elitza,Riccardo}. This
procedure is equivalent to the {\it warning propagation} (WP) 
update (\ref{WP}) which we outlined in sec.~\ref{sec:RSeq}.
Whitening is able to identify if a solution belongs to a cluster with 
frozen variables or not. Particularly, the result of the whitening is a 
set of hard cavity fields.


Since the survey propagation algorithm is computing statistics over the states
that contain hard fields, then the solution found after decimating
the survey propagation result should a priori also contain hard fields. However,
recent works show that if one applies the whitening procedure starting from
solutions found by SP on large graphs, whitening converges every time to the trivial fixed point (see 
detailed studies for K-SAT in~\cite{Elitza,Riccardo}). A possible solution to
this apparent paradox is discussed in sec.~\ref{sec:algo}.

\subsubsection{Hard fields in the simplest case, $m=0$}
\label{hard_0}

Let us now consider the survey propagation equations originally derived
in~\cite{Coloring} from the energetic zero temperature limit (\ref{energetic})
when $y\to \infty$. For simplicity we will write them only for the 3-coloring.
We consider the 1RSB cavity equation (\ref{1RSB}) for $m \to 0$, then the
reweighting factor $(Z_0^{i\to j})^m$ is equal to zero when the arriving hard
fields are contradictory, and equal to one otherwise. The update of
probability $\eta_s$ that a field is frozen in direction $s$ is then written from
eq.(\ref{1RSB}):
\begin{equation}
    \eta^{i \to j}_s= \frac{
    \prod_{k\in i - j} (1- \eta^{k\to i}_s) - \sum_{p \neq s} \prod_{k\in i - j} (\eta^{k\to i}_0 + \eta^{k\to i}_p) +  \prod_{k\in i - j} \eta^{k\to i}_0
     }{
  \sum_{p} \prod_{k\in i - j} (1- \eta^{k\to i}_p) -   \sum_{p} \prod_{k\in i - j} (\eta^{k\to i}_0 + \eta^{k\to i}_p) +  \prod_{k\in i - j} \eta^{k\to i}_0
     } \, .\label{SP_color}
\end{equation}
In the numerator there is a telescopic sum counting the probability that color
$s$ and only color $s$ is not forbidden by the incoming fields. In the
denominator the telescopic sum is counting the probability that there is at
least one color which is not forbidden. If we do not want to actually find a
proper coloring on a single graph but just to compute the replicated free
energy/entropy, we can further simplify eq. (\ref{SP_color}) by imposing the
color symmetry.  Indeed, the probability that in a given state a field is hard
in direction of a color $s$ has to be independent of $s$ (except $s=0$ which
corresponds to a soft field).  Then (\ref{SP_color}) becomes, now for
general number of colors $q$:
\begin{equation}
         \eta^{i \to j} = w(\{ \eta^{k\to i} \}) =
         \frac{ \sum_{l=0}^{q-1} (-1)^l {q-1 \choose l} 
         \prod_{k\in i-j} \left[1-(l+1)\eta^{k\to i} \right]
         }{ \sum_{l=0}^{q-1} (-1)^l {q \choose l+1} 
         \prod_{k\in i-j} \left[ 1-(l+1)\eta^{k\to i} \right] } 
         \label{SP}\, .
\end{equation}

Note that since $\partial \Sigma(s)/\partial s = -m$, the value 
$m=0$ corresponds to the point where the function $\Sigma(s)$ has a zero slope. 
If a nontrivial solution of (\ref{SP}) exists, then $\Sigma(s)|_{m=0}$
is the maximum of the curve $\Sigma(s)$ and is counting the total
log-number of clusters of size $s$, which is due to the exponential dependence, also the total log-number of all clusters, regardless their size.
There are two points that we want to emphasize:  
\begin{itemize}
\item{Suppose that a nontrivial solution of (\ref{SP}) exists, {\it i.e.}
many clusters exist and their number can be computed with the
energetic zero temperature limit calculations. Then the clusters 
might be very small and contain very few solution in comparison to bigger 
less numerous clusters; or in comparison to a giant single cluster which 
might still exist. 
This situation cannot be decided by the energetic formalism that
weights clusters equally independently of their size.}
\item{Suppose, on contrary, that a nontrivial solution of (\ref{SP})
does not exist. It might still well be that many clusters exist, but the
$\Sigma(s)$ curve has no part with zero slope.}
\end{itemize}
We will see that these two cases are actually observed. The energetic method,
that can locate the coloring threshold and from which the survey propagation
can be derived, is therefore not a good tool to study the clustering
transition.

\subsubsection{Generalized survey propagation recursion}
\label{generalized}

Let us compute how the fraction of hard fields $\eta$ evolves after one
iteration of equation (\ref{1RSB}) at general $m$. There are two steps 
in each iterations of (\ref{1RSB}).  
In the first step, $\eta$ iterates via eq. (\ref{SP}). In the
second step we re-weight the fields.
Writing $P^{\rm hard}_m(Z)$ the ---unknown--- distribution of the reweightings $Z^m$  for
the hard fields, one gets
\be \eta^{i \to j} =
\frac{1}{\cal{N}} \int {\rm d}Z \, P^{\rm hard}_m(Z) \,  Z^m  w(\{ \eta^{k\to i} \}) = 
\frac{w(\{ \eta^{k\to i} \})}{\cal{N}} \int {\rm d}Z \, P^{\rm hard}_m(Z)\,  Z^m  =
\frac{ w(\{ \eta^{k\to i} \})}{\cal{N}}\,  \overline{Z^m_{\rm hard}}.
\label{eta_prelim}
\ee
A similar equation can formally be written for the soft fields 
\be 1 - q \eta^{i \to j} =
\frac{ 1-q w(\{ \eta^{k\to i} \})}{\cal{N}}\, \overline{Z^m_{\rm soft}}.
\label{eta_prelim2}
\ee
Writing explicitly the normalization ${\cal N}$, we finally obtain the
generalized survey propagation equations:
\be
\label{hard_m}
 \eta^{i \to j} =
\frac{ w(\{ \eta^{k\to i} \})} {q
w(\{ \eta^{k\to i} \})
+\left[1-q
w(\{ \eta^{k\to i} \})
\right] r(\{ \eta^{k\to i} \})}, \text{~~~~~~~with~~} r(\{ \eta^{k\to i} \})=
\frac {\overline{Z^m_{\rm soft}}}{\overline {Z^m_{\rm      hard}}}\, .
\ee 
In order to do this recursion, the only information needed is the ratio $r$
between between soft- and hard-field reweightings, which is in general 
difficult to compute since it depends on the full distribution of soft fields. 

There are two cases where eq. (\ref{hard_m}) simplifies so that the 
hard-field recursion become {\it independent} from the soft-field 
distribution. The first case is, of course, $m=0$ then $r=1$ independently 
of the edge $(ij)$, and the equation reduces to the original SP. The second case 
arise for $m=1$, where one can use the so-called reconstruction formalism 
and obtain again a closed set of equations.  The computation is done in appendix
\ref{app:m1}, and the SP equations at $m=1$ read
\be
       \eta_s^{i\to j} = \frac{1}{q} \sum_{l=0}^{q-1} \left[
        (-1)^l \sum_{s_1,\dots,s_l \neq s} \prod_{k\in i -j} 
        \left(1-\frac{q}{q-1}\sum_{\alpha=1}^l \eta_{s_\alpha}^{k\to i}\right) \right]\, .
        \label{SP_m1_app}
\ee 

It would be interesting in the future to use eq.~(\ref{hard_m}) in an
algorithm to find proper graph colorings, as it has been done with the
original SP equation~\cite{Coloring}. As an approximation one might also use a
value $r$ independent of the edge $(ij)$, but different from one.

For the purpose of the present work, it is important to notice that it is also
possible to use eq.~(\ref{hard_m}) in the population dynamics to simplify the
numerical evaluation of the 1RSB solution by separating the hard-field and the
soft-field contributions. Indeed, it gives the exact density of hard fields
provided the ratio $r$ is calculated, which is doable numerically (see
appendix~\ref{app:num}). This allows us to monitor precisely the hard-field
density and only the soft-field part is given by the population dynamics. This
turns out to greatly improve the precision of the numerical solution of the
cavity equations and to considerably fasten the code.

\subsubsection{The presence of frozen variables}
\label{sec:hard_sol}

A natural question is: ``When are the hard fields present?'' or more precisely:
``When does eq. (\ref{hard_m}) have a nontrivial solution $\eta>0$?''
First notice that in order to constrain a node into one color, one needs at
least $q-1$ incoming fields that forbids all the other colors. It means that
function $w(\{ \eta^{k\to i} \})$ defined in eq.~(\ref{SP}) is identically
zero for $k<q-1$ and might be non-zero only for $k \ge q-1$, where $k$ is the number of
incoming fields.

In the limit $r \to 0$ (which corresponds to $m \to
-\infty$) eq.~(\ref{hard_m}) gives $\eta=1/q$ if $w(\{ \eta^{k\to i}
\})$ is positive, and $\eta=0$ if $w(\{ \eta^{k\to i}
\})$ is zero. Updating eq. (\ref{hard_m}) on a given graph, from 
initial conditions $\eta=1/q$ everywhere, is equivalent to recursive 
removing of all the nodes of connectivity smaller than $q$. 
This shows that the first nontrivial solution with hard fields 
exists if and only if the $q$-core~\cite{q_core} of
the graph is extensive. For regular graphs it is simply at connectivity 
$c=q$ while for Erd\H{o}s-R\'enyi graphs these critical connectivities 
can be computed exactly and read, for small $q$, $c_3=3.35$, $c_4=5.14$, 
$c_5=6.81$~\cite{q_core}. Indeed we see
that the first nontrivial solution to the 1RSB equation appears much before
those of the original SP equation at $m=0$. 

\begin{figure}[!ht]
 \resizebox{18cm}{!}{
\includegraphics[angle=270]{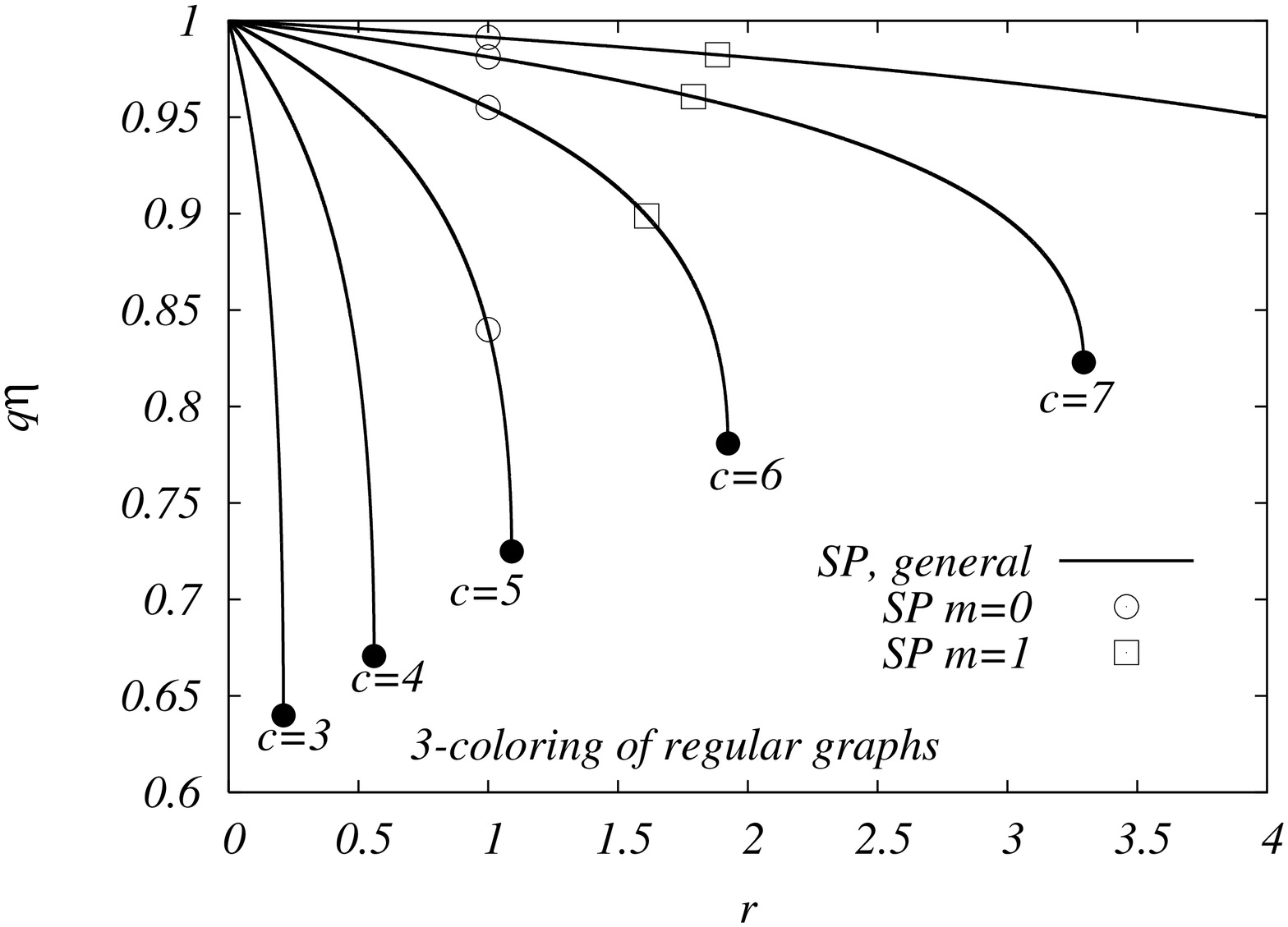}
\includegraphics[angle=270]{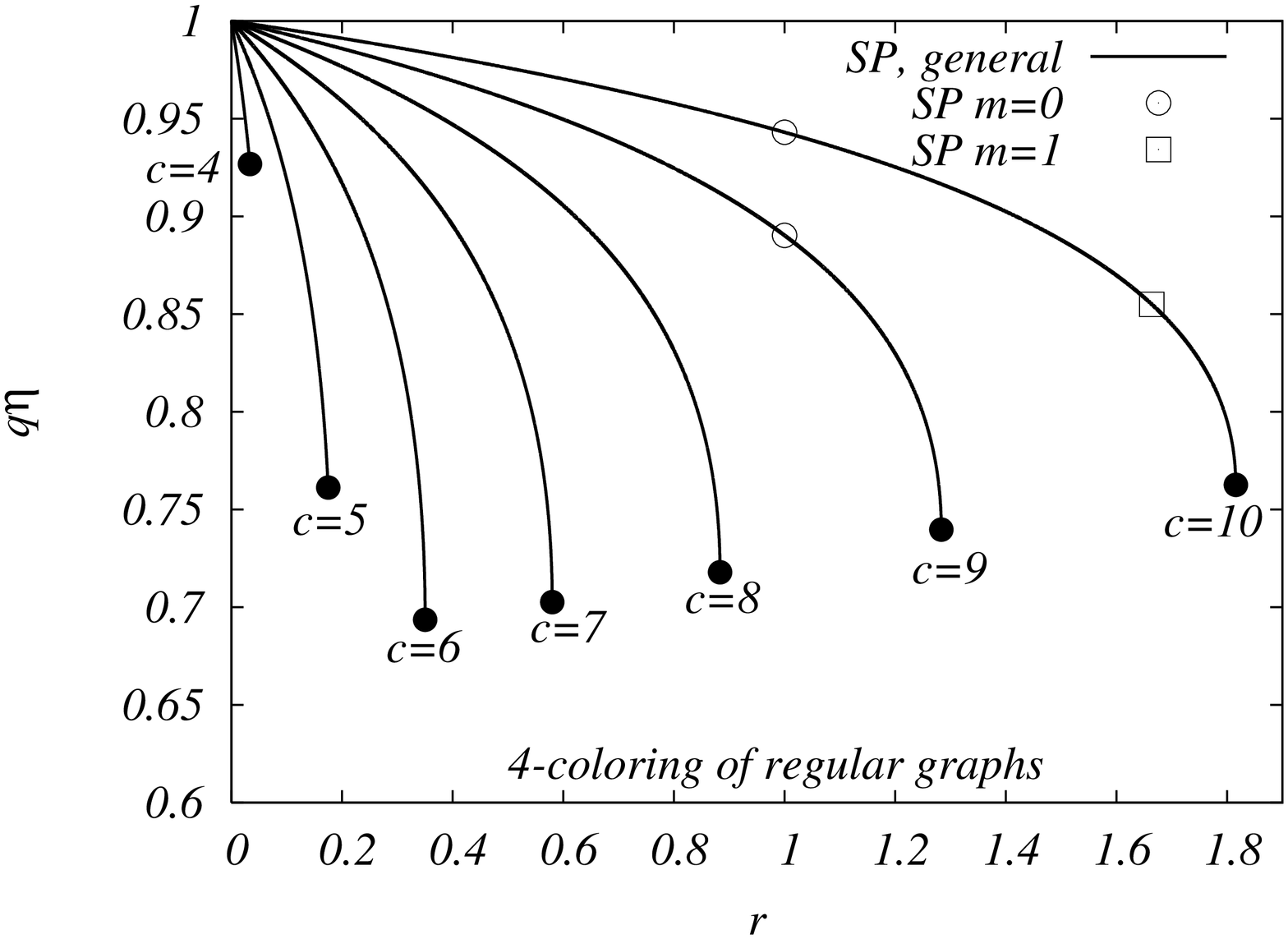}}
\caption{\label{fig_eta} The lines are solutions of eq. (\ref{hard_m_reg})
  and give the total fraction $q\eta$ of hard fields for a given value of
  ratio $r=\overline{Z^m_{\rm soft}}/{\overline {Z^m_{\rm hard}}}$ for
  $q=3$ (left) and $q=4$ (right) in regular random graphs.
  There is a critical value of the ratio (full point) beyond which only the
  trivial solution $\eta=0$ exists. Note that the solutions at $m=0$ and $m=1$
  only exist for a connectivity large enough.}
\end{figure}

On a regular graph, the equations
further simplify as $\eta$ factorizes (is edge independent) and follows 
a simple self-consistent equation
\be
\label{hard_m_reg}
 \eta = w(\eta)
\frac{1} {q
w(\eta)
+\left[1-q
w(\eta)
\right] r
} \, .
\ee 
This equation can be solved for every possible ratio $r$ so that for all $c
\ge q$, we can compute and plot the curve $\eta(r)$.  We show the results in
fig.~\ref{fig_eta} for different numbers of colors $q$. On this plot, we
observe that $\eta=1/q$, as predicted, for $r=0$. It then gets smaller for
larger value of the ratio and, at a critical value $r_{\rm crit}$, the
solution disappears discontinuously and only the (trivial) solution $\eta=0$
exists.  The values $r_{\rm crit}$ correspond to a critical value of $m_r$.
For all $m>m_r$ no solution containing frozen variables can exist.

\subsection{Validity conditions of the 1RSB solution}

Now that we have
discussed in detail the 1RSB formalism the next question is: ``Is 
this approach correct?''  To answer this question, one has to test if the Gibbs
measure is extremal {\it within} the thermodynamically dominating pure states.
This is equivalent to checking if the two-step Replica Symmetry Breaking
(2RSB) solution is non trivial. Computing explicitly the 2RSB solution
is however very complicated numerically, especially for Erd\H{o}s-R\'enyi graphs.
Instead, the local stability of the 1RSB solution towards 2RSB should be
checked, in analogy with the RS stability in sec.~\ref{sec:stab}. It is indeed
a usual feature in spin glass physics to observe that the 1RSB glass phase
become unstable at low temperatures towards a more complex RSB phase and this
phenomenon is called the Gardner transition~\cite{Gardner}.

To perform the stability analysis~\cite{MR03,MPR03,ColoringFlo,Rivoire}, one
first writes the 2RSB recursion ---where the order parameter is a distribution
of distributions of fields on every edge $P_1(P_2(\psi))$--- and then two types of 1RSB
instabilities have to be considered depending on the way the 2RSB arises from
the 1RSB solution.  The first type of instability ---called {\it states
  aggregation}--- corresponds to $\delta(P(\psi)) \rightarrow P_1(P_2(\psi))$
while the second type ---called {\it states splitting}--- corresponds to
$P(\delta(\psi)) \rightarrow P_1(P_2(\psi))$. A complete stability analysis is
left for future works,
but it is worth to discuss the relevance of the results derived over the last few
years~\cite{MR03,MPR03,ColoringFlo}.


The 1RSB stability was studied for the coloring problem in
\cite{ColoringFlo} but only for the energetic zero temperature 
limit (\ref{energetic}).  In this case the 
parameter $y=\beta m$ is conjugated to the energy.  
The results derived in \cite{ColoringFlo} 
---as well as those previously derived for other
problems \cite{MR03,MPR03}--- thus concern {\it only the clusters of sizes
corresponding to $m= 0$} at zero (for $y=\infty$) or at positive 
(for finite $y$) energy. 
The main result of \cite{ColoringFlo,MR03,MPR03} was that the 1RSB approach
was stable in vicinity of the coloring threshold $c_s$. As we shall see the
clusters corresponding to $m=0$ are those dominating the total entropy at the
coloring threshold and as a consequence its location is thus {\it exact}
within the cavity approach.  The states of the lowest energy (the ground states) in
the uncolorable phase also correspond to $m=0$, and thus the conclusions of
\cite{ColoringFlo,MR03,MPR03} concerning the uncolorable phase are also
correct. In particular, a Gardner transition towards further steps of RSB
appears in the uncolorable phase beyond a connectivity denoted $c_G$ in
\cite{ColoringFlo}.

On the other hand  in the colorable phase the stability of the entropically 
dominating clusters that correspond to $m>0$ should be investigated. 
Some more relevant information can be, however, already
drawn from known results.  It was indeed found that the 1RSB approach at $m=0$
is type I stable for all $y$, and type II stable for all $y>y_I$ in vicinity
of the coloring threshold.  These results concerns the states of positive
energy, but keeping in mind the interpretation of $y$ as a slope in $T,m$
diagram, we see that the clusters of zero energy corresponding to small but
nonzero positive $m$ and zero temperature are also stable with respect to both
types of stabilities.  Near
the colorable threshold, the value of $m^*$ which describes the dominating clusters
is close to zero and as a consequence {\it all the dominating clusters are
  1RSB stable in vicinity of the coloring threshold}.  Far from the coloring
threshold, the stability analysis of~\cite{MPR03,ColoringFlo} is irrelevant.
In particular, the predictions of a full-RSB colorable phase made in
\cite{ColoringFlo,MR03,MPR03} is {\it not correct}. Quite the contrary, our
preliminary results indicate that {\it all the dominating clusters are 1RSB
  stable for $q>3$}.

The $3-$coloring is however a special case as the clustering transition is
continuous. Although the type II instability seems irrelevant in this case as
well, we cannot at the moment dismiss a type I instability close to the
clustering transition. Indeed the entropically relevant clusters correspond to values of
$m^*$ close to one in this case, and it is simple to show that the clusters at
$m=1$ are type I unstable in the case there is a continuous transition: this
is because the type I stability is equivalent to the convergence of the 1RSB
update on a single graph. Since for $m=1$ the averages of the 1RSB fields
satisfy the RS belief propagation equations, and since we know from the RS
stability analysis in section~\ref{sec:stab} that those equations do not
converge in the RS unstable region ({\it i.e.}  for $c>c_{\rm RS}=4$ in 3-coloring
of Erd\H{o}s-R\'enyi graphs), it then follows that the 3-coloring is
unstable against state aggregation at $m=1$ for all connectivities $c>4$.  Therefore, it is possible
that the 1RSB result for 3-coloring are only approximative for what concerns
the number and the structure of solutions close to the clustering
transition (note that the critical values for the phase transition are however
correct and do not depend on that). This, and related issues~\cite{SUSY}, will
be hopefully clarified in future works.

To conclude, we believe that all the transition points we discuss in this
paper (and those computed in the K-SAT problem in \cite{US}) as well as the
overall picture, are exact and would not be modified by considering further 
steps of replica symmetry breaking.

\section{The coloring of random graphs: cavity results}
\label{sec:result}

We now solve the 1RSB equations, discuss and interpret the results.  
We solve the 
equation (\ref{1RSB_pop}) by the population dynamics technique, the technical 
difficulties and the precision of the method are discussed in appendix 
\ref{app:num}. Let us stress at this point that the correctness of 
eq. (\ref{1RSB_pop}) is guaranteed only in the limit of large graphs 
($N\to \infty$), unfortunately the cavity method does not give us any 
direct hint about the finite graph-size corrections. 
We start by the results for the regular random graphs, then consider 
ensemble of bi-regular graphs
and after that we turn towards Erd\H{o}s-R\'enyi graphs. 
Finally, we discuss the limit of large number of colors.

\subsection{Regular random graphs}

\begin{figure}[!ht]
\begin{minipage}{0.49\linewidth}
\begin{center}
  \resizebox{9cm}{!}{\includegraphics[angle=270]{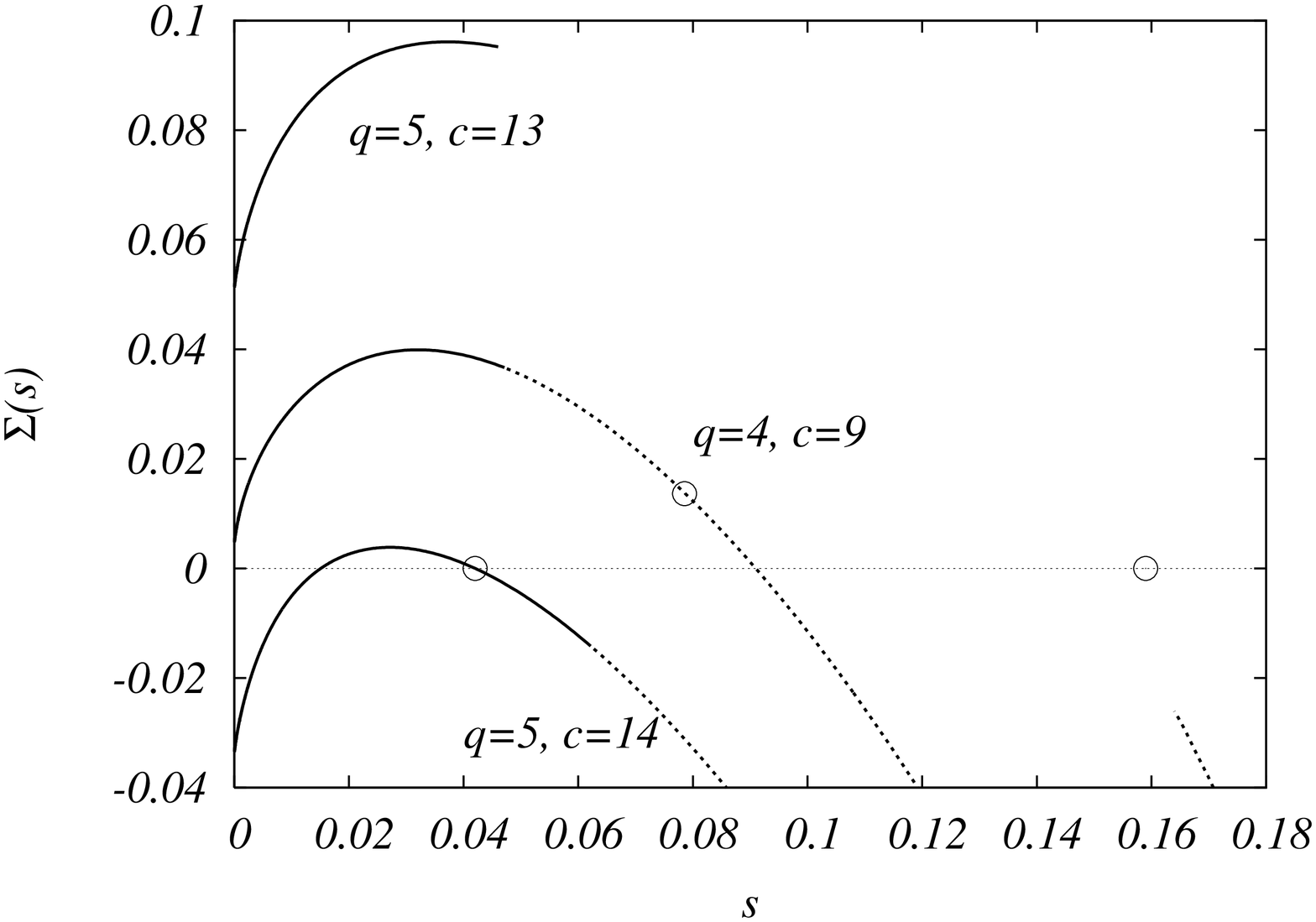}}
\end{center}
\end{minipage}
\begin{minipage}{0.49\linewidth}
\begin{center}
  \resizebox{9cm}{!}{\includegraphics[angle=270]{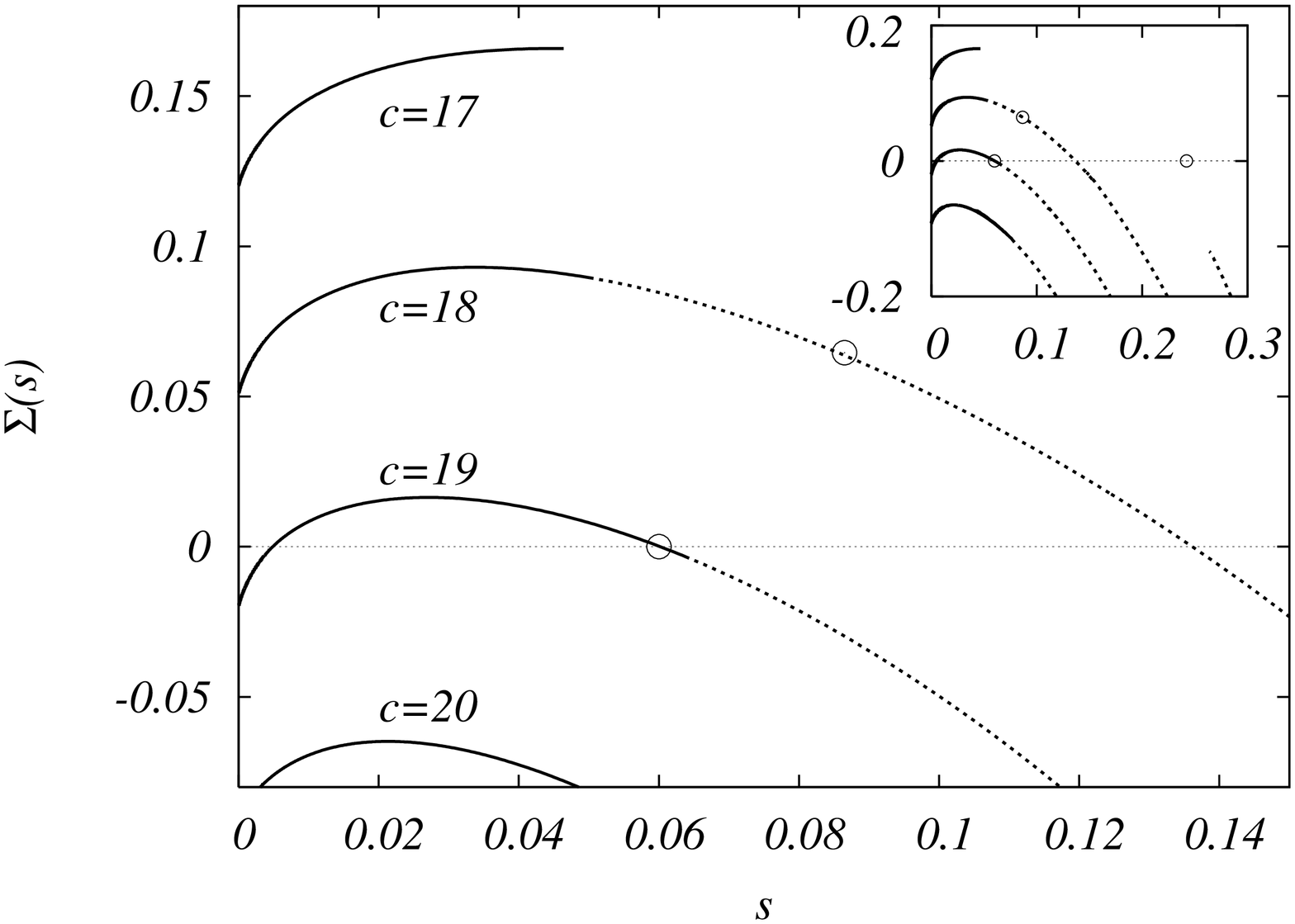}}
\end{center}
\end{minipage}
\caption{\label{fig:reg_all} Complexity as a function of the internal state
  entropy for the q-coloring problem on random regular graphs of connectivity
  $c$.  The full line corresponds to the clusters where a finite fraction of
  hard fields (frozen variables) is present and the dotted line to the clusters
  without hard fields. The circle signs the entropically dominating clusters.
  Left: ($q=4$, $c=9$) is in the clustered phase; ($q=5$, $c=13$) is in a
  simple replica symmetric phase and ($q=5$, $c=14$) is in the condensed
  clustered phase. Right: results for 6-coloring for connectivities 17 (RS),
  18 (clustered), 19 (condensed) and 20 (uncolorable).  For 4-,5- and
  6-coloring all the smaller connectivities are in the RS phase while all the
  larger one are uncolorable.}
\end{figure}

Let us fix the number of colors $q$, vary the connectivity, and identify
successively all the transitions that we shall encounter. For the sake of the
discussion, we choose as a typical example the 6-coloring and we discuss later
in details the cases, for different number of colors, where some transitions
are missing or are arriving in a different order. 
We solved the 1RSB equation (\ref{1RSB}) for regular graphs, where the
distribution $P^{i\to j}(\psi)$ is the same for every edge $(ij)$ (see
appendix \ref{app:num}) and  plot the curves for $\Sigma(s)$ we obtained doing
so in fig.~\ref{fig:reg_all}. We now describe the phase space of solutions
when the connectivity is increased:
\begin{enumerate}
\item[1)]{At very low connectivities $c<q$, only the paramagnetic RS solution is found at
  all $m$. {\it i.e.}  $P(\psi)=\delta(\psi-1/q)$. The
  phase space is made of a single RS cluster.}
\item[2)]{For larger connectivities $c \ge q$, we saw in section~\ref{sec:hard_sol} that the 1RSB equations start to have 
  nontrivial solutions with hard fields in an interval
  $[-\infty,m_r]$.  Interestingly, another nontrivial solution, 
  without hard fields, can be found numerically in an interval 
  $[m_s,\infty]$, and we shall call this one the soft-field solution.  
  As the connectivity increases, we find that $m_r$ increases while
  $m_s$ decreases, so that the gap $[m_r,m_s]$ 
  where no nontrivial solution exists it getting continuously smaller. 
  
  However, there is no nontrivial solution at $m=1$ for connectivities smaller then $c_d$
  (see fig.~\ref{fig:reg_all} for the example of the $6$-coloring at $c=17$). 
  This means that the Gibbs measure (\ref{Gibbs}) is
  still extremal. In other words the large RS state still exists
  and is entropically dominant (its entropy (\ref{S_RS}) is noted by a circle in
  fig.~\ref{fig:reg_all}).
  Despite the fact that an exponential number of clusters of solutions exist
  and that the SP equations converge to a nontrivial result, a
  random proper coloring will almost surely belong to the large RS
  cluster.
}
\item [3)]{If the connectivity is increased at and above the clustering
    threshold $c_d$, a nontrivial solution with positive complexity $\Sigma$
    is found at $m=1$. In fig. \ref{fig:reg_all}, we see that this happens at
    $c_d=18$ for the regular $6$-coloring.  At this point, the RS Gibbs
    measure (\ref{Gibbs}) ceases to be extremal and the single large RS
    cluster splits into exponentially numerous components. To cover almost all
    proper colorings we need to consider exponentially many clusters ${\cal N}
    \sim \e^{N\Sigma(m^*=1)}$.  The probability that two random proper
    colorings belong to the same cluster is going exponentially to zero with
    the system size.  The connectivity $c_d$ is thus the true clustering
    (dynamic) transition.  This is not, however, a thermodynamic phase
    transition because the 1RSB total entropy reduces to the RS entropy
    (\ref{S_RS}) at $m=1$ which is analytical in $c$. 
  Thus the RS approach gives a correct number of solution and 
  correct marginals as long as the complexity function at $m=1$ 
  is non-negative. 
}
\item [4)] For even larger connectivities $c \ge c_c$, the complexity at 
  $\Sigma(m=1)$ becomes negative, e.g. $c_c=19$ for 6-coloring. It means 
  that the clusters corresponding to $m=1$ are absent with probability one. 
  The total entropy is then smaller than the RS/annealed one and is dominated 
  by clusters corresponding to $m^*<1$ such that $\Sigma(m^*)=0$.
  The ordered weights of the entropically dominating clusters 
  follow the Poisson-Dirichlet process (explained in 
  appendix~\ref{app:PD}).  
  As a consequence, the probability that two random proper colorings 
  belong to the same cluster is finite in the thermodynamic limit. 
  Another way to describe the situation is that the entropy condenses into
  a finite number of clusters. 
  This condensation is a true thermodynamic transition, since 
  the total entropy is non-analytical 
  at $c_c$ (there is a discontinuity in its second derivative with respect to 
  connectivity). The condensation is analogous to the static (Kauzmann) glass 
  transition observed in mean field models of glasses~\cite{REM,GLASSTHEORY}. 
\item [5)] For connectivities $c\ge c_s$ ($c_s=20$ for $6$-coloring) even the
  maximum of the complexity $\Sigma(m=0)$ becomes negative. In this case
  proper colorings are absent with probability going to one exponentially fast
  with the size of the graph, and we are in the uncolorable phase.
\end{enumerate}

It is useful to think of the growing connectivity as additions of the
constraints into a fixed set of nodes. From this point of view the set of
solutions which exists at connectivity $c$ gets smaller when new edges are
introduced and the connectivity increased. This translates into the cartoon in
the introduction (fig.~\ref{fig:zero_T_3}) where all the successive
transitions are represented.  Finally, another important transition has to be
considered:
\begin{enumerate}
\item [6)] There is a connectivity $c_r$ beyond which the measure is dominated by
  clusters that contain a finite fraction of {\it frozen variables}. For the regular $6-$coloring,
  $c_r=19$.  We refer to this as the {\it rigidity transition}, by
  analogy with~\cite{Jamming,Jorge}.
\end{enumerate}
The presence or the absence of hard fields inside a given cluster is crucial:
if a cluster contains only soft fields, then after the addition of a small but finite 
fraction of new constraints, its size will get smaller (or it will split). 
If, however, a cluster contains a finite fraction of frozen variable, 
then after adding a small but finite fraction of links the cluster will 
almost surely disappear.

Since the connectivities of regular graphs are integer numbers, we define the 
dynamical threshold $c_d$ as the smallest connectivity where a nontrivial 
1RSB solution exists at $m=1$, the
condensation transition $c_c$ as the smallest connectivity where complexity at
$m=1$ is negative, $c_r$ the smallest connectivity where hard fields are
present at $m^*$ and the coloring threshold $c_s$ as the first uncolorable
case.  The scenario described here is observed for all cases of the regular 
ensemble, although, since connectivities are integer, the transitions are not
very well separated at small $q$.  We summarize the results in table~\ref{tab:regular}. 

Note that for $q>3$, the local RS stability discussed in section 
\ref{sec:stab} is irrelevant in the colorable regime. The only subtle 
case being for $3-$coloring of 5-regular graphs where the RS solution 
is only marginally stable, {\it i.e.} 
the spin glass correlation function goes to zero only algebraically instead 
of exponentially (from this point of view $c=5$ would correspond to the critical point well known in the second order phase transitions). More
interesting cases will arise in the other ensembles of random graphs.

\begin{table}[!ht]
\begin{minipage}{0.38\linewidth}
\begin{tabular}{|c|c|c|c|c|c|}
\hline
q & $c_{SP}$~\cite{ColoringFlo} & $c_d$~\cite{MM05} & $c_r$ & $c_c$ & $c_s$~\cite{ColoringFlo} \\
\hline
3 & 5 & $5^+$ & - & 6 & 6 \\
\hline
4 & 9 & 9 & - & 10 & 10 \\
\hline
5 & 13 & 14  & 14 & 14 & 15 \\   
\hline
6 & 17 & 18 & 19 & 19 & 20  \\   
\hline
7 & 21 & 23 & - & 25 & 25  \\   
\hline
8 & 26 & 29 & 30 & 31 & 31  \\   
\hline
9 & 31 & 34 & 36 & 37 & 37  \\   
\hline
10 & 36 & 39 & 42 & 43 & 44  \\   
\hline
20 & 91 & 101 & 105 & 116 & 117  \\   
\hline 
\end{tabular}
\end{minipage}
\begin{minipage}{0.38\linewidth}
\begin{tabular}{|c|c||c|c|c|}\hline
q & $ c $ & $m^*$ & $m_r$ & $m_s$ \\
\hline 
5 & 3 & RS$^+$ &  0.12  & 1.2(1) \\
\hline
4 & 8 & RS &  -0.03  & 3.4(1)\\
4 & 9 & 1  &   0.41  & 0.41 \\
\hline
5 & 12 & RS & -0.02  & 3.7(1)\\    
5 & 13 & RS &  0.20  & 2.0(1) \\    
5 & 14 & 0.50 & 0.90 & 0.90 \\    
\hline
6 & 16 & RS & -0.02  & 4.3(1)\\   
6 & 17 & RS &  0.05  & 3.2(1)\\   
6 & 18 & 1  &  0.40  & 0.40 \\   
6 & 19 & 0.92 & 0.96 & 0.96 \\    
\hline
7 & 21 & RS &  0.01  & 4.7(1)\\
7 & 22 & RS &  0.17  & 3.2(1) \\
7 & 23 & 1  &  0.60  & 0.60\\
7 & 24 & 1  &  0.95  & 0.95\\  
\hline 
\end{tabular}
\end{minipage}
\caption{ \label{tab:regular} Left: The transition thresholds for regular random graphs: $c_{\rm
    SP}$ is the smallest connectivity with a nontrivial solution at $m=0$; the
  clustering threshold $c_d$ 
  is the smallest connectivity with a nontrivial solution at $m=1$; the
  rigidity threshold $c_r$ is
  the smallest connectivity at which hard fields are present in the dominant
  states, the condensation $c_c$ is the smallest connectivity for which the complexity at $m=1$ is
  negative and $c_s$ the smallest UNCOL connectivity. Note that $3-$coloring of
  $5-$regular graphs is exactly
  critical for that $c_d=5^+$. The rigidity transition may not
  exist due to the discreteness of the connectivities. Right: Values of $m^*$
  (corresponding to the dominating clusters), and in the range of $[-\infty,m_r]$ 
  the hard-field solution exists, in the range $[m_s,\infty]$ the
  soft-field solution exists.}
\end{table}

\subsection{Results for the bi-regular ensemble}

The bi-regular ensemble allows us to fine-tune the connectivity while
preserving the factorization of the 1RSB solution, which is crucial for the
numerical precision. It is actually more correct to say that 
the solution is ``bi-factorized'',
as all the messages going from the nodes with connectivity $c_1$ to $c_2$ are
the same and the other way around. The bi-regular ensemble allows us to
describe with large precision two interesting cases, which reappear in the
Erd\H{o}s-R\'enyi ensemble and which are not present in the regular ensemble
(again, due to the discrete nature of the connectivity).
Let us remind here that bipartite graphs are always 2-colorable, but we
consider only the color symmetric cavity solutions and that is why we get a
nontrivial result from this ensemble.

\begin{figure}[!ht]
\begin{minipage}{0.49\linewidth}
\begin{center}
  \resizebox{9cm}{!}{\includegraphics[angle=270]{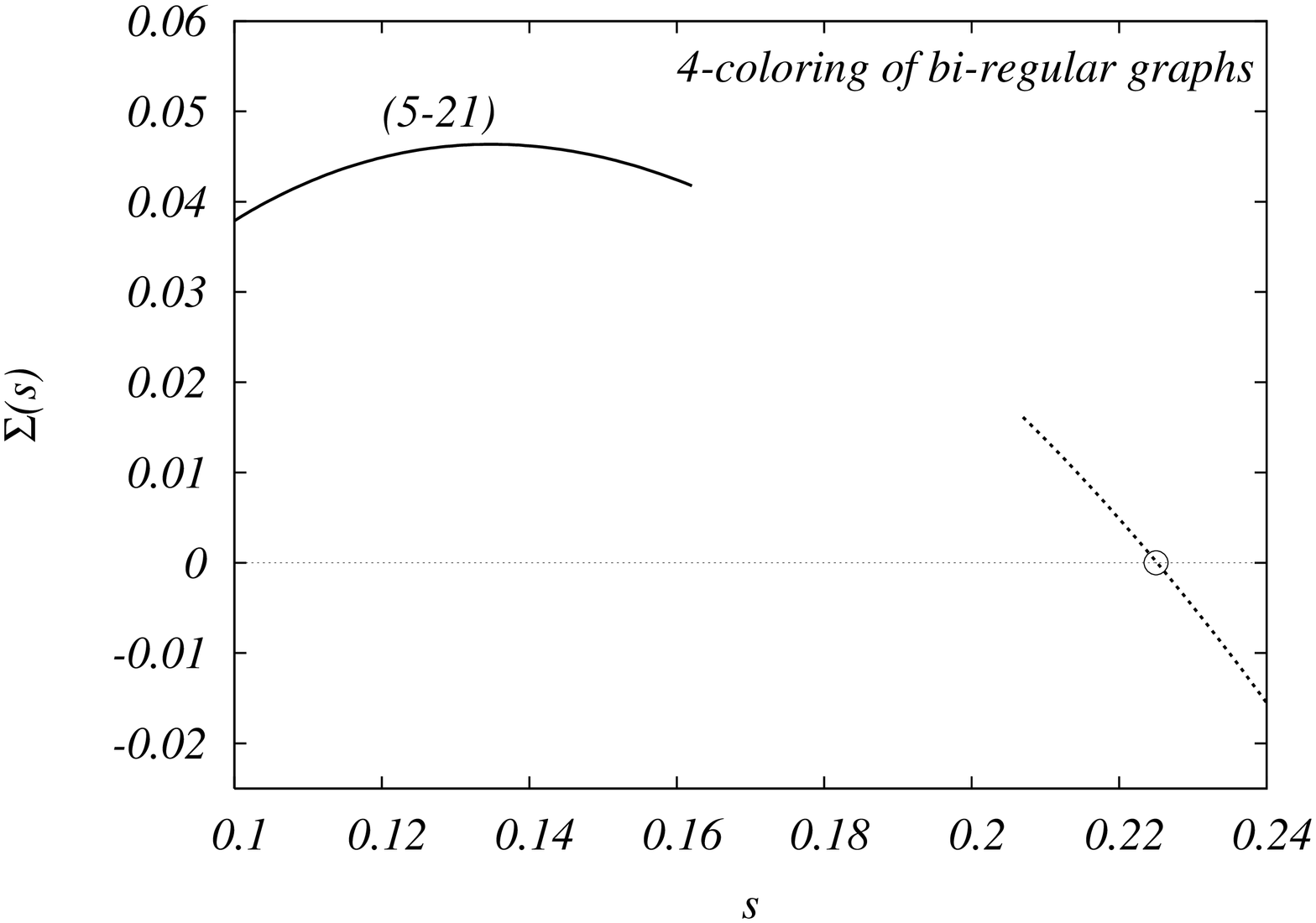}}
\end{center}
\end{minipage}
\begin{minipage}{0.49\linewidth}
\begin{center}
   \resizebox{9cm}{!}{\includegraphics[angle=270]{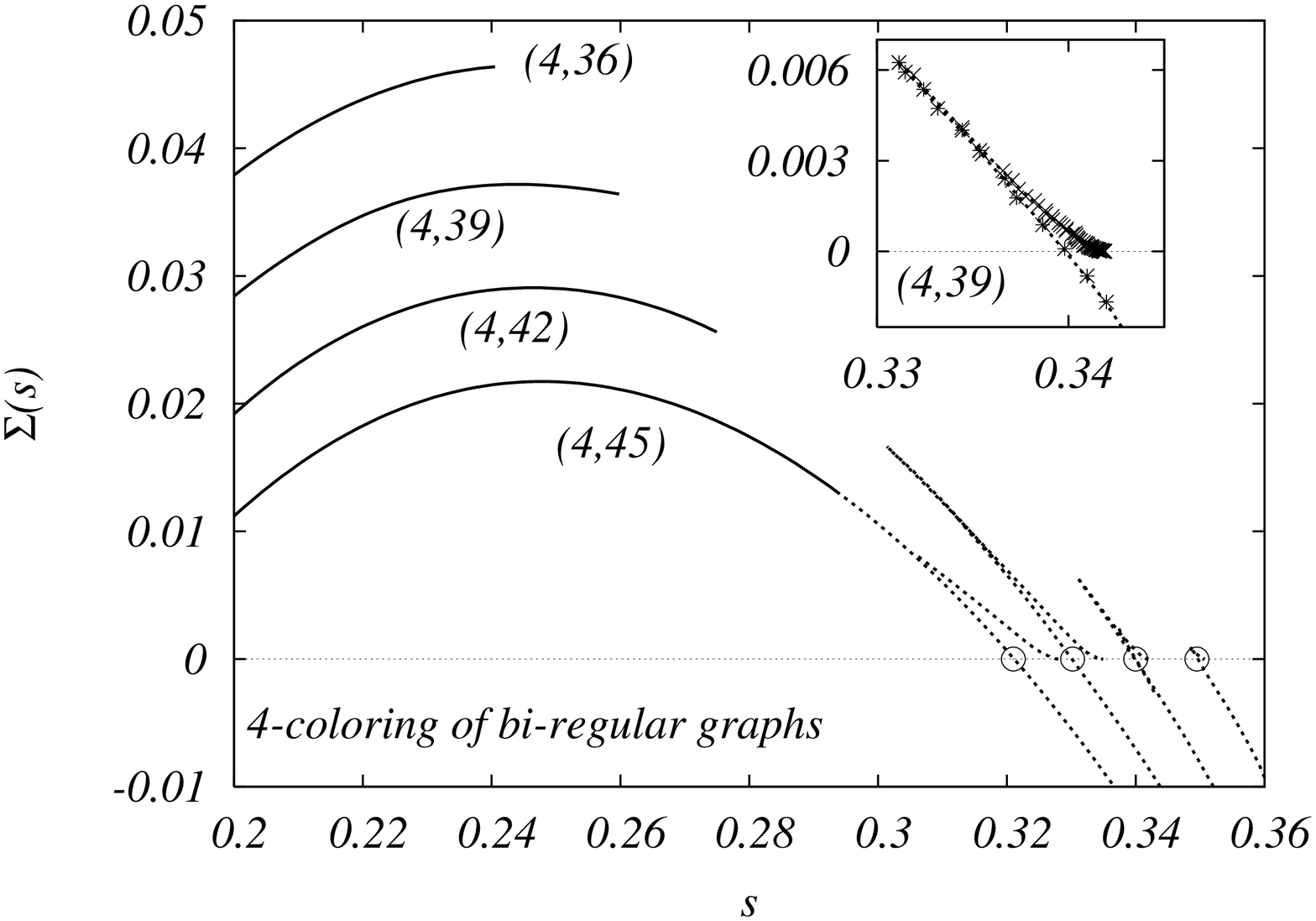}}
\end{center}
\end{minipage}
\caption{\label{fig:2ref}
  The complexity as a function of entropy for 4-coloring or bi-regular graphs. 
  Left: 5-21-bi-regular graph, an example where the entropy is dominated by 
  clusters with soft fields while the gap in the curve $\Sigma(s)$ still 
  exists. 
  Right: 4-c-bi-regular graphs for c=36, 39, 42, 45. In all these cases the 
  replica symmetric solution is locally unstable. In the dependence 
  $\Sigma(s)$ we see an unphysical branch of the complexity 
  which is zoomed in the inset for $c=39$. }
\end{figure}

In fig.~\ref{fig:2ref}, the left picture is the result for the complexity as a
function of entropy $\Sigma(s)$ for 4-coloring of 5-21-bi-regular graphs.  The
replica symmetric solution on this case is locally stable.  We see clearly
the gap between the hard-field and the soft-field solution, and yet we are
already beyond the clustering transition $c_d$; actually the system is in the
condensed phase.
This example is similar to what happens for
the 4-coloring of Erd\H{o}s-R\'enyi graphs. 

The second interesting case, the right hand side of fig.~\ref{fig:2ref}, is given
by the results for $\Sigma(s)$ for the 4-coloring of 4-$c$-bi-regular graphs,
which are RS unstable for $c > 28$.  Both the clustering and the condensation
transitions coincide with the RS instability $c_d=c_c=28$. The survey
propagation equations have a nontrivial solution starting from $c_{\rm SP}=
37$. The rigidity transition is at $c_r=49$.  Finally the coloring threshold
is $c_s=57$.  Qualitatively, the results for this 4-$c$-bi-regular ensemble
are the same as those for the 3-coloring of Erd\H{o}s-R\'enyi random graphs.

We see that for $c\le 42$ the gap between the hard-field (full line) and 
soft-field (dotted line) solution exists. For $m>m_s$ there is a non-physical 
nontrivial soft-field solution, the convex part of the line in the figure, 
zoomed in the inset. 
It means that for $m<m_r$ we actually can find two solutions depending if we 
start or not with a population containing enough hard fields. The unphysical 
branch survives even when the gap $[m_r,m_s]$ closes, see the example of $c=45$
in the figure. 

We would like to stress at this point 
the enormous similarity of the soft-field part of the curve 
$\Sigma(s)$ to the one in fig. 4 in ref.~\cite{GiulioRemiMartin}. Actually the 
variational results of~\cite{GiulioRemiMartin} should be very precise and 
relevant near to the continuous clustering transition (this is also case for 
the 3-coloring of Erd\H{o}s-R\'enyi graphs or for 3-SAT). 

\subsection{Results for Erd\H{o}s-R\'enyi random graphs}

For Erd\H{o}s-R\'enyi random graphs obtaining the solution of eq. (\ref{1RSB_pop})
is computationally more involved as the solution is no longer factorized. In the
population dynamics a population of populations has to be updated, which is
numerically possible only for small populations, and so one has to be
careful that the finite population-size corrections are small enough, see details in appendix \ref{app:num}.  
However, all the
computations can be done with the same computational complexity as for the
regular graphs for $m=0$, the energetic zero temperature limit
(section~\ref{hard_0}), and for $m=1$ (appendix~\ref{app:m1}). That is enough
to obtain the SP, clustering, condensation and COL/UNCOL transitions (from
which the first and last one was computed in~\cite{Coloring}). We can also
compute exactly when hard fields appear for $m=1$, eq. (\ref{hard_m1_app}),
this transition is further studied in~\cite{Guilhem_rear}. Finally, using the
generalized survey propagation equation introduced in section
\ref{generalized}, the rigidity transition can be computed quite precisely.

\subsubsection{The general case for $q>3$, discontinuous clustering transition}
\begin{figure}[!ht]
\begin{minipage}{0.49\linewidth}
\begin{center}
  \resizebox{9cm}{!}{\includegraphics[angle=270]{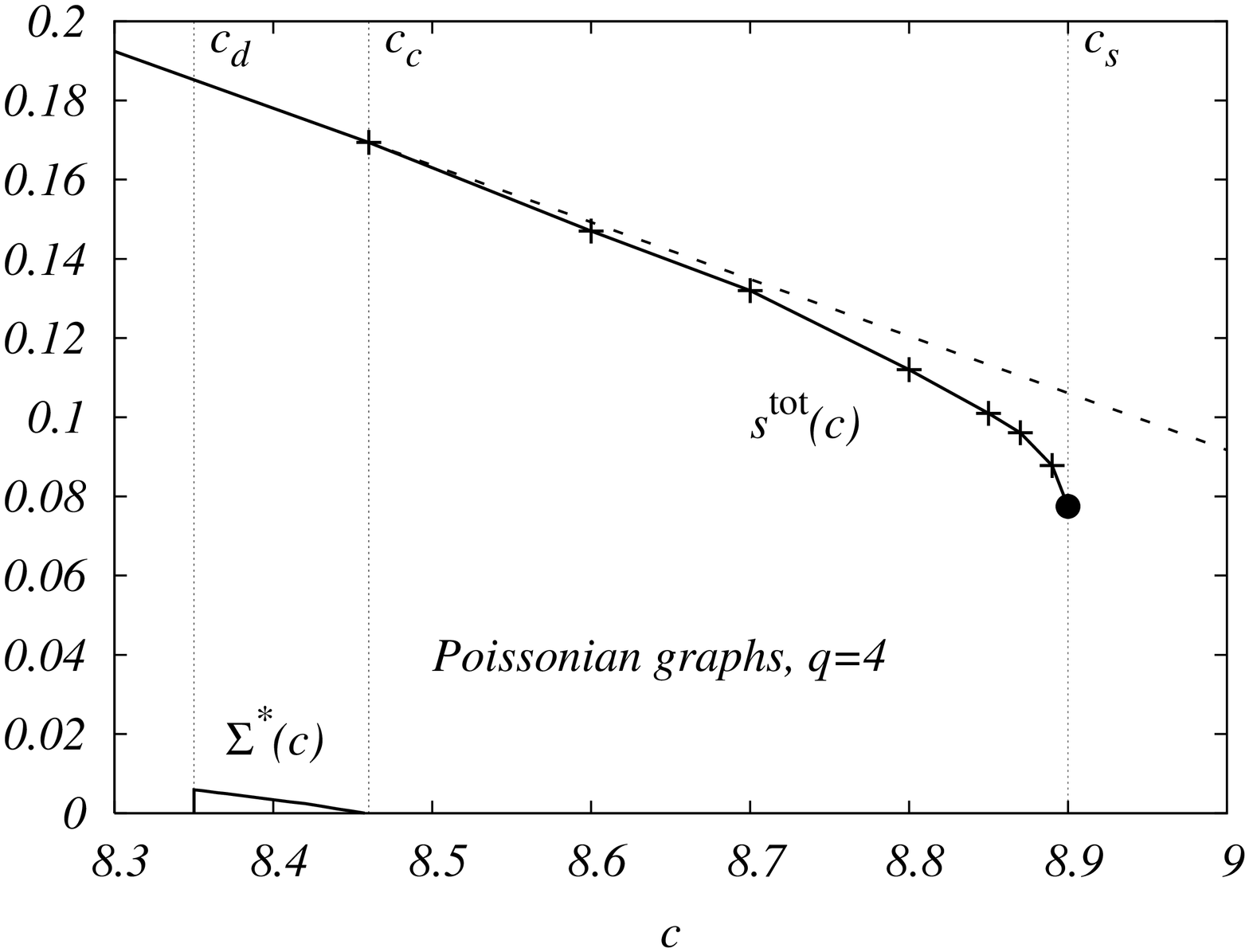}}
\end{center}
\end{minipage}
\begin{minipage}{0.49\linewidth}
\begin{center}
  \resizebox{9cm}{!}{\includegraphics[angle=270]{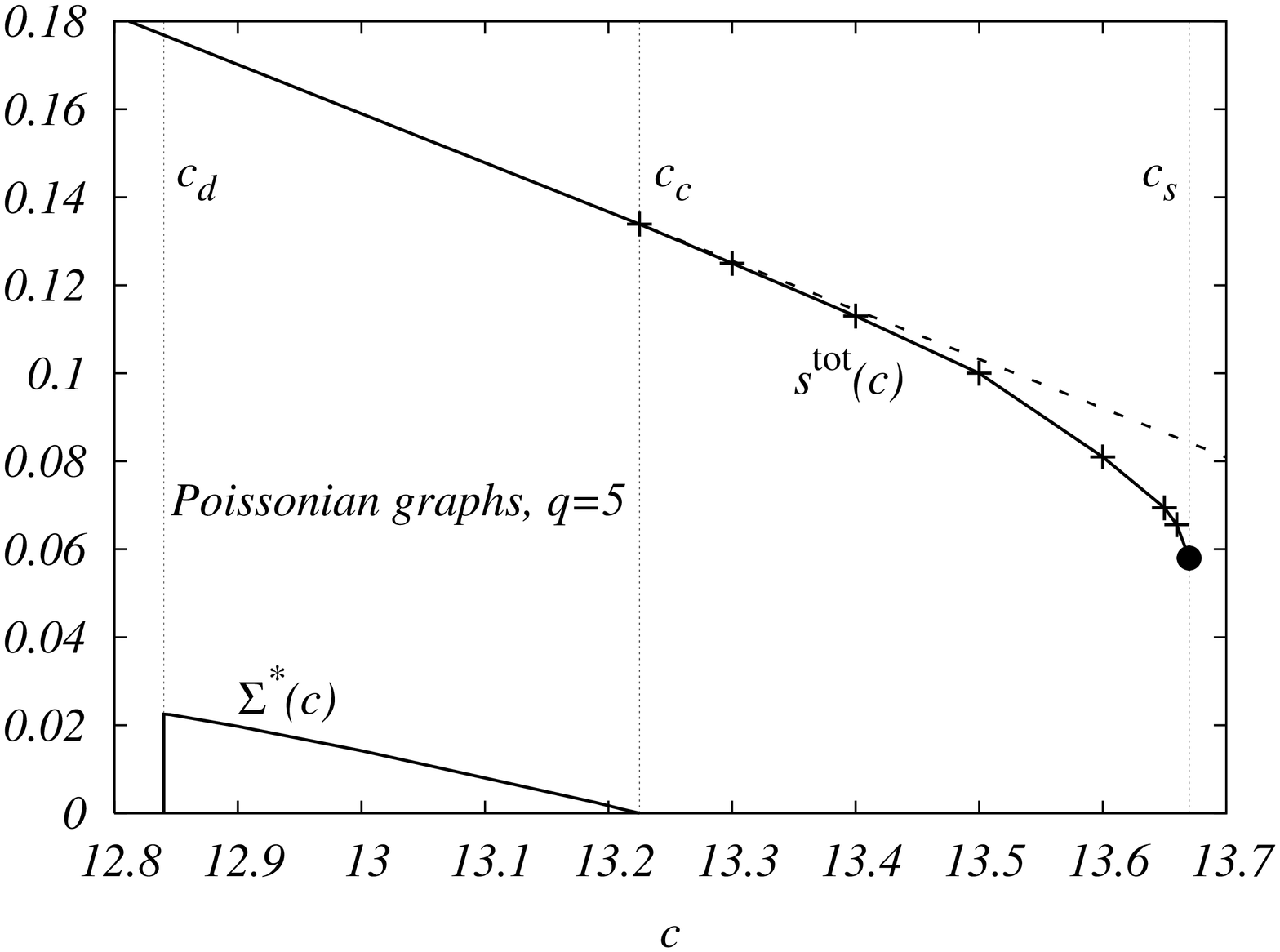}}
\end{center}
\end{minipage}
\caption{\label{fig:er_sigma} The 1RSB total entropy and complexity of the dominating 
  clusters for 4- and 5-coloring of Erd\H{o}s-R\'enyi random graphs. The complexity jumps discontinuously at the clustering transition $c_d$ while the total entropy stays analytical. The complexity disappears at the condensation transition $c_c$ causing a non-analyticity in the total entropy. Finally the total entropy discontinuously disappears at the coloring threshold. Dashed is the RS entropy left for comparison.
}
\end{figure}

The phase transitions in $q$-coloring of random Erd\H{o}s-R\'enyi graphs 
for $q>3$ are qualitatively identical to those discussed in the case of random 
regular graphs. We plot the results for the total entropy (number of solutions) 
and complexity (number of clusters which dominate the entropy) in
the $4-$ and $5-$ coloring in fig.~\ref{fig:er_sigma}.
 
At the clustering transition $c_d$ the complexity becomes discontinuously 
positive, the large RS cluster suddenly splits in an exponential number of smaller ones. 
The total entropy  $\Sigma^*+s^*$ is given by the RS formula (\ref{S_RS}) up to 
the condensation transition $c_c$. 
At the condensation transition the complexity of the dominating clusters becomes zero,   
the total entropy $s^{\rm tot}=s^* < s_{\rm RS}$ is given by the point where $\Sigma(s^*)=0$.
The function $s^{\rm tot}(c)$ is non-analytical at the point $c_c$, 
it has a discontinuity in the second derivative. 
At the coloring threshold $c_s$ all the clusters of solutions disappear, 
note, however, that the total entropy of the last existing clusters is strictly positive 
(about a half of the total entropy at the condensation transition).
That means that the COL/UNCOL transition is not only sharp but
also discontinuous in terms of entropy of solutions. Note that the positive 
entropy has two contributions: the trivial and smaller one coming from 
presence of leaves and other small subgraphs, and the nontrivial 
and more important one connected with the fact, that the ground state 
entropy is positive, even in the uncolorable phase or for the random regular 
graphs.  

Finally we located the rigidity transition, when frozen variables appears in the 
dominating clusters. For $3 \le q \le 8$ this transition appears in the condensed phase.
As the number of colors grows it approaches the clustering transition.
All the four critical values $c_d$, $c_r$, $c_c$ and $c_s$ are summarized 
in table~\ref{tab:results_complete}, values of $c_{\rm SP}$ and $c_r(m=1)$ are given for 
comparison.

\begin{table}[!ht]
\begin{tabular}{|c||l|l|l|l| |l|l|}\hline
q & $c_d$ & $c_r$ &  $c_c$ & $c_s$ & $c_{\rm SP}$ & $c_{r(m=1)} $ \\
\hline \hline
3 & 4 & 4.66(1) & 4 & 4.687(2) & 4.42(1) & 4.911 \\
\hline
4 & 8.353(3) & 8.83(2) & 8.46(1) & 8.901(2) & 8.09(1) &  9.267 \\
\hline
5 & 12.837(3) & 13.55(2) & 13.23(1) & 13.669(2) & 12.11(2) &  14.036 \\
\hline
6 & 17.645(5) & 18.68(2) & 18.44(1) & 18.880(2) & 16.42(2) &  19.112 \\
\hline
7 & 22.705(5) & 24.16(2) & 24.01(1) & 24.455(5) & 20.97(2) &  24.435 \\
\hline
8 & 27.95(5) & 29.93(3) & 29.90(1) & 30.335(5) & 25.71(2) & 29.960 \\
\hline
9 & 33.45(5) & 35.658 & 36.08(5) & 36.490(5) & 30.62(2) & 35.658 \\
\hline
10 & 39.0(1) & 41.508 & 42.50(5) & 42.93(1) & 35.69(3) & 41.508 \\
\hline
\end{tabular}
\caption{\label{tab:results_complete} Critical connectivities $c_d$ (dynamical, clustering), $c_r$ 
(rigidity, rearrangments), $c_c$ (condensation, Kauzmann) and $c_s$ (COL/UNCOL) for the phase transitions in the coloring problem on
Erd\H{o}s-R\'enyi graphs. The connectivities  $c_{SP}$ (where the first non
trivial solution of SP appears) and $c_{r(m=1)}$ (where hard fields appear at
$m=1$) are also given. 
The error bars consist of the numerical precision on evaluation of the critical connectivities by the population dynamics technique, details are given in appendix \ref{app:num}.
}
\end{table}

\subsubsection{The special case of $3-$coloring, continuous clustering transition}

The only case which is left to be discussed is the 3-coloring of
Erd\H{o}s-R\'enyi graphs. It is different from $q>3$ because the replica
symmetric solution is locally unstable in the colorable phase (see
section~\ref{sec:stab}).  The extremality condition underlying the RS
assumption ceases to be valid because of the mechanism discussed in
section~\ref{sec:stab}, with a divergence of the spin glass correlation
length: the main difference with the previous cases is therefore that 
the clustering transition is continuous and coincide with the condensation transition.

\begin{figure}[!ht]
\begin{minipage}{0.49\linewidth}
\begin{center}
  \resizebox{9cm}{!}{\includegraphics[angle=270]{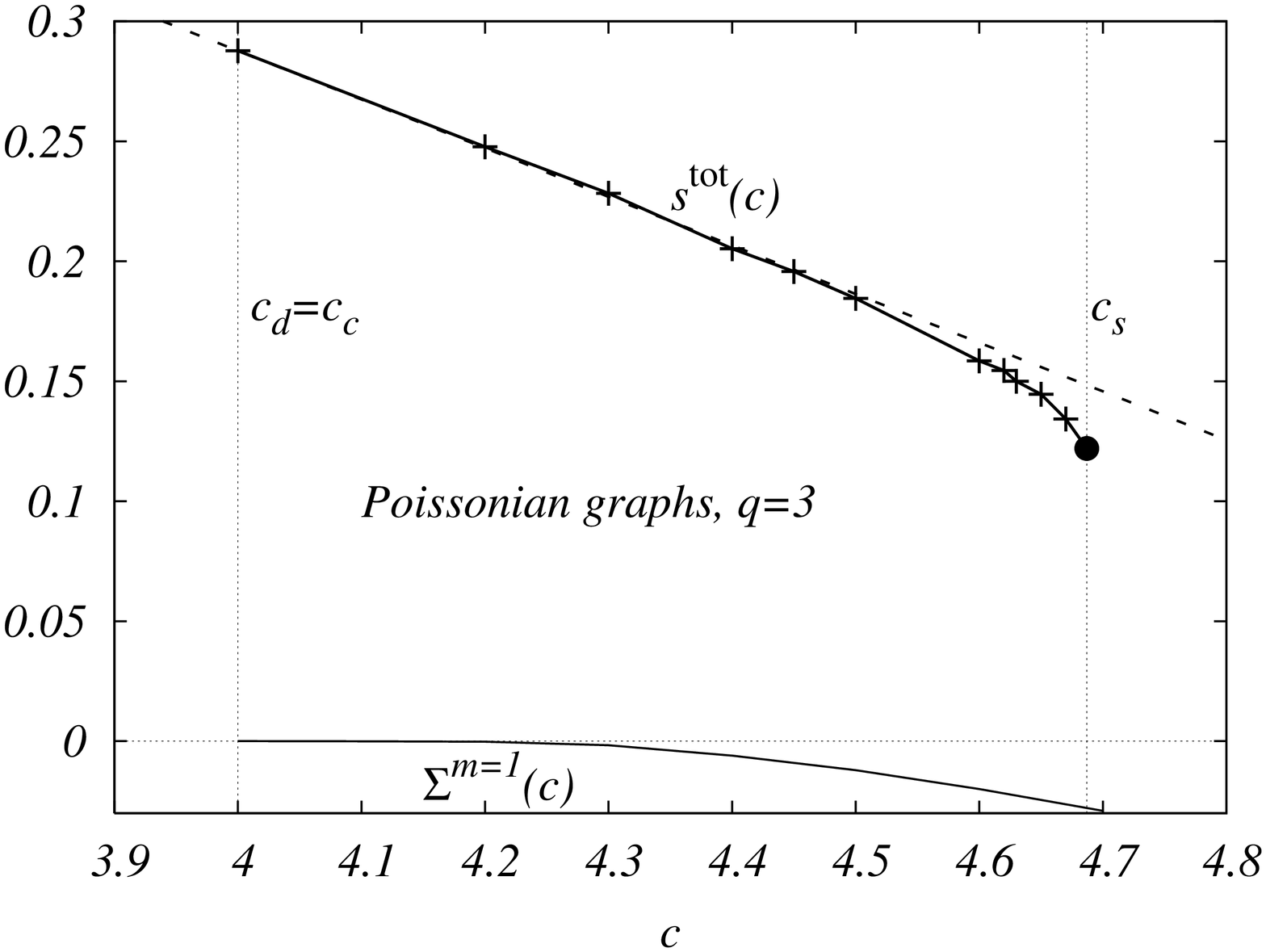}}
\end{center}
\end{minipage}
\begin{minipage}{0.49\linewidth}
\begin{center}
  \resizebox{9cm}{!}{\includegraphics[angle=270]{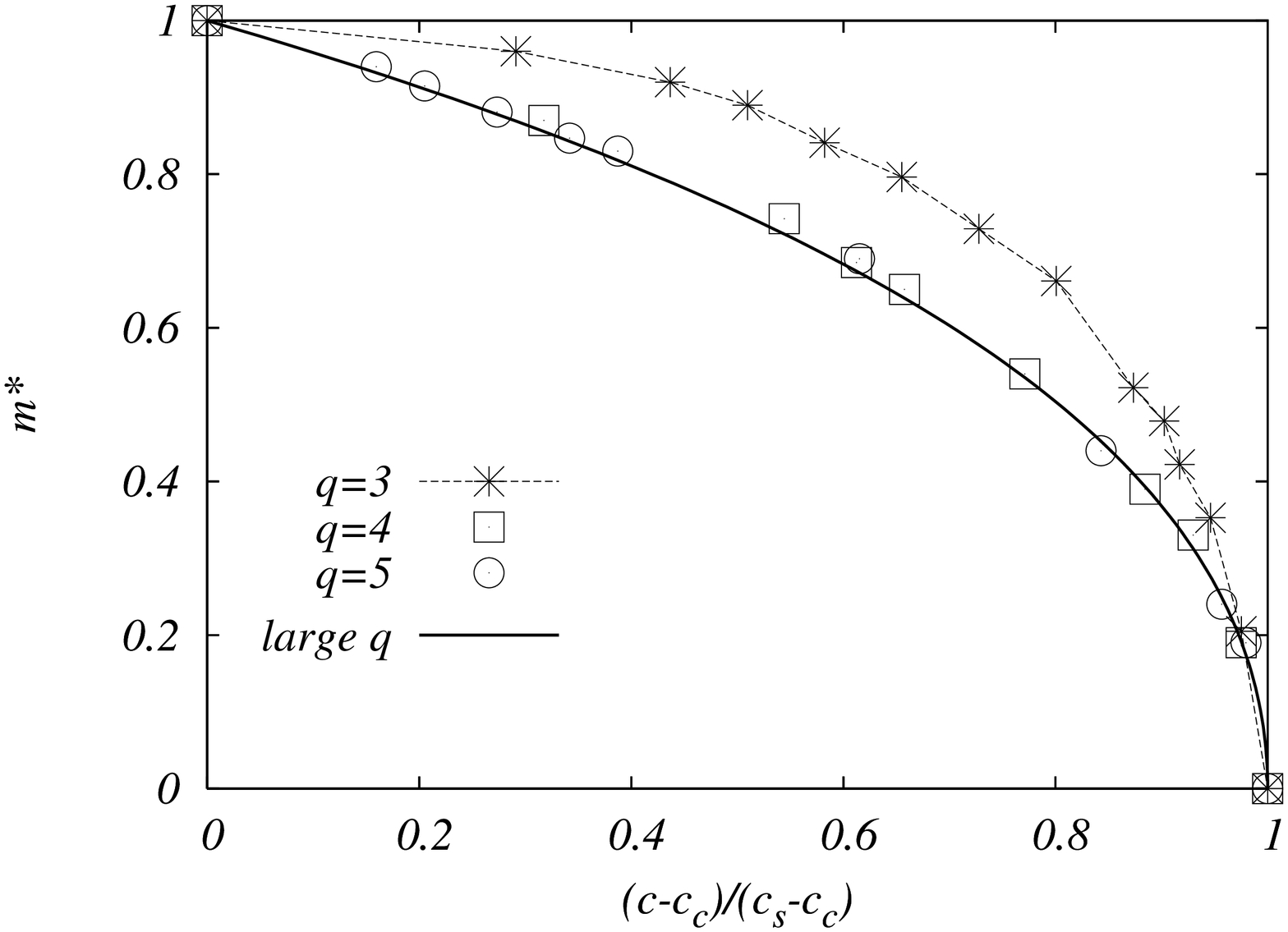}}
\end{center}
\end{minipage}
\caption{\label{fig:3_comp} Left: The total entropy for 3-coloring of Erd\H{o}s-R\'enyi random graphs. The dashed line is the replica symmetric (also the annealed) entropy, left for comparison. The complexity at $m=1$ is shown, it is negative for $c>4$, however, for connectivity near to four it is very near to zero. Right: The values of parameter $m^*$ 
  ($\Sigma(m^*)=0$) as a function of connectivity for $q=3,4,5$ and in the
  large $q$ limit.  The connectivity $c$ is rescaled as $(c-c_c)/(c_s-c_c)$.
  It is striking that for $q>3$ the curves are so well fitted by the large $q$
  limit one. We are even not able to see the difference due to the error bars
  which are roughly of the point size.}
\end{figure}

However, the phenomenology does not differ too much from the other cases: $c_{\rm RS}=c_d=c_c=4$; the phase where
the entropy is dominated by exponential number of states is thus missing and the
complexity corresponding to $m=1$ is always negative (see
fig.~\ref{fig:3_comp} left together with the dependence of the total entropy
on the connectivity).  Note that the curves $\Sigma(s)$ for the 3-coloring
have been already studied in~\cite{ColoringMarc,OlivierThesis} where the authors considered
however only the range of connectivities $c=[4.42,4.69]=[c_{\rm SP},c_s]$.

All the results derived for the 4-coloring of 4-$c$-bi-regular bipartite
graphs are quantitatively valid also here. We are thus not surprised by the
fact that in interval $c=[4,4.42]$ the survey propagation algorithm gives us a
trivial result: simply the maximum of the curve $\Sigma(s)$ does not exist yet
there is no nontrivial solution at $m=0$. Yet, the entropy is dominated by
finite number of largest clusters which do not contain hard fields. The two
solutions (hard-field and soft-field) join at a connectivity around $4.55$.
Finally at $c_r=4.66$ the hard fields arrive to the dominating states (and in
consequence to all others). 

\subsubsection{The overlap structure}

\begin{figure}[!ht]
\begin{minipage}{0.49\linewidth}
\begin{center}
  \resizebox{9cm}{!}{\includegraphics[angle=270]{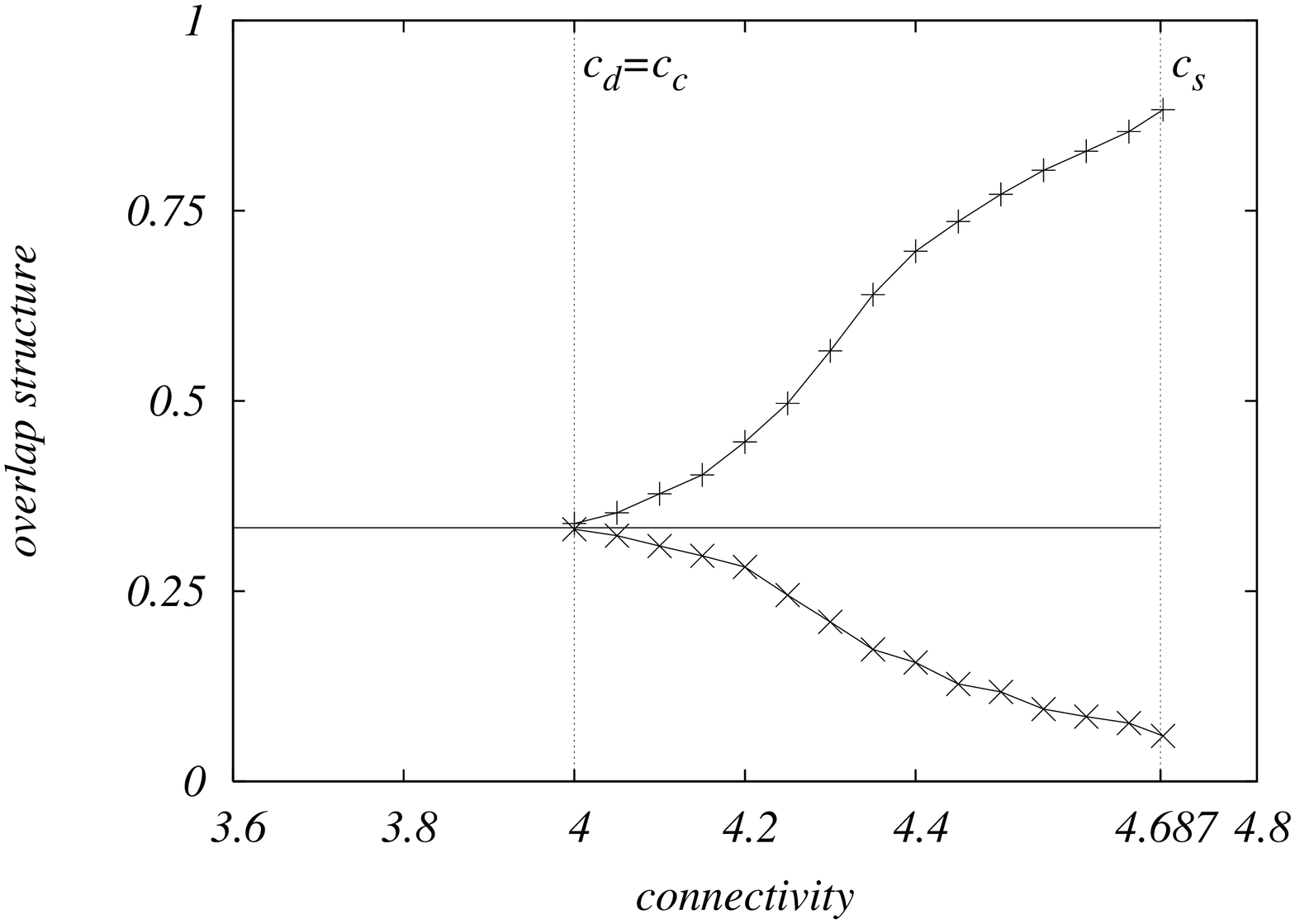}}
\end{center}
\end{minipage}
\begin{minipage}{0.49\linewidth}
\begin{center}
  \resizebox{9cm}{!}{\includegraphics[angle=270]{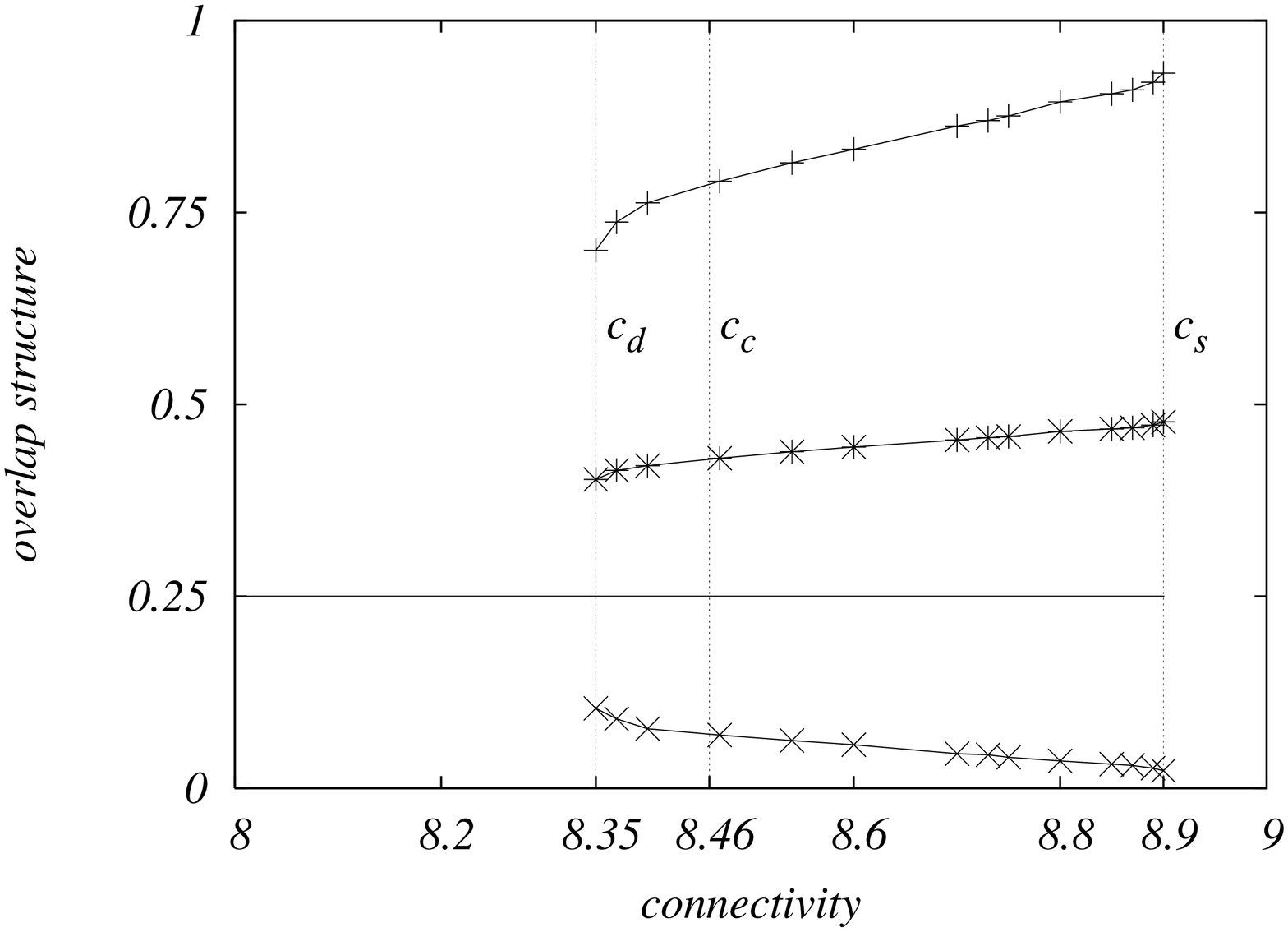}}
\end{center}
\end{minipage}
\caption{\label{fig:overlaps} Left: Overlaps structure in 3-coloring of random graphs as a function of connectivity. The intra-cluster overlap (upper curve) grows continuously from 1/3 at the clustering transition $c=4$. In the figure from up there are $\delta=\delta_3$, $\delta_1$ and $\delta_0$. Right: Overlaps structure in 4-coloring of random graphs as a function of connectivity. The intra-cluster overlap (upper curve) 
jumps discontinuously from 1/4 at the clustering transition $c=8.35$.
The probability that two random solutions belong to the same cluster, however, is zero between the clustering and condensation transition $[8.35,8.46]$.  
In the figure from up there are $\delta=\delta_4$, $\delta_2$, $\delta_1$ and $\delta_0$.}
\end{figure}

We now give some results about the overlap structure in the random coloring to
elaborate the intuition about clusters. First, consider marginal probabilities
$\psi^{i,\alpha}_{s_i}$ within a cluster $\alpha$. Note that due to the color
symmetry there exist another $q!-1$ clusters different only in the permutation
of colors. We define the intra-cluster overlap of two solutions (averaged over
states) as
\begin{equation}
     \delta = \frac{1}{N} \sum_i \sum_{s_i} \langle (\psi^{i,\alpha}_{s_i})^2 \rangle_{\alpha}\, .
\end{equation}
In the paramagnetic phase $\delta=1/q$, otherwise we have to compute it from 
the fixed point of equation (\ref{1RSB}).
The overlap between two solutions which lie in two clusters, which differ 
just by permutation $\pi$ of colors is 
\begin{equation}
\delta_j=  \delta  \frac{j-1}{q-1}+ \frac{q-j}{q(q-1)} \, ,
\end{equation} 
where $j$ is the number of fixed positions in the permutation $\pi$ (in particular $\delta_q=\delta$, and $\delta_1=1/q$).
In fig.~\ref{fig:overlaps} we show the overlap structure for 3- and 4-coloring.
The probabilities that two random solutions have one of the overlaps can be computed from the Poisson-Dirichlet process described in appendix~\ref{app:PD}, in fact this is not a self-averaging quantity \cite{MPVkniha}. 

\subsection{Large q Asymptotics}

We give here the exact analytical large $q$ expansion of the previous results.
In the asymptotic computations the regular and Erd\H{o}s-R\'enyi ensembles are
equivalent (the corrections are of smaller order in $q$ that the orders we
give). We refer to the appendix~\ref{app:largeq} for the explicit derivation
of the formulae.

At large $q$ a first set of transitions arises for connectivities scaling as
$q \log {q}$: 
\bea
c_{\rm SP} = c_r(m=0) &=& q\ \left[ \log q + \log \log q +1 -\log 2 + o(1)\right],\\
c_r \mathop{=}_{q \to \infty} c_r(m=1) &=& q [\log q + \log\log q + 1 +o(1)].
\eea
$c_{\rm SP}$ was already computed in~\cite{ColoringFlo} and $c_r$ is the
rigidity transition. The clustering transition has to appear before the
rigidity one $c_d<c_r$.  For all the finite $q$ cases we looked at, $c_d$ was
between $c_{\rm SP}$ and $c_r$.

A second set of transitions arises for connectivities scaling as $2q
\log {q}$:
\bea
c_c &=& 2q \log{q} - \log{q} - 2 \log{2} + o(1)\, , \\
c_s &=&  2q \log{q} -\log{q} -1 + o(1)\, . 
\eea
The condensation thus appears very close the COL/UNCOL transition and both are
very far from the clustering and rigidity transitions (those are on a half way
in the phase diagram).

We show also in appendix \ref{app:largeq} that for connectivity $c=2q \log {q} - \log{q} + \alpha$,
one has 
\bea
2 q s(m) &\simeq& 2^m \log{2}\, , \\
2 q \Sigma(m) &\simeq& 2^m -2 -m2^m \log{2} - \alpha \, .  
\eea 
Since the RS free energy is correct until $c_c$, which differers just by
constant from $c_s$, that means that for all connectivities bellow $c_c$ the
number of solutions is correctly given by the replica symmetric entropy
(\ref{S_RS}). Indeed, the value $s(m=1)$ can be obtain by a large expansion of
eq.~(\ref{S_RS}).

\begin{figure}[!ht]
\begin{minipage}{0.49\linewidth}
\begin{center}
  \resizebox{9cm}{!}{\includegraphics[angle=270]{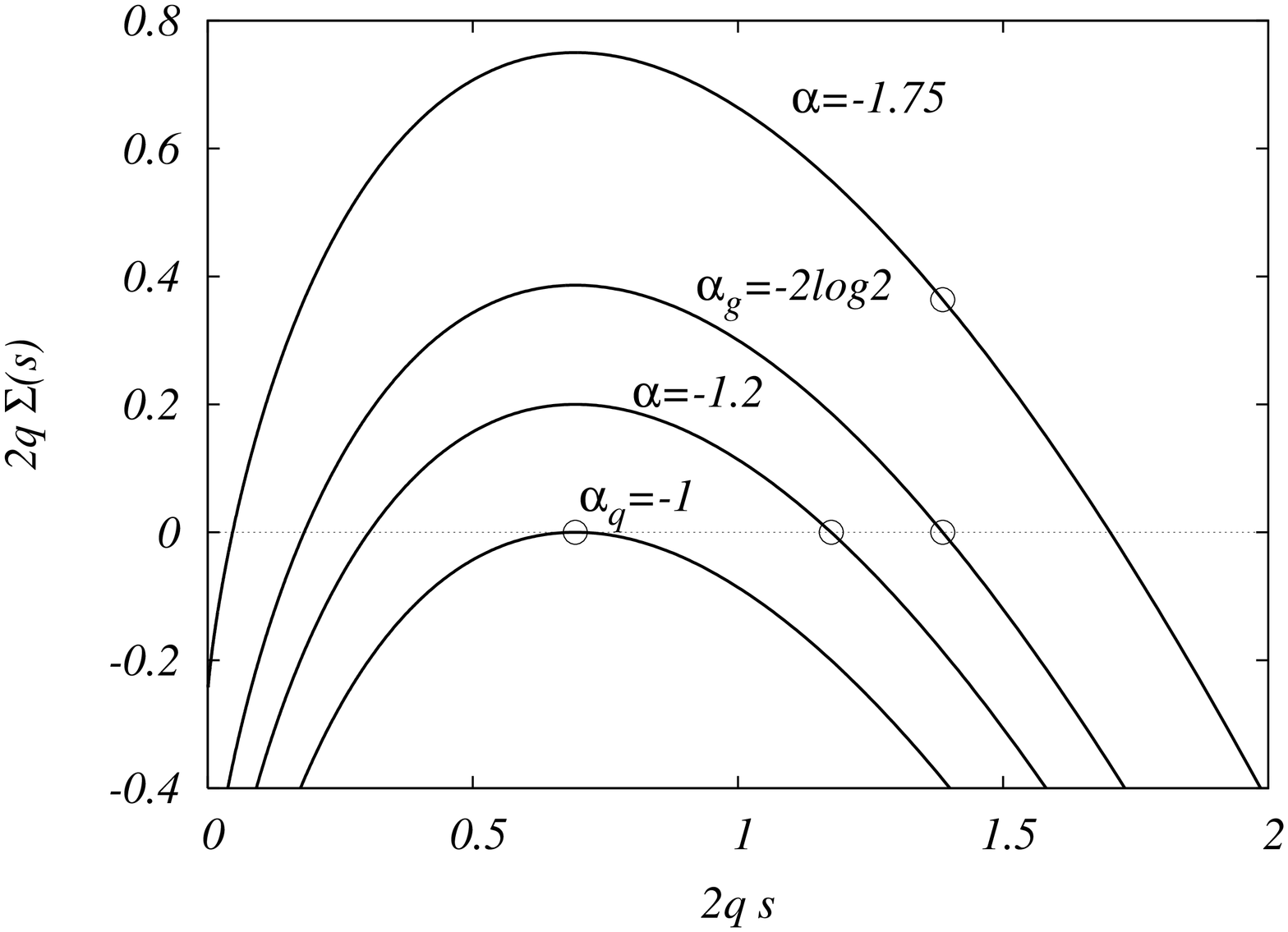}}
\end{center}
\end{minipage}
\begin{minipage}{0.49\linewidth}
\begin{center}
  \resizebox{9cm}{!}{\includegraphics[angle=270]{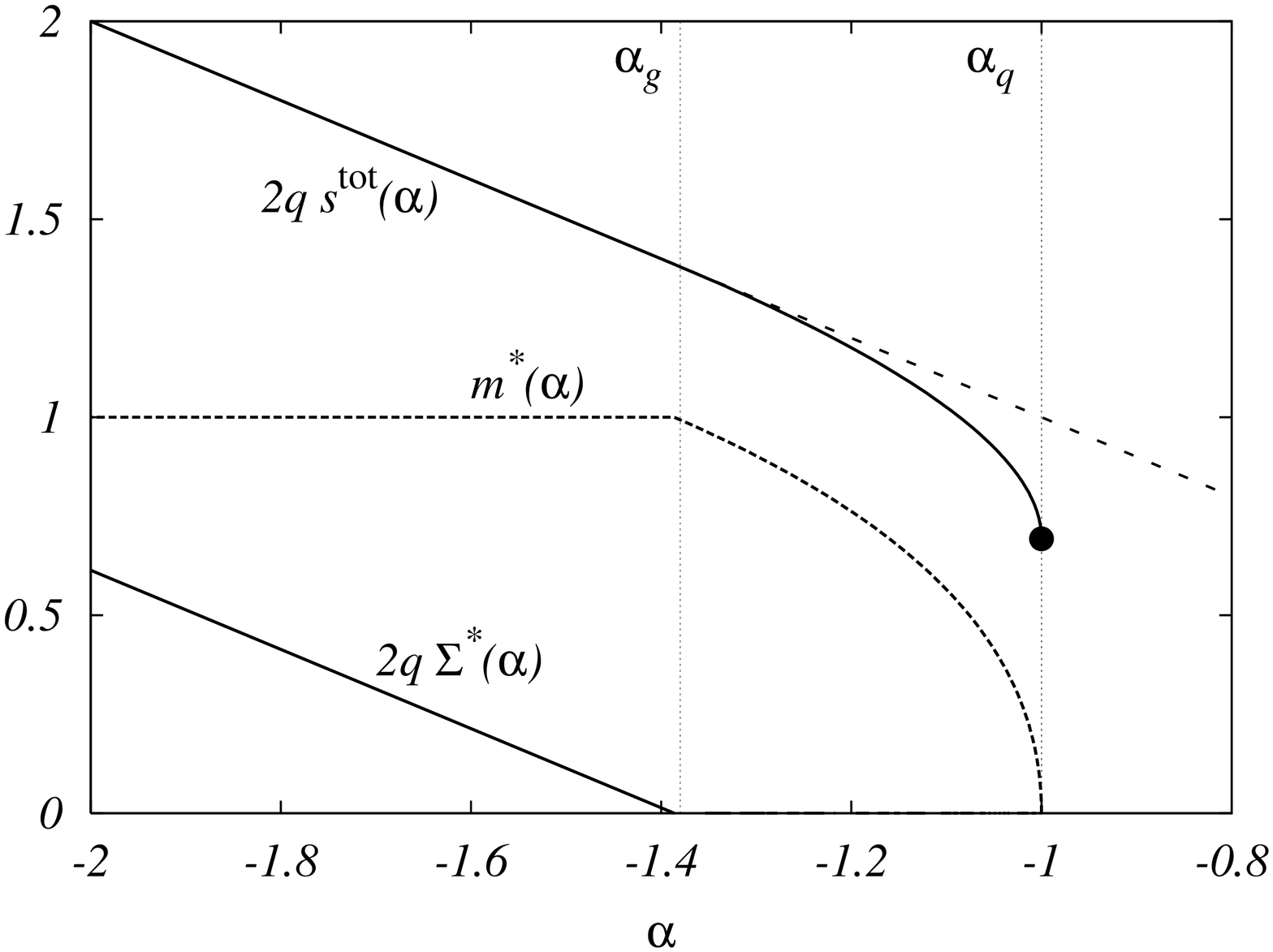}}
\end{center}
\end{minipage}
\caption{\label{fig:largeq} Analytical result for the large $q$ asymptotics close to the COL/UNCOL
  transition for $c=2q\log{q}-\log{q}+\alpha$. Left: (rescaled) complexity
  versus (rescaled) internal entropy for different connectivities. The
  condensation transition appears for $\alpha=-2\log{2}$. The maximum
  of the complexity becomes zero at $c_s$ for $\alpha=-1$. Right: Total entropy
  ($s^{\rm tot}$), complexity ($\Sigma^*$) and the parameter $m^*$ versus
  $\alpha$.  Notice how the values for the total number of solution are
  already very close to those for finite low $q$ in 
figs.~\ref{fig:er_sigma},~\ref{fig:3_comp}.}
\end{figure}

In fig.~\ref{fig:largeq} we plot the complexity of dominating clusters
$\Sigma^*= \Sigma(m^*)$, the total entropy $s^{\rm tot}=\Sigma^*+s^*$, and the
physical value of $m^*$ as a function of connectivity
$c=2q\log{q}-\log{q}+\alpha$. Note that the properly scaled values of the total 
number of solutions at $c_c$ and $c_s$, and the values $c_c$, $c_s$ themselves, 
are already very close to those at $q=3,4,5$ 
(see figs.~\ref{fig:er_sigma} and fig.~\ref{fig:3_comp} left).
The closeness is particularly striking for the values $m^*$ 
for $q=4$ and $q=5$ (see fig. \ref{fig:3_comp} right).

These formulae show that in the large $q$ limit, near to the coloring threshold,
it is the number of clusters which change with connectivity ({\it i.e.} $\alpha$), 
and not their internal entropy (size).  In the
leading order, adding constraints near to the COL/UNCOL transition thus
destroys clusters of solutions, but do not make
 them smaller: this is due to
the fact that these clusters are dominated by frozen variables so that adding a
link kills them most of the time. We also computed the
entropy value at the condensation transition, and found $s(m=1)=\log{2}/q$.
The entropy of the last cluster
(exactly at the COL/UNCOL transition) is $s=\log{2}/2q$.

\section{Algorithmic consequences}
\label{sec:algo}

In this section we give some algorithmic consequences of our findings.
First, we discuss the whitening procedure. 
We then introduce a random walk algorithm adapted from the Walk-SAT strategy
and study its performance. We show in particular that the
clustering/dynamical transition {\it does not} correspond
to the onset of hardness in the problem and argue that it is instead the 
rigidity transition. Finally, we discuss the performance of the belief
propagation algorithm in counting and finding solutions, and show that is
works much better than previously anticipated. 

\subsection{The whitening procedure}
\label{sec:White}

The whitening procedure as introduced in~\cite{ParisiWhit} can distinguish
between solutions which belong to a cluster containing hard fields and those
which do not. Generally whitening is equivalent to the warning propagation
(version of belief propagation which distinguish only if a field is hard or
not). Warning propagation for coloring was derived in \cite{Coloring}. Let us
call $u^{i\to j}=(1,0,0,\dots,0)$ the hard field in the direction of the first
color, {\it i.e.} in absence of node $j$ the node $i$ takes only the first color in all
the colorings belonging to the cluster in consideration, and similarly for
other colors. Denote $u^{i\to j}=(0,0,0,\dots,0)$ if $\psi^{i\to j}$ is not
frozen in the cluster, we say that the oriented edge $i \to j$ is then
``white''. The update for $u$'s follows from (\ref{update})
\begin{equation}
      u^{i\to j}_s = \min_{r}{\left(\sum_{k\in i -j} u^{k\to i}_r+\delta_{r,s}
      \right)} - \min_{r}{\left(\sum_{k\in i -j} u^{k\to i}_r\right)} \, .
      \label{WP}
\end{equation}

To see if a solution $\{s_i\}$ belongs to a 
cluster with frozen variables or not we initialize warning propagation with 
$u^{i\to j}_s=\delta_{s,s_i}$, and update iteratively according to 
(\ref{WP}) until a fixed point is reached (the update every time converge, 
because starting from a solution we are only adding white edges). 
In the fixed point or all edges are white, then the solution  $\{s_i\}$ 
does not belong to a frozen cluster, or some of the edges stay colored 
(non-white), then the solution $\{s_i\}$ belongs to a frozen cluster. 
Note that in the K-SAT problem (but not in general), whitening is equivalent 
to a more intuitive procedure, where the directed edged are not considered
\cite{Elitza,Riccardo}.

We wish to offer here an explanation of a paradox observed in 
\cite{Elitza,Riccardo}. The SP algorithm 
gives information on the frozen variables in the most numerous 
clusters ($m=0$). Yet, the solutions which are
found by the standard implementation (decimation and SP plus Walk-SAT)
do not belong to clusters with frozen
variables, since they always give a trivial whitening result (all 
directed edges are white)\cite{Elitza,Riccardo}. We suggest that the
decimation strategy
drives the system towards a solution belonging to a large cluster, 
which does not contain frozen variables. In this case, it is logical that
the result of the whitening is trivial, as it is observed.
We believe this is reason why no nontrivial whitenings are
observed so far in the study of the K-SAT problem on large graphs. 

Note that beyond the rigidity transition this argument does not work 
anymore, since there all the clusters (for all $m$ such that $\Sigma(m)>0$) 
contain frozen variables.
More precisely, for $q\ge 9$ we could in principle end up in
soft-clusters even beyond the rigidity transition (since that one concerns
only the dominant states), if this is possible is let for further
  investigation. Interestingly, in the coloring problem we have not been able
to find solutions beyond the rigidity transition even with survey propagation
algorithm (compare $c_r$ with the performance of SP in~\cite{Coloring}).
Further, more systematic, investigations have to be done about
these issues, employing other strategies for the use of the survey propagation
equations (for example the reinforcement \cite{reinforcement}).

\subsection{A Walk-COL algorithm to color random graph}
\label{WalkCol}

In this paper, we have computed the correct clustering transition $c_d$ for
the random coloring problem. Beyond this transition, Monte Carlo algorithms
are proven not to reach equilibrium as their time of equilibration diverges
\cite{MS05,MS06}. It was often claimed, or assumed, that this point
corresponds to the onset of hardness of the problem. However, the fact that
the physical dynamics does not equilibrate just means that the complete set of
solutions will not be correctly sampled ---indeed Monte-Carlo experiments
clearly display slow relaxation \cite{MonteCarlo}--- but not that no solutions
can be eventually found. This simple fact explains the results of
\cite{ColoringSaad} where a simple annealing procedure was shown to 
3-color a ER graph beyond $c_d=4$.

In this section, we use a local search strategy which does not satisfy the
detailed balance condition. Therefore, we do not expect to be able to find
{\it typical} solutions, however it might be possible to find {\it some}
solutions to the problem.  The Walk-COL algorithm~\footnote{A slightly different
  adaptation, closer to~\cite{walksat}, was performed few years ago by Andrea
  Pagnani and Martin Weigt (private communication).} is a simple adaptation
of the celebrated Walk-SAT \cite{walksat}. More precisely, we adapted the
method designed for satisfiability in~\cite{John}.  Given a graph, and
starting from an initial random configuration, we recursively apply the
following procedure:
\begin{enumerate}
\item[1)] Choose at random a spin which is not satisfied ({\it i.e.} at least one 
     of its neighbors has the same color).
\item[2)] Change randomly its color. Accept this change with probability one
  if the number of unsatisfied spin has been lowered, otherwise accept it with
  probability $p$.
\item[3)] If there still are unsatisfied nodes, go to step 1) unless the
  maximum running time is reached
\end{enumerate}

The probability $p$ has to be tuned in each different case for a better
efficiency of the algorithm. Typically, values between $0.01-0.05$ give good
results. We shall now briefly discuss the performance of the algorithm, to
illustrate the two following points: (a) When the phase space is RS, we
observe that Walk-COL finds a solution in linear time. (b) Even in the
``complex'' phase for $c>c_d$, the algorithm can find in some cases solutions
in linear time.

Concerning the first point, we tested the algorithm in the RS phase of regular
random graphs for $q=3,4,5,6,7$. In all these cases, we were able to color in
linear time all the graphs of connectivities that correspond to a replica symmetric
solution. In particular, the cases
$(q=3,c=5)$, $(q=5,c=13)$, $(q=6,c=17)$, $(q=7,c=21)$, $(q=7,c=22)$ are found to
be colorable with the Walk-COL algorithm even if a nontrivial solution to the
SP equations exists.

\begin{figure}[!ht]
\begin{minipage}{0.49\linewidth}
\begin{center}
  \resizebox{9cm}{!}{\includegraphics[angle=270]{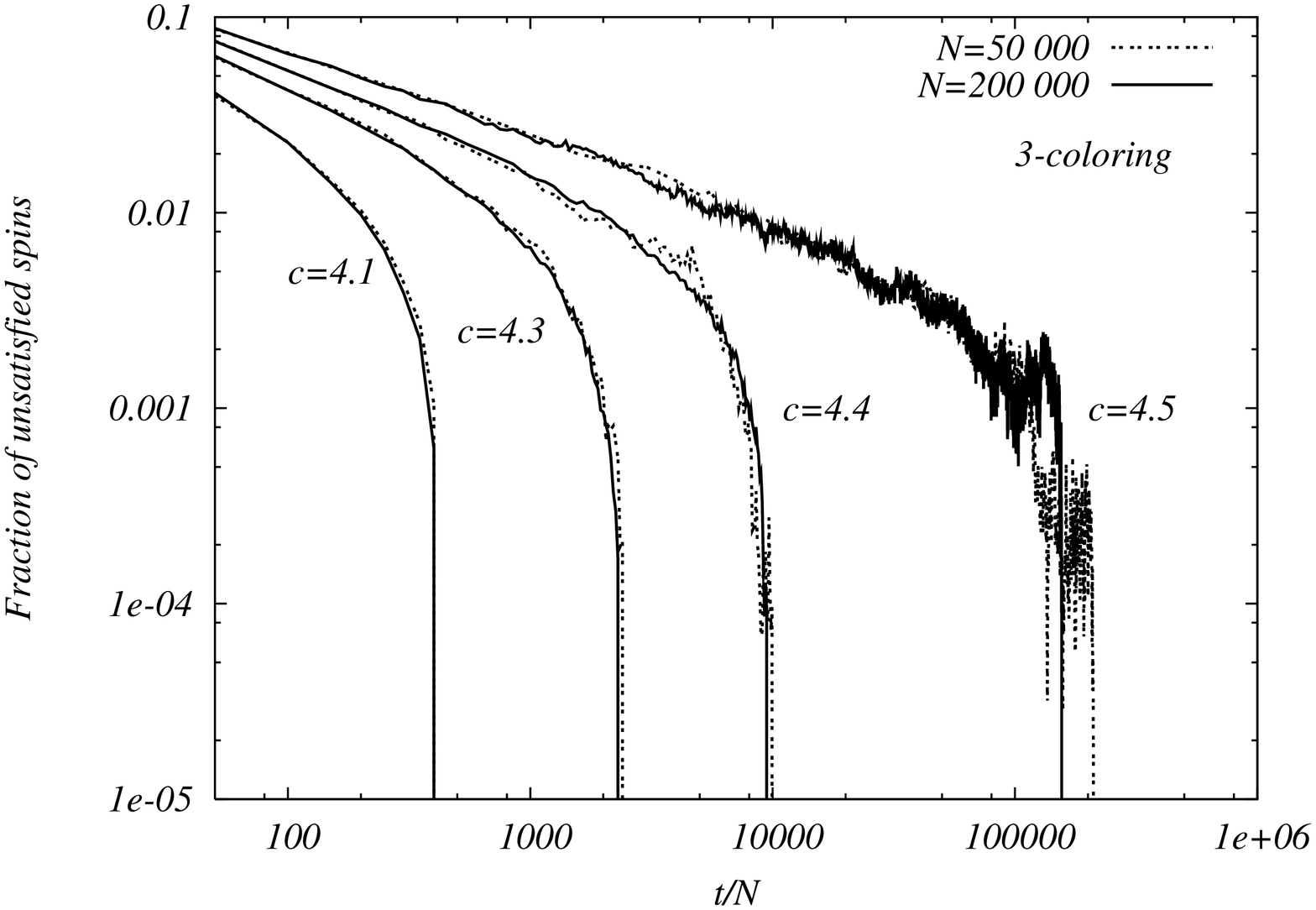}}
\end{center}
\end{minipage}
\begin{minipage}{0.49\linewidth}
\begin{center}
  \resizebox{9cm}{!}{\includegraphics[angle=270]{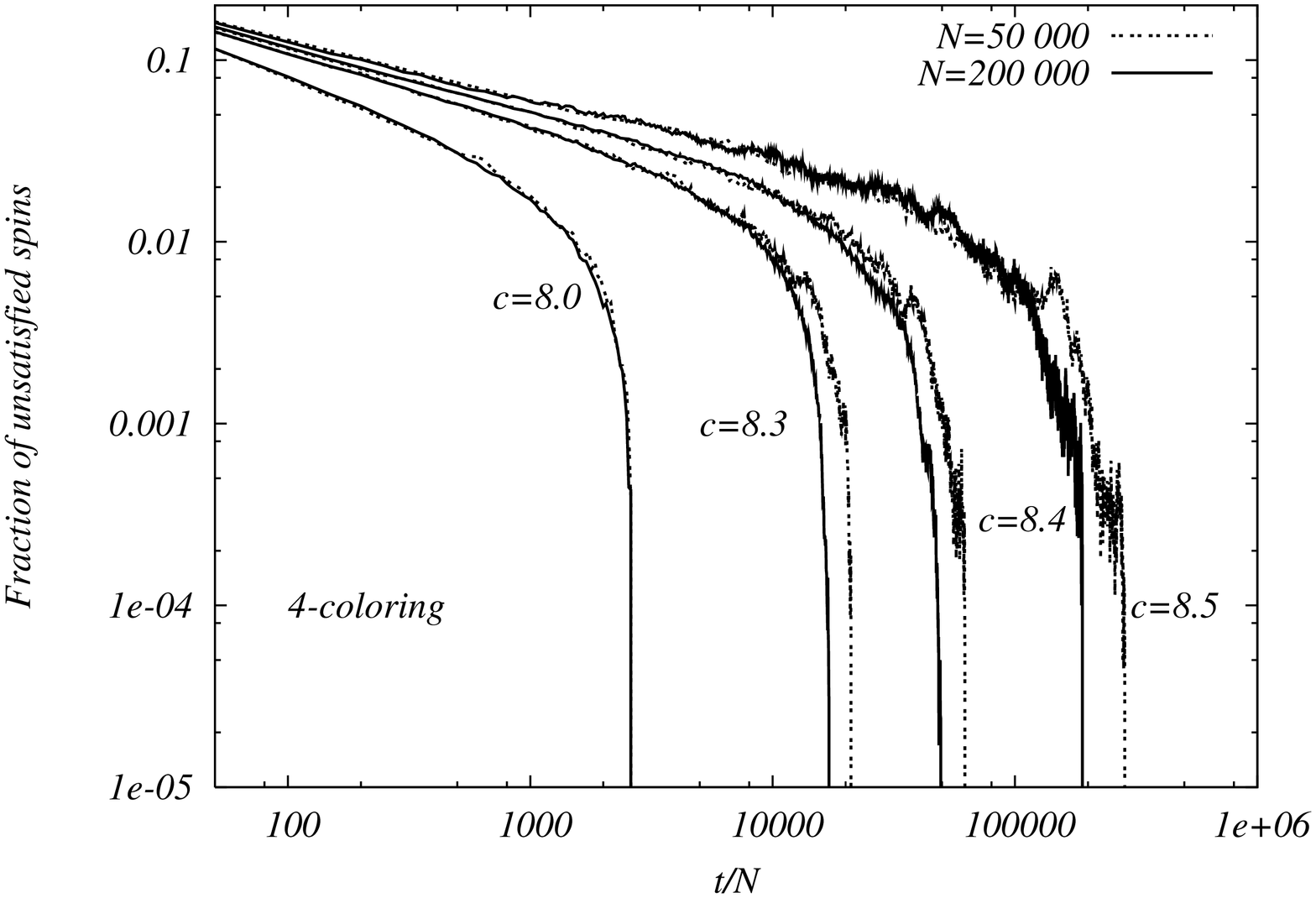}}
\end{center}
\end{minipage}
\caption{ \label{fig:walkcol} Performance of the Walk-COL algorithm in coloring random graphs for
  $3-$coloring (left) and $4-$coloring (right). We plot the rescaled time
  (averaged over $5$ instances) needed to color a graph of connectivity $c$.
  The strategy allows one to go beyond the clustering transition ($c_d=4$ for
  3-coloring and $c_d=8.35$ for 4-coloring) in linear time with respect to the
  size of the graph.}
\end{figure}

Concerning the second point, we considered the $3-$ and $4-$coloring of
Erd\H{o}s-R\'enyi random graphs. The results are shown in fig. \ref{fig:walkcol}
where the percentage of unsatisfied spins versus the number of attempted flips
(averaged over $5$ different realizations) divided by $N$ is plotted.  We
observe that the curves corresponding to different values of $N$ superpose
quite well (and that actually the results for $N=2\cdot 10^5$ are systematically
lower than those for $N=5\cdot 10^4$) so that an estimation of the time needed to color a
graph can be obtained. The connectivities of these graphs are beyond the
dynamical transition ($c_d=4$ for 3-coloring and $c_d=8.35$ for 4-coloring).
It would be interesting to systematically test Walk-COL, as it has been 
done for Walk-SAT in \cite{John}, to derive the precise connectivity at 
which it ceases to be linear.

Already these results show that the dynamical transition is not
a problem for the algorithms. This can also be observed in a number
of numerical experiments for the satisfiability \cite{walksat_whitening,John}
and the coloring \cite{Jorge,Stefan} problems.

We believe, however, that the rigidity transition plays a fundamental role for
the average computational complexity.  A first argument for this is that, 
for large graphs, it seems that 
all the known algorithms are only able to find solutions with a
trivial whitening, {\it i.e.} solutions that belong to clusters without hard fields.
Beyond the rigidity transition however, the clusters without hard fields
become very rare (in the sense that the dominating clusters and all the
smaller, more numerous ones, contain hard fields). For $q \ge 9$ maybe the
connectivity where hard fields appear in clusters corresponding to
$\Sigma(m)=0$, $m>1$ should be considered. This suggests that the known 
algorithms will not be able to find a solution beyond this point. 

A second argument is the following: local search algorithms are either
attracted into a solution or stucked in a metastable state. These metastable
states, in order to be able to trap the dynamics, have to contain a finite
fraction of hard fields. Given an algorithm, determining which of these two
situations happens is not only a question of existence of states, but also a
question of basins of attraction and a theoretical analysis of such basins is a
very difficult task so that the precise analysis of the behavior of local
algorithms remains a hard problem. However, the metastable states are known,
from the cavity formalism, to be much more numerous than the zero-energy
states. Moreover the basin of attraction of a zero-energy state that contains
hard fields does not probably differ much from those of the metastable state
(while, on the other hand, the basin of attraction of a zero-energy state which
does {\it not} contain hard fields might be slightly different and arguably
relatively larger). It thus seems to us reasonable that local algorithms will get 
trapped by the metastable states beyond the rigidity transition.

A similar conclusion was reached recently in \cite{Jorge} where the recursive
implementation of the Walk-COL algorithm was studied and found to be somehow
simpler to analyze. Again the strategy was found to be efficient (with linear time with respect to the size of the graph) beyond $c_d$
{\it but bellow} $c_r$. The precise algorithmic implications of the rigidity
transition thus require further investigation, maybe in the lines of
\cite{walksat_whitening,Jorge}.

\subsection{A belief propagation algorithm to color random graph}
\label{sec:BP}

Another consequence of our results, that we already discussed shortly in~\cite{US}, is that the standard belief propagation (BP) algorithm gives
correct marginals until the condensation transition. It is actually a simple
algebraic fact that the 1RSB approach at $m=1$
gives the same results for the marginals (average probabilities over all
clusters) as the simple RS approach (see appendix \ref{app:m1}).
Moreover the log-number of solutions in clusters corresponding to $m=1$ is
also equal to the RS
one. This suggests to use the BP marginals (as was already suggested in \cite{BP-Pearl})
and a decimation procedure to find proper colorings.
Compared with the SP algorithm which has computational
complexity proportional to $q!$ (factorial) the BP is only $q$. We have seen moreover that
for large numbers of colors, the condensation point is very close to the
COL/UNCOL transition, so that BP could be used in a large interval of connectivities.

As a simple application, we tested how the straightforward implementation of
the BP algorithm plus decimation allows one to find solutions of $3-$ and
$4-$coloring of random Erd\H{o}s-R\'enyi graphs. Note that in the $3-$coloring
the clustered but not condensed phase is missing, so the argumentation above
does not concern this case. The algorithm works by iterating the following
procedure:
\begin{itemize}
\item[(i)] Run BP on the graph for a given number $l$ of iterations.
\item[(ii)] Consider the most biased variable, and color it with its most
  probable color.
\end{itemize}

Two problems have to be mentioned. The first one is rather trivial: since at
the beginning all colors are symmetric, the first color had to be put at
random. The second one is more serious and concerns the convergence of BP.
Indeed, we saw that there is local instability in the BP (replica symmetric)
equations at connectivity $c=4$ for the $3-$coloring of random graphs, so that
the BP equations {\it do not converge}. This seems to be a problem restricted
to the $3-$coloring, but even in the case of $4-$ or more coloring, the BP equations do not converge on the decimated graph
when a finite fraction (typically few percent) of variables is fixed.
The reason or that is to be understood.

\begin{figure}
\begin{minipage}{0.49\linewidth}
\begin{center}
  \resizebox{9cm}{!}{\includegraphics[angle=0]{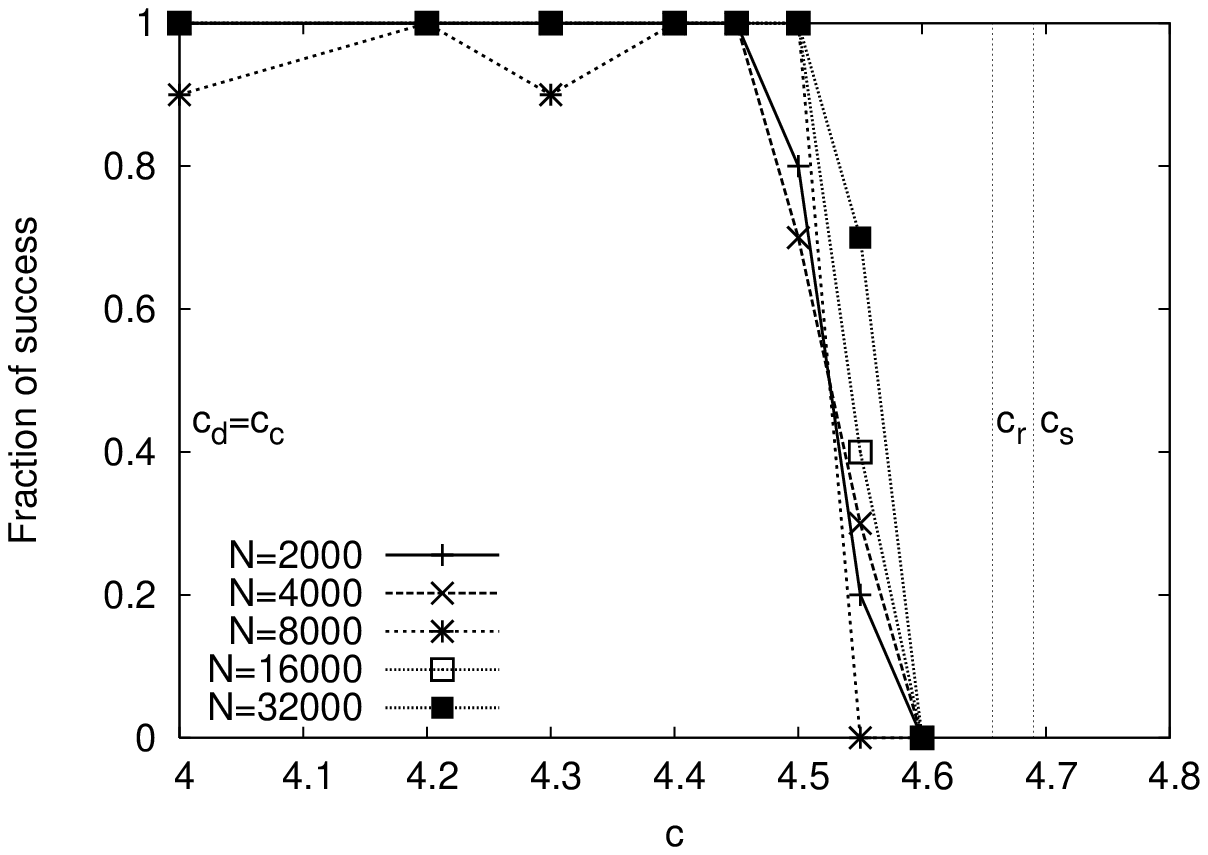}}
\end{center}
\end{minipage}
\begin{minipage}{0.49\linewidth}
\begin{center}
  \resizebox{9cm}{!}{\includegraphics[angle=0]{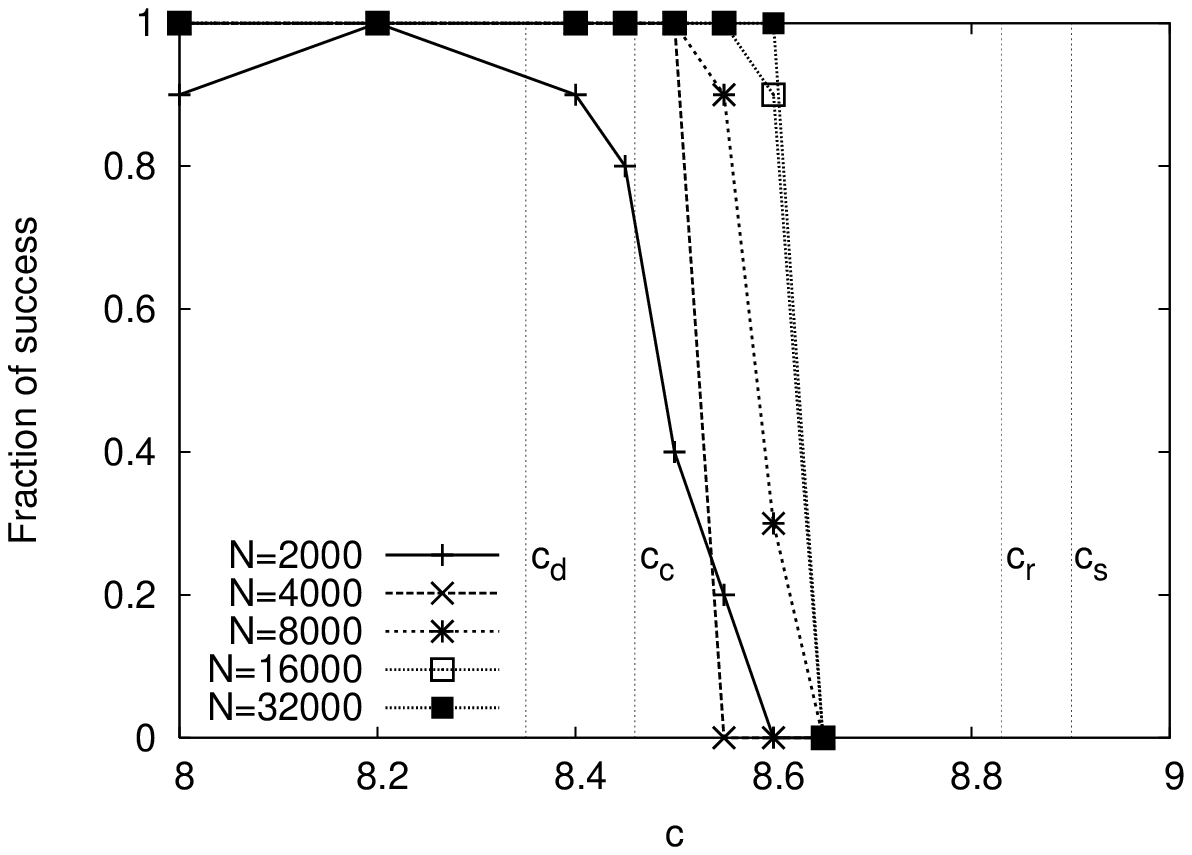}}
\end{center}
\end{minipage}
\caption{ \label{fig:BP} Performance of the BP algorithm plus decimation 
in coloring random graphs for  $3-$coloring (left) and
$4-$coloring (right). The strategy described in the text allows 
to color random graphs beyond the clustering and even the condensation 
transitions.}
\end{figure}

Nevertheless, since we merely want to design an effective tool to solve the
coloring problem, we choose to avoid this problem by fixing the number of
iteration $l$ at each step and thus ignore the non-convergence.  We tried the
method on both the $3-$ and $4-$coloring and obtained unexpectedly good
results. We used the following protocol in the code: We first try to find a
solution with $l=10$. If we do not succeed, we restart with $l=20$ and once
more with $l=40$. We tried that on $10$ different samples for different
connectivities. The probability to find a proper coloring with these
conditions is shown in fig.\ref{fig:BP}. 

We thus observe that the BP strategy is able to find solutions, even beyond
the condensation transition. This shows clearly that the decimation procedure
is a nontrivial one, and that the problem is not really hard in that region of
connectivities. Note that the SP algorithm plus decimation has been shown to
work in the $3-$coloring very well until about $4.60$ \cite{Coloring}: 
our results are thus very close to those obtained using SP. This rises again the question
of the rigidity transition $c_r=4.66$, which might also be problematic for the
decimated survey propagation solver.

\section{Conclusions}
\label{sec:conclustions}

Let us summarize the results. They are perhaps best illustrated looking back
to the cartoon in fig.~\ref{fig:zero_T_3}, where the importance of the size of
clusters is evidenced.  We find that the set of solutions of the $q$-coloring
problem undergoes the following transitions as the connectivity is increased:
\begin{enumerate}
\item[(i)] At low connectivity, $c<c_d$, many clusters might exist but 
  they are very small and the measure over the set of solutions is dominated
  by the single giant cluster described by the replica symmetric 
  approach. 
\item[(ii)] Only at the {\it dynamical} transition $c_d$ the giant cluster 
  decomposes abruptly into an exponentially large number of clusters (pure states). 
  For connectivities $c_d<c<c_c$, the measure is dominated 
  by an exponential number of clusters. Yet, the total number of solutions is 
  given by the replica symmetric entropy (\ref{S_RS}), and the marginals 
  are given by the fixed point of the replica symmetric equations (belief 
  propagation) (\ref{update}). Starting from this transition the uniform 
  sampling of solution becomes hard. 
\item[(iii)] At connectivity $c_c$ the set of solutions undergoes a
  {\it condensation} transition, similar to the one appearing in mean field spin
  glasses. In the condensed phase the measure is dominated by finite 
  number of the largest clusters. The total entropy is strictly smaller 
  than the replica symmetric one and has a discontinuity in the second 
  derivative at $c_c$. 
\item[(iv)] When connectivity $c_s$ is reached, no more clusters exist: 
  this is the COL/UNCOL transition. Note that the entropy of last existing 
  clusters is strictly positive, and not given only by the contribution of 
  the isolated nodes, leaves and other small subgraphs, the COL/UNCOL 
  transition is thus discontinuous in entropy.   
\end{enumerate}.

This picture is very similar to the well-known scenario of the glass
transition in temperature, with the dynamical and glass (Kauzmann) transition
\cite{GLASSTHEORY}. In some cases, the main one being 
the 3-coloring of Erd\H{o}s-R\'enyi graphs, the clustering and the condensation 
transition merge and a continuous transition take place 
at $c_d=c_c$, which then coincide with the local instability of 
the replica symmetric solution. Interestingly the variational approach of
\cite{GiulioRemiMartin} is very precise near to the continuous clustering 
transition. Since the 3-SAT problem behaves in the same manner, 
this solves the apparent contradictions between the results of
\cite{GiulioRemiMartin} and \cite{Science}.

In addition to the transitions describing the organization of clusters,
another important phenomenon concerning the internal structure of clusters
takes place. A finite fraction of frozen variables can appear in the clusters
(a frozen variable takes the same color in all the solutions that belong to
the cluster). We found that the fraction of such variables in each cluster
undergoes a first order transition and jumps from zero to a finite fraction at
a connectivity that depends on the size of the cluster.  In particular:
\begin{enumerate}
\item[(v)] There exists a critical connectivity $c_r$ (rigidity/freezing) at
  which the thermodynamically relevant clusters ---those that dominate the
  Gibbs measure--- start to contain a finite fraction of frozen variables.
\end{enumerate}

The results above were obtained within the 1RSB scheme, but should 
not change when considering further steps of RSB (an exception might be 
the 3-coloring near to the clustering transition). 

We discussed some algorithmic consequences of these transitions. First, 
the belief propagation algorithm is efficient in counting
solution and estimating marginals until the condensation transition. More
interestingly, it can also be used, just like survey propagation, together with a 
decimation procedure in order to find solutions as we numerically demonstrated.
Secondly, the dynamical transition is not the one at which 
simple algorithms fail as we illustrated using the Walk-COL
strategy.  For the 3-coloring of ER graph, 
there is even a  rigorous proof of algorithmic performance beyond $c_d=4$ and
until $c=4.03$ \cite{AchlioptasMoore03}.
We argued that, instead, the rigidity phenomenon
is responsible for the
onset of computational hardness.
This is a major point that we hope to see more investigated in the future.

Our study opens a way to many new and promising investigations and
developments. For instance, we wrote the equivalent of the survey propagation
equations for general value of $m$, which has a particularly simple form
for $m=1$ (\ref{SP_m1_app}). It would be interesting to use these equations to find
solutions. The behavior at finite temperature and the performance of the
annealing procedure are also of interest. It would furthermore be interesting
to re-discuss other finite connectivity spin glass models like for instance
the lattice glass models \cite{Rivoire} in the light of our findings. 
The stability towards more
steps of replica symmetry breaking, or the super-symmetric approach
\cite{SUSY}, should be further investigated. Finally, it would be interesting to combine
the entropic and energetic approach to investigate the frozen variables in the
meta-stable states. We hope that our results will stimulate the activity in
these lines of thoughts.

\begin{acknowledgments}
  We thank Jorge Kurchan, Marc M\'ezard, Andrea Montanari, Federico
  Ricci-Tersenghi, Guilhem Semerjian and Riccardo Zecchina for cheerful and
  very fruitful discussions concerning these issues. The numerical
  computations were done on the cluster EVERGROW (EU consortium FP6 IST) at
  LPTMS, Orsay, and on the cluster DALTON at ESPCI, Paris. This work has been
  partially supported by EVERGROW (EU consortium FP6 IST).
\end{acknowledgments}


\appendix

\section{Stability of the paramagnetic solution}
\label{app:stab}

In this appendix, we show how to compute the stability of the paramagnetic 
solution towards the continuous appearance of a 1RSB solution. 
This happens, as usual
for continuous transition, when the spin glass correlation length, or
equivalently, the spin glass susceptibility, diverges. Obviously, the presence
of the diverging correlation length invalidate the premise of the RS cavity
method.  Recall that the spin glass susceptibility is defined as
\be  \chi_{\rm SG} = \frac{1}{N} \sum_{i,j} \langle s_i s_j\rangle^2_c \, . \ee
and can be rewritten for the present purpose as 
\begin{equation}
  \chi_{\textrm{SG}}=\sum_{d=0}^\infty \gamma^d {\mathbb{E}}(\langle s_0 s_d \rangle^2_c)\, ,
  \label{chi2}
\end{equation}
where we consider the average over graphs, in the thermodynamic limit, 
where spins $s_0$ and $s_d$ are at distance $d$. The factor $\gamma^d$ 
stands for the average number of neighbors at distance $d$, when $d \ll \log N$.
Assuming that the limit for large $d$ of the summands in (\ref{chi2})
exists (with the limit $N \to \infty$ performed first), we relate it
to the \emph{stability parameter}:
\begin{equation}
   \overline \lambda = \lim_{d\to \infty}
   \gamma \Big(
   {\mathbb{E}}(\langle s_0 s_d \rangle^2_c)
   \Big)^{\frac{1}{d}} \, .
\label{stab}
\end{equation}
Then the series in (\ref{chi2}) is essentially geometric, and converges if
and only if $\overline{\lambda}<1$.

Using the fluctuation-dissipation theorem we relate the
correlation $\langle s_0 s_d \rangle_c$ to the variation of
magnetization in $s_0$, caused by an infinitesimal field in
$s_d$. 
Finally, using the fact that we perform the large-$N$ limit first, 
the variation above is dominated 
by the direct influence through the length-$d$ path between the two nodes,
and this induces a ``chain'' relation: if the path involves the nodes
$(d, d-1, \ldots, 0)$ we have
\begin{equation}
{\mathbb{E}}(\langle s_0 s_d \rangle^2_c)=
{\cal C} \cdot
\sum_{\substack{
a \in \partial d \\
b \in \partial 0 }}
{\mathbb{E}}
\left[\left( \frac{\partial \psi^{a \to 0} }
  { \partial \psi^{b \to d} }\right)^2 \right]
=
{\cal C} \cdot
{\mathbb{E}}
    \left[ \prod_{\ell=1}^d
\left( \frac{\partial \psi^{l \to l-1} } 
    {\partial \psi^{l+1 \to l} } \right)^2 \right]
\, .
\label{fdt}
\end{equation}

The stability parameter of the paramagnetic solution of the cavity
equations towards small perturbations can be computed from the following
Jacobian
\begin{equation}
  \label{first_kind}
  T^{\tau\sigma} = \frac {\partial \psi^{1\to 0}_\tau} {\partial
    \psi^{2\to 1}_\sigma} \bigg|_{\rm RS}\, ,
\end{equation}
which gives the infinitesimal probability that a change in the input
probability $\psi^{2\to 1}_\sigma$ will change the output probability $\psi^{1\to 0}_\tau$.
The index RS says that the expression has to be evaluated at the RS
paramagnetic solution. 

This matrix has only two
different entries, all the diagonal elements are equal, and all the non-diagonal
elements are also equal. As an immediate consequence all Jacobians commute and
are thus simultaneously diagonalizable so that it will be sufficient to study
the effect after one cavity iteration (one step in the chain). The matrix $T$ has only two distinct
eigenvalues,
\begin{equation}
  \label{jac_eig}
  \begin{cases}
    \lambda_1 = \((  \frac {\partial \psi^{1\to 0}_1} {\partial \psi^{2\to 1}_1} -
    \frac {\partial \psi^{1\to 0}_1} {\partial \psi^{2\to 1}_2} \)) \Big|_{\rm RS} \, ,
    \\  \\
    \lambda_2 =\(( \frac {\partial \psi^{1\to 0}_1} {\partial \psi^{2\to 1}_1} +
    (q-1)\ \frac {\partial \psi^{1\to 0}_1} {\partial \psi^{2\to 1}_2} \))
    \Big|_{\rm RS}  \, .
\end{cases}
\end{equation}
The second eigenvalue corresponds to the homogeneous eigenvector
$(1,1,...,1)$ and describes a fluctuation changing all $\psi^{2\to 1}_\tau,
\tau=1,...,q,$ by the same amount, and maintains the
color symmetry. It is thus not likely to be the relevant
one and we will see that indeed $\lambda_2=0$. The first eigenvalue,
however, is $(q-1)$-fold degenerate and its eigenvectors are spanned by
$(1,-1,0,...,0),\ (0,1,-1,0,...,0),\ \dots,\ (0,...,0,1,-1)$. The
corresponding fluctuations explicitly break the color symmetry, and
are in fact the critical ones. Using the cavity recursion (\ref{update}), the two
derivatives simply read
\be
\begin{cases}
  \frac {\partial \psi^{1\to 0}_1} {\partial \psi^{2\to 1}_2} =(1-e^{-\beta}) \frac
  {(\psi^{1\to 0}_1 )^2}{1-(1-e^{-\beta})
    \psi_1^2} \, ,\\  \\
  \frac {\partial \psi^{1\to 0}_1} {\partial \psi^{2\to 1}_1} = (1-e^{-\beta}) \(( \frac
  {(\psi^{1\to 0}_1 )^2}{1-(1-e^{-\beta})\psi_1^{2\to 1}} -
  \frac{\psi^{1\to 0}_1}{1-(1-e^{-\beta})\psi_1^{2\to 1}}\)) \, ,
\end{cases}
\ee so that the values of the two eigenvalues evaluated at the RS solution,
where all $\psi$ are equal to $1/q$, are 
\be
  \lambda_1 = \frac 1 {1-\frac q{1-e^{-\beta}}},
\text{~~~~~~~}
  \lambda_2 =  0.
\ee
The stability parameter (\ref{stab}) is thus $\overline \lambda = \gamma \lambda_1^2$ and
the critical temperatures bellow which the instability sets in are
\bea
 T_c^{\rm reg} (q,c) = - 1/{\log\(( 1-\frac q{\sqrt{c-1}+1}\))} \text{,~~~~~}
T_c^{\rm ER} (q,c) = - 1/{\log \((1-\frac q{\sqrt{c}+1}\))}. \label{stab_KS}
\eea
For regular and Erd\H{o}s-R\'enyi graphs respectively. Thus at zero temperature the critical connectivities reads
\bea 
c^{\rm reg}_{\rm RS~stab}= q^2 -2q+2\, \text{,~~~~}c^{\rm ER}_{\rm RS~stab}= q^2 -2q+1. 
\eea
These results coincide perfectly with the numerical simulations of the cavity
recursion of~\cite{ColoringSaad}. The analytical expressions equivalent to
(\ref{stab_KS}) were in fact first obtained in~\cite{KS66} in the context of the
reconstruction problem on trees as an upper bound for the Gibbs measure
extremality, and its connection with the statistical physics approach was
explained in~\cite{MM05}. The case of bi-regular random graphs can
be easily understood by noticing that two recursions should be considered, one with 
$\gamma=c_1-1$, and one with $\gamma=c_2-1$. As a consequence, the stability point 
is equivalent in this case to the one of a regular random graph with an effective 
connectivity equal to $c=1+\sqrt{\((c_1-1\))\((c_2-1\))}$.

Another instability appears when $\gamma |\lambda_1| > 1$. This has been refered to
as the {\it modulation} instability in~\cite{Rivoire}. Actually, this is
the continuous instability towards the appearance of the
anti-ferromagnetic order. Since at zero temperature $\lambda_1=(1-q)^{-1}$,
then for connectivities larger than $c_{mod}=q$ for random regular graphs (and
$c_{mod}=q-1$ for Erd\H{o}s-R\'enyi) the paramagnetic solution
becomes unstable towards the anti-ferromagnetic order.  However, this is
correct if we study a tree with some given (and well chosen) boundary
condition, but as noted in~\cite{Rivoire}, the anti-ferromagnetic solution 
in impossible on random graphs because of the existence of frustrating loops 
of arbitrary length. The cavity equations (\ref{update}) can actually
never converge towards an anti-ferromagnetic solution of a random graph.
Instead, when iterating, the fields oscillate between different solutions (thus
the name {\it modulation}). 

In other words, although on a random tree with special boundary conditions
there exists for $c>c_{mod}$ a nontrivial solution to the cavity recursion
(for the Gibbs state is no longer unique (\ref{Gibbs_uni})), this solution
does not exist on a random graph (and the Gibbs state is still extremal
(\ref{extremality})).  Note that this instability could anyway be a source of
numerical problems that can be overcome considering that the distribution of
cavity fields ${\cal P}(\psi)$ over the ensemble of random graphs
has to be symmetric in the color permutation.  Another possibility is
to randomly mix the new and old populations in the population dynamics so that
the anti-ferromagnetic oscilations are destroyed.

\section{The relative sizes of clusters in the condensed phase
}
\label{app:PD}

In this section, we introduce
the Poisson-Dirichlet point process and we shortly 
review some of its important properties. We also sketch its deep 
connection with the size of clusters in the condensed phase. 
Poisson-Dirichlet (PD) point process is a set of points $\{x_i\}$, 
$i=1,\dots,\infty$ such that $x_1>x_2>x_3 > \dots $ and $\sum_{i=1}^\infty x_i=1$.
To construct these points we consider a Poisson process $\{y_i\}$, 
$i=1,\dots,\infty$ of intensity measure $y^{-1-m^*}$, $0<m^*<1$ (note that this measure 
is not a probability measure). We order the sequence $\{ y_i\}$ in such a way that
$y_1 > y_2 > y_3 > \dots$ and define the PD point process as
\begin{equation}
      x_j = \frac{y_j}{\sum_{i=1}^\infty y_i}.
\end{equation}

If we identify the parameter $m^*$ with the value for which the 
complexity is zero $\Sigma(m^*)=0$ then $y_i$ is proportional to the 
number of solutions in cluster $i$ (or to $e^{-\beta F}$ for 
non-zero temperature), and $x_i$ is the size on that cluster relative to
the total number of solutions. This connection was (on a non-rigorous level) 
understood in~\cite{MPV}, for more mathematical review see~\cite{Talagrand}. 
Note that that due to the
permutation symmetry in graph coloring there are every time $q!$ copies of one
clusters (different in the color permutation).

\begin{figure}[!ht]
\begin{center}
  \resizebox{9cm}{!}{\includegraphics[angle=270]{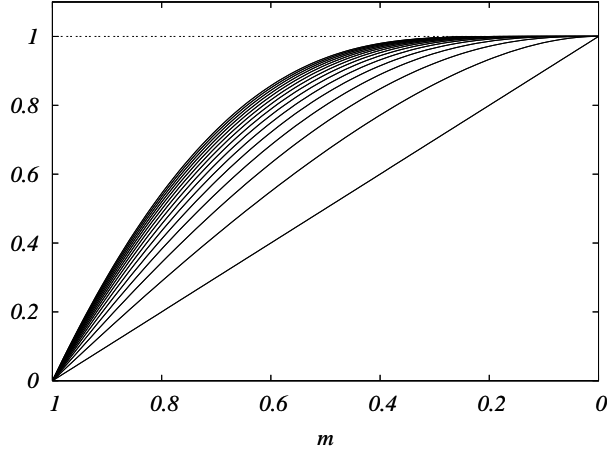}}
\end{center}
\caption{ \label{fig:PD} The sketch of size of the largest clusters for given value 
of parameter $m^*$. The lower curve is related to the average size of 
the largest clusters as $1/ \mathbb{E}[1/x_1]=1-m^*$. The following curves are 
related to the size of $i$ largest clusters, their distances 
are $ \mathbb{E}[R_i]  \mathbb{E}[R_{i-1}] \dots  \mathbb{E}[R_1] (1-m^*)$.}
\end{figure}

To get feeling about the PD statistics let us answer in fig. \ref{fig:PD} to the following 
question: Given the value $m^*$ how many clusters do we need to cover
fraction $r$ of solutions, in other words what is the smallest $k$ such 
that $\sum_{i=1}^k x_i >r$?  
The mathematical properties of the PD process are very clearly 
reviewed in~\cite{PitmanYor}. 
To avoid confusion, note at this point that the PD process 
we are interested in is the ${\rm PD}(m^*,0)$ in the notation of~\cite{PitmanYor}. 
In the mathematical literature, it is often referred to the ${\rm PD}(0,\theta)$ 
without indexing by the two parameters.
Let us remind two useful results. Any moment of any $x_i$ can
be computed from the generating function 
\begin{equation}
          \mathbb{E} [\exp{(-\lambda/x_i)}] = e^{-\lambda} \phi_{m^*}(\lambda)^{i-1} 
          \psi_{m^*}(\lambda)^{-i} \, ,
\end{equation} 
where $\lambda \ge 0$ and the functions $\phi_{m^*}$ 
and $\psi_{m^*}$ are defined as
\begin{eqnarray}
     \phi_{m^*}(\lambda) &=& m^* \int_1^\infty  e^{-\lambda x} x^{-1-m^*} {\rm d} x\, , \\
     \psi_{m^*}(\lambda) &=& 1+m^* \int_0^1(1 -e^{-\lambda x}  )  x^{-1-m^*} {\rm d} x\, .
 \end{eqnarray}

Another relation is that the ratio of two consequent points $R_i=x_{i+1}/x_i$,
$i=1,2,\dots$ is distributed as $im^* R_i^{im^*-1}$.  In particular its 
expectation is $ \mathbb{E}[R_i]=im^*/(1+im^*)$ and the random variables $R_i$ are mutually independent.
We used these relation to obtain data in figure~\ref{fig:PD}.

\section{The 1RSB formalism at $m=1$ and the reconstruction equations}
\label{app:m1}

In this appendix we discuss the considerable simplification of eq. (\ref{1RSB_pop}) 
that is obtained by working directly at $m=1$. This was first remarked and proved in
\cite{MM05} when dealing with the tree reconstruction problem (for a 
discussion of a case where the RS solution is not paramagnetic 
see~\cite{THEM}). We first introduce the probability distribution of fields 
(\ref{1RSB}) averaged over the graph
\begin{eqnarray}
        \overline P(\psi) &\equiv& \int {\rm d}P \, {\cal P}[P(
        \psi)] \, P(\psi) =
        \sum_k {\cal Q}_1(k) \frac{1}{Z_1} \int \prod_{i=1}^k {\rm d}\psi^i
         \, \overline P^i(\psi^i) \,  \delta \[[ 
         \psi-{\cal F}(\{\psi^i\}) \]]  \, Z_0, \label{1RSB_m1_pom}
\end{eqnarray}
where $Z_1$ is computed from (\ref{1RSB}) as 
\begin{eqnarray}
          Z_1 \equiv \int \prod_{i=1}^k {\rm d}\, \psi^i \, P^i(\psi^i)\,  Z_0 =
          \sum_{j=1}^q \prod_{i=1}^k \left(1- \overline \psi^i \right)\, ,
\end{eqnarray}
where $\overline \psi=\int {\rm d}\, \psi \, P(\psi)\, \psi$.
Generally, $\overline \psi$ is a solution of the RS equation (\ref{RS_pop}),
which is easily seen from (\ref{1RSB}), (\ref{1RSB_pop}).  Since the 
RS solution is the paramagnetic one $\overline \psi=1/q$ the form of $Z_1$
is particularly simple.

In the next step, we want to get rid of the term $Z_0$ in eq.
(\ref{1RSB_m1_pom}). We thus introduce $q$ distribution functions
$\overline P_s$
\begin{equation}
              \overline P_s(\psi) = q \psi_s \overline P(\psi)\, .
\end{equation} 
It is then easy to show that if $\overline \psi=1/q$ then $\overline
P_s(\psi)$ satisfies 
\begin{eqnarray}
      \overline P_{s}(\psi) &=& \sum_k {\cal Q}_1(k)
      \sum_{s_1\dots s_k} \prod_{i=1}^k \pi(s_i|s) \int 
      \delta\[[\psi- {\cal F}(\{\psi^i\}) \]] 
      \prod_{i=1}^k {\rm d} \psi^{i} \overline P_{s_i}(\psi^{i}) \, ,
      \label{1RSB_m1}
\end{eqnarray}
where  
\begin{equation}
      \pi(s_i|s)=\frac{1-(1-\e^{-\beta})\delta(s_i,s)}{q-(1-\e^{-\beta})}.
\end{equation}
We solve eq.~(\ref{1RSB_m1}) by population dynamics. In order to do this, one
needs to deal with $q$ populations of $q$-component fields, and to update them
according to (\ref{1RSB_m1}). Is is only a functional equation and not a
double-functional as the general 1RSB equation (\ref{1RSB_pop}).  Moreover the
absence of the reweighting term $Z_0^m$ simplifies the population dynamics
algorithm significantly. Finally, it is important to note that the
computational complexity here is the same as the one for regular and ER random
graphs.

A crucial theorem is also proven in~\cite{MM05}: the population dynamics of
eq. (\ref{1RSB_m1}) has a nontrivial solution if and only if it converge to a
nontrivial solution starting from initial conditions:
\begin{equation}
      \overline  P_{r}^{\rm init}(\psi_s)=\delta(r,s).   \label{init}
\end{equation}
This shows that when a paramagnetic solution is found, 
then no other solutions exist.

Similar manipulations allow us to obtain the replicated free energy (\ref{Psi})
which is equal in this case to the replica symmetric free energy (\ref{free_en}),
and the free energy (\ref{free_energy}) inside the corresponding states as 
\bea -\beta f(\beta) &=& \sum_k {\cal
  Q}(k) \sum_{s_1\dots s_k} \sum_s \frac{1}{q} \prod_{i=1}^k \pi(s_i|s) \int
\left(\log{ Z_0^i}\right) \, \prod_{i=1}^k {\rm d} \psi^{i} \overline
P_{s_i}(\psi^{i}) \, ,\\ \nonumber &-& \frac{c}{2} \sum_{s_1, s_2} \frac{\pi(s_1|s_2)}{q} \int
\left(\log{ Z_0^{12}}\right) \, {\rm d} \psi^{1} {\rm d} \psi^{2}\, 
\overline P_{s_1}(\psi^{1}) \overline P_{s_2}(\psi^{2})\, , \eea 
where the normalization factors $Z_0^i$, $Z_0^{ij}$ are defined by (\ref{free_i}) and
(\ref{free_ij}). The complexity follows from (\ref{comp}).
Since the replicated free energy $\Phi(\beta,1)$ is equal,
according to (\ref{phi_m1}), to the total free energy, we showed the statement 
used several time in the paper, {\it i.e.} the total free energy (entropy) at $m=1$ 
is equal to the replica symmetric free energy (\ref{free_en}).

Another important point is that one can write the recursion separating the
 hard and soft fields. In general, at zero temperature,
 we can write the distribution $\overline P_{q}(\psi_{s})$ in eq.
(\ref{1RSB_m1}) as 
\begin{equation}
      \overline  P_{r}(\psi)=\sum_{s}   \mu_{r,s} \delta(\psi_s-1) + (1-\sum_{s}
        \mu_{r,s}) {\tilde P}_{r}(\psi).
\end{equation}
Plugging this to eq.~(\ref{1RSB_m1}) and taking into account the initial
condition (\ref{init}) and color symmetry, we see that $\mu_{q,s}=q\eta
\delta(q,s)$, where $\eta$ satisfies 
\begin{equation}
        q\eta = \sum_k {\cal Q}_1(k)  \sum_{m=0}^{q-1} (-1)^m 
        {q-1 \choose m} \left(1- \frac{mq}{q-1}\eta\right)^k\, .
        \label{hard_m1_app}
\end{equation}
On ER graphs, the sum can be performed analytically and one finds 
\be q\eta=\left(1-\e^{-\frac{q \eta
      c}{q-1}} \right)^{q-1}\, . \ee

This equation can be solved iteratively starting from $\eta^{\rm init}=1/q$. It
is a very simple equation, as the one obtained for $m=0$, which gives us a
very efficient way to compute the fraction of hard fields at $m=1$ for both
regular and ER graphs. Indeed $\eta$ is larger that zero only above a certain
average connectivity $c_r(m=1)$.

\section{Numerical Methods}
\label{app:num}

In this section, we detail the numerical methods we used to solve the
1RSB equations (\ref{1RSB},\ref{1RSB_pop}), and the procedures used to generate the data.
We use a population dynamics method, as introduced in~\cite{MP99,MP03}, and
model the distribution $P^{i\to j}(\psi^{i\to j})$ by a population of $N$
vectors $\psi^{i\to j}$.  To compute $P^{i\to j}(\psi^{i\to j})$ knowing the
$P^{k\to i}(\psi^{k\to i})$ for all incoming $k$ we perform the 1RSB
recursion in eq. (\ref{1RSB}) in two steps: (i) first we compute the new
vectors $\psi^{i\to j}$ using the simple RS recursion in
eq. (\ref{update_zero}) (this is the iterative step) and (ii) we
take into account the weight $(Z_0^{i\to j})^m$ for each of the vectors
(this is the reweighting step). For the reweighting we tried different strategies, two of them perform very
well.
\begin{itemize}
\item[a)]{For every field $\psi$ in the population, we keep its weight $Z_0$.
    We then compute the cumulative distribution of weights $Z_0$ and sample
    uniformly the incoming fields. Using dichotomy we generate a random fields with its proper
    weight in $O(\log(N))$ steps. A complete iteration thus takes $O(N\log{N})$ steps.}
\item[b)]{We compute $N$ new vectors and then we make a new population when we
    clone some of them while erasing others so that in this new population
    each field is present according to its weight (in principle, one can even
    change the size of the population, although we have not implemented this
    strategy).  This second approach can be implemented in linear time
    (generating an ordered list of random numbers is a linear problem, see
   ~\cite{GENERATION}), but is a bit less precise as we introduce redundancy
    in the population.}
\end{itemize}
We finally choose to use the second strategy, as we observed that it performs almost
as good at the first one (for a given size of population) while it was much
faster, so that, for given computer time, it allows a better representation of
the population. We also force the population to be color-symmetric by adding
a random shift of colors in the incoming messages. This is needed in order to
avoid the anti-ferromagnetic solution. The learned reader will notice that
this is equivalent to solving a {\it disordered} Potts glass instead of a
anti-ferromagnet model. Indeed the fact that an Ising anti-ferromagnet on a
random graph is equivalent to an Ising spin glass was already noticed
\cite{FedeKrzakala}.

\begin{figure}[!ht]
\begin{minipage}{0.49\linewidth}
\begin{center}
  \resizebox{9cm}{!}{\includegraphics[angle=270]{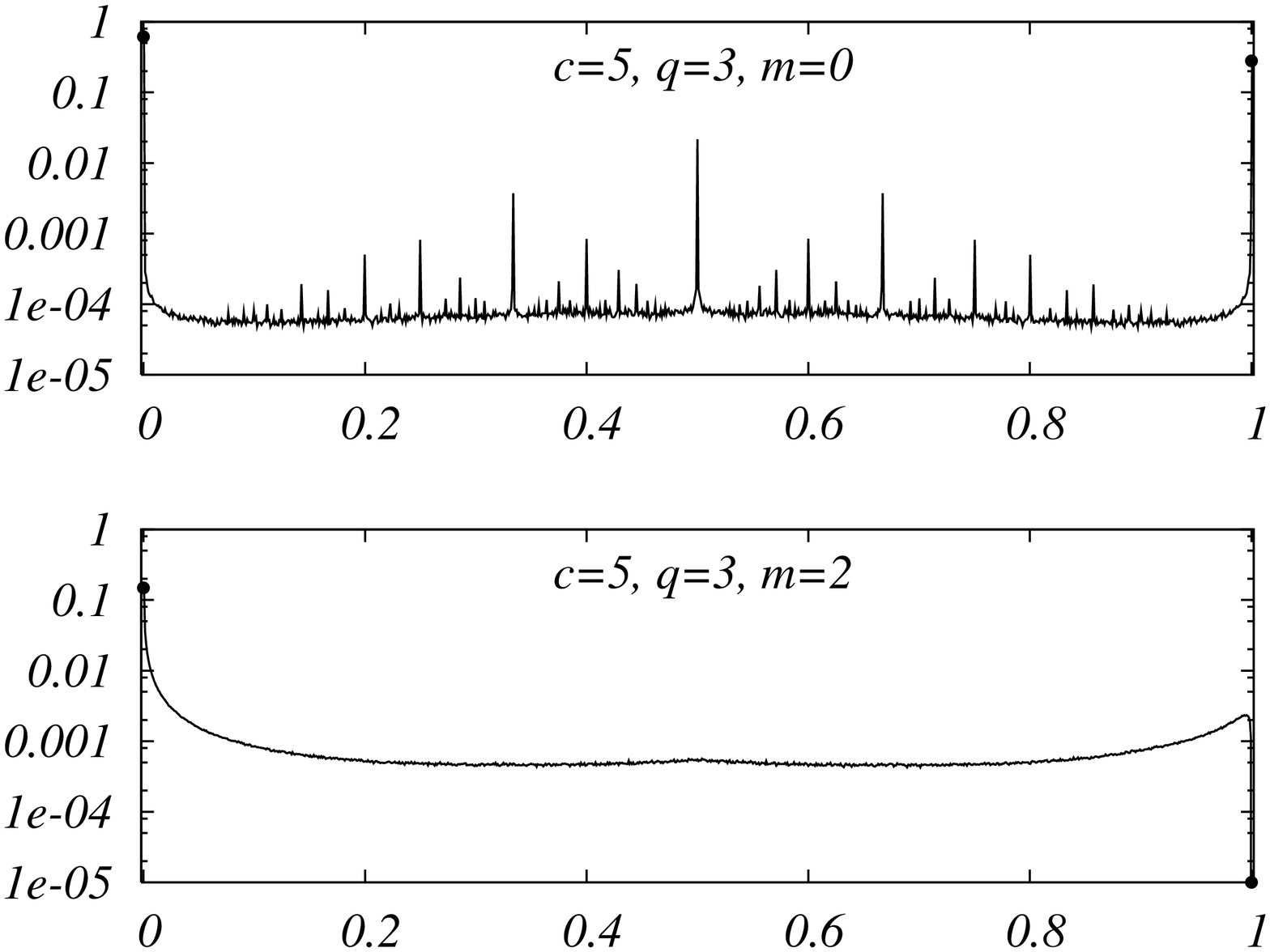}}
\end{center}
\end{minipage}
\begin{minipage}{0.49\linewidth}
\begin{center}
  \resizebox{9cm}{!}{\includegraphics[angle=270]{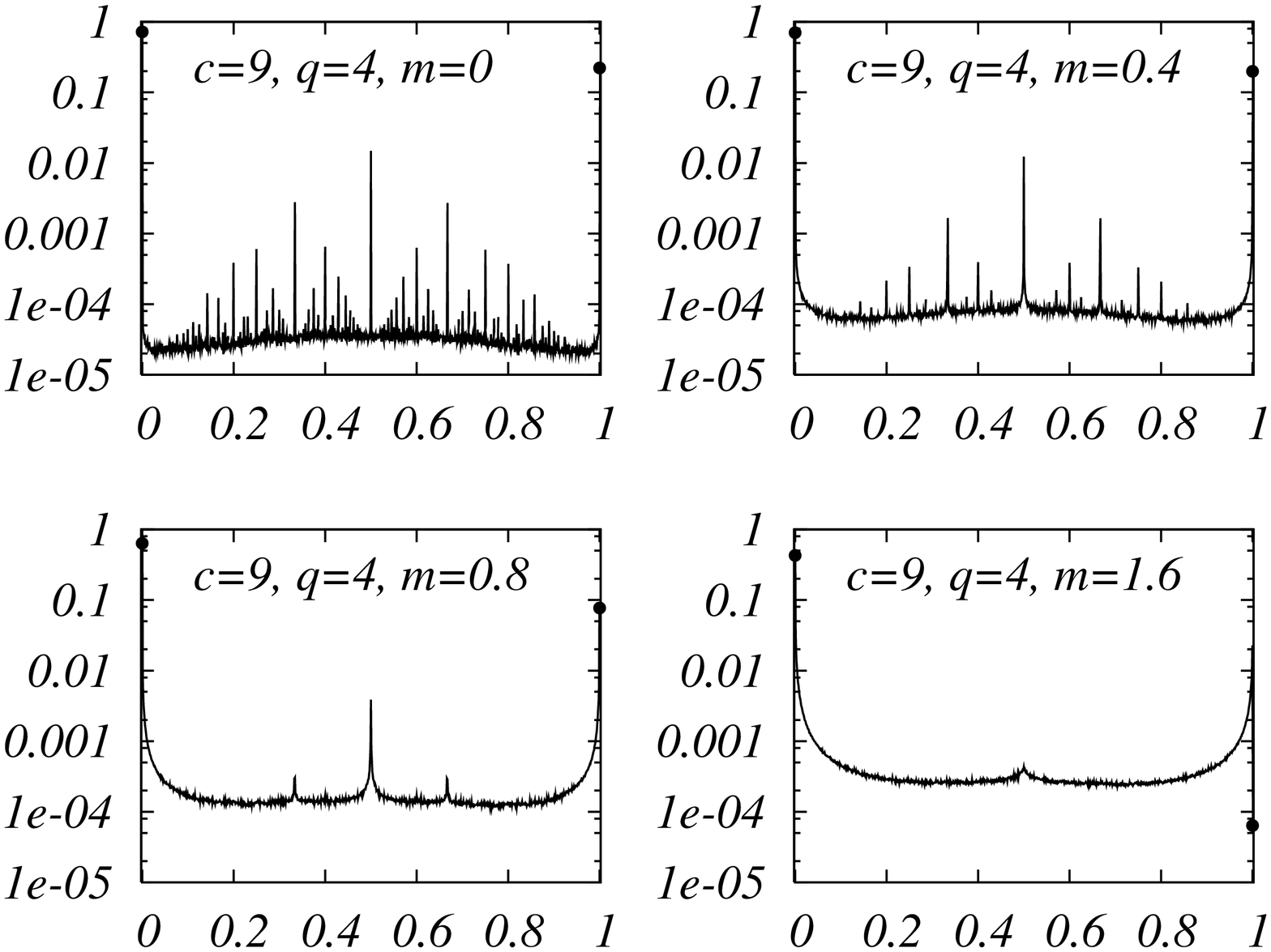}}
\end{center}
\end{minipage}
\caption{ \label{fig:reg_histo} Histograms of the first component of the cavity field  $\psi_1$, 
{\it i.e.} the probability that a node takes color one. Notice the logarithmic y-scale.
  Frozen fields ($\psi_1=1$) are present in the solution for the three upper
  cases; there are delta-peaks on 0,1,1/2 and other simple fractions depending
  on $q$. Notice that even when frozen fields are not present, there are many 
  almost frozen fields (the distributions only concentrate around
  $0$,$1$,$1/2$ and other fractions).}
\end{figure}

Another important issue is the presence of hard fields. In
fig.~\ref{fig:reg_histo} are histograms of the first component of the vectors
in the population for $3$- and $4$-coloring of $5$- and $9$-regular random
graphs respectively. It is interesting to see how they peak around fractional values due to
the presence of hard fields (see the three upper one). Maybe even more
interesting are the lower one where {\it no hard fields are present}. However,
since there are soft fields with values $1-\epsilon$, where $\epsilon$ can be
almost arbitrary small, one cannot see from these picture the absence of
frozen variables. For the case $c=9,q=4,m=0.8$ for instance, the
presence of the quasi-hard fields makes the distribution clearly concentrate on
values around one, zero and half (note however that the amplitude ---on a
logarithmic scale--- is far less important).

\begin{figure}[!ht]
\begin{center}
  \resizebox{9cm}{!}{\includegraphics[angle=270]{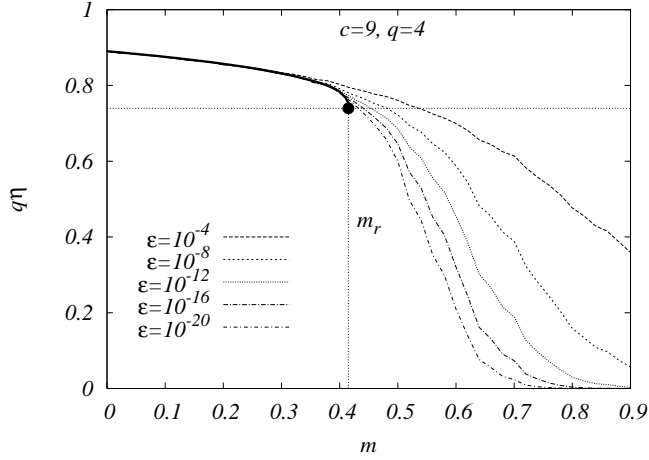}}
\end{center}
\caption{Fraction of the hard and the quasi-hard cavity fields $q\eta$ (a field is quasi-hard if 
  $\psi>1-\epsilon$) in the $4-$coloring of $9-$regular graph.
  The bold line is obtained with the analytical computation of the fraction
  of hard fields and the dot corresponds to the threshold $m_r$.
  \label{fig:quasihard}}
\end{figure}

The quasi-hard fields are therefore very hard to distinguish numerically from the true
hard ones. This is evidenced on fig.~\ref{fig:quasihard} where we plot the
fraction of hard fields computed using the expression (\ref{hard_m}) together with a
numerical estimate made by population dynamics without the separate hard/soft
implementation. We show that the fraction of fields of value $1-\epsilon$ 
is not zero in regions where we know that there are no hard fields even for
$\epsilon=10^{-20}$. This demonstrates the presence of quasi-hard fields, with
$\epsilon$ going to zero as the critical $m$ is approached. This transition 
is further studied in \cite{Guilhem_rear}.

An important simplification of the 1RSB update (\ref{1RSB}) arises when we
consider the soft and the hard fields separately.  The fraction of hard fields
can be computed using the generalized SP equation (\ref{hard_m}),
provided the ratio ${\overline{Z^m_{\rm soft}}}/{\overline {Z^m_{\rm hard}}}$
is computed. This considerably reduced the size of the population as only soft
field has to be kept in memory. Another way to further speed up
the code is to generate directly soft fields with a uniform
measure instead of waiting for them to come out from the 1RSB recursion. Indeed
they might be quite rare in the region of small $m$ and one can spend a 
considerable amount of time before being able to sample them correctly. 
Generating the soft fields with a
uniform weight turns out to be rather easy using the following method: (i)
Choose two random colors $q_1$ and $q_2$. (ii) Perform the usual recursion
(\ref{update_zero}) in order to have a new vector but forbid incoming hard
fields to $q_1$ and $q_2$. (iii) To obtain a uniform soft field generator, the
resulting field should be weighted by $1/{s\choose 2}$ where $s$ is the
number of non-null components in the vector. This is specially useful in the
case of Erd\H{o}s-R\'enyi graphs.

The formula for the free entropy $\Phi_s(m)$ (\ref{free_entropy}) also
simplifies in this case.  Consider a given site $i$; the site free entropy
term can be split into three parts when (i) the field is hard, (ii) the total
field is soft and (iii) the field is contradictory.  Then
\be 
\label{hard_site_phi}
\Phi^i_s = \log {\((p_{\rm hard} \overline {(Z_{\rm hard})^m}+ p_{\rm soft}
  \overline {(Z_{\rm soft})^m} \))} \, ,
\ee
where $p_{\rm hard}$/$p_{\rm soft}$ are the probabilities that the total field is
frozen/soft, and are given by the SP recursion. Indeed the probability that the
total field is not contradictory ($p_{\rm hard}+p_{\rm soft}$) is the
denominator in eq. (\ref{SP}) while $p_{\rm hard}$ is the numerator of eq.
(\ref{SP}). The link part can also be simplified using the fact that
contradictions arise when two incoming frozen messages of the same color are
chosen, so that
\be
\label{hard_link_phi}
\Phi^{ij}_s = \log  {\((p_{\rm no~contr} \overline
  {(Z_{\rm no~contr})^m}\))} \, ,
\ee
where $p_{\rm no~contr}$ is simply $(1-q \eta^{i\to j} \eta^{j \to i})$.

For $m=0$ the formula further simplify as
$Z_{\rm hard}^m=Z_{\rm soft}^m=Z_{\rm no~contr}^m=1$ so that 
\bea 
  \Phi^i_s &=&
\log {\((\sum_{l=0}^{q-1} (-1)^l {q \choose l+1}
  \prod_k   \left[1-(l+1)\eta^{k \to i}\right] \))}\, , \\
\Phi^{ij}_s &=& \log {\((1-q \eta^{i\to j} \eta^{j \to i}\))}\, . \eea
This is precisely what was obtained within the energetic cavity approach in
\cite{Coloring}. The numerical population dynamics implementation with mixed
hard/soft strategy is therefore as precise as it could be since we obtain the
{\it exact} evaluation in the $m=0$ case. This simple computation also
demonstrates how one can recover the energetic zero temperature limit from the
generic formalism.

\begin{figure}[!ht]
\begin{minipage}{0.49\linewidth}
\begin{center}
  \resizebox{9cm}{!}{\includegraphics[angle=270]{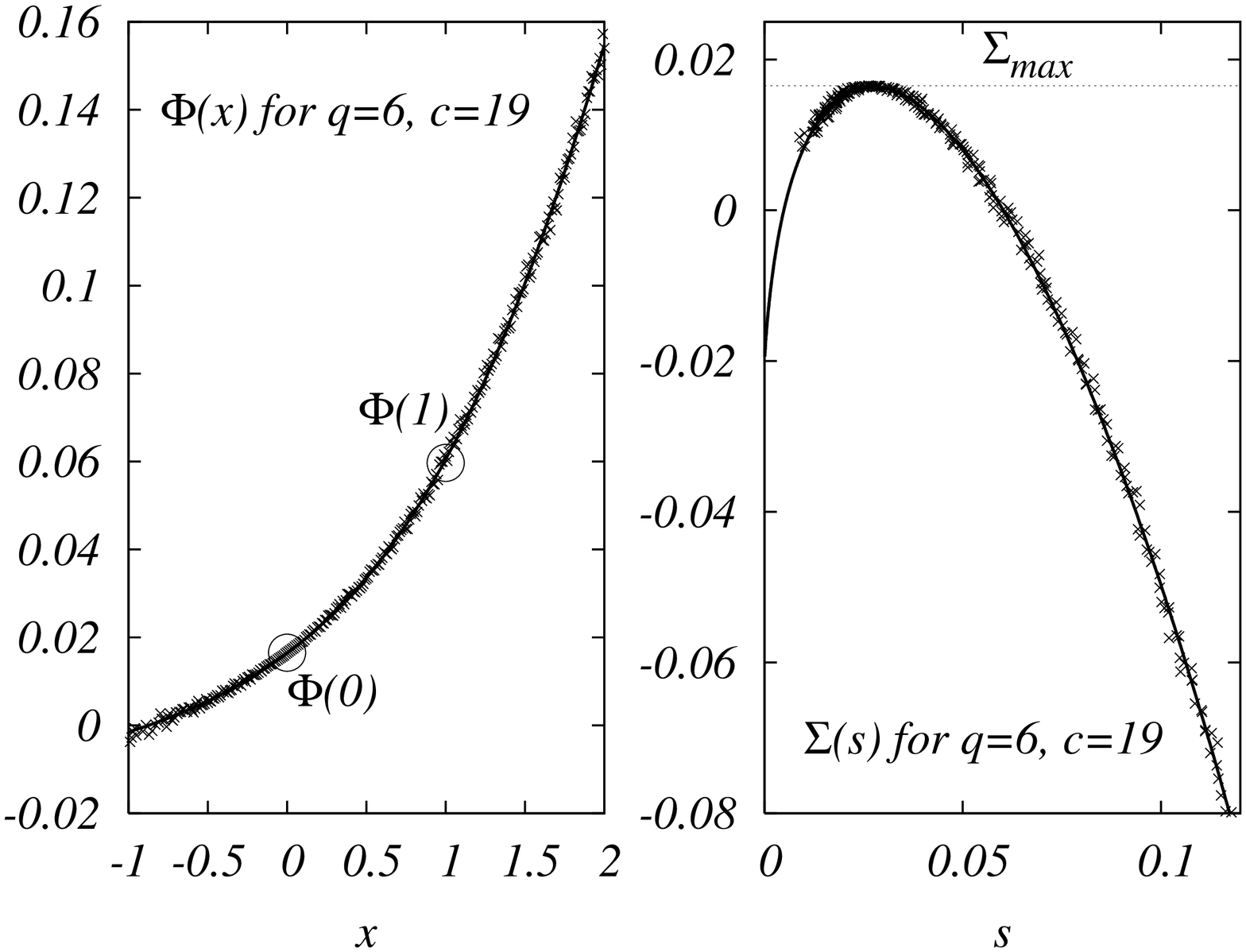}}
\end{center}
\end{minipage}
\begin{minipage}{0.49\linewidth}
\begin{center}
  \resizebox{9cm}{!}{\includegraphics[angle=270]{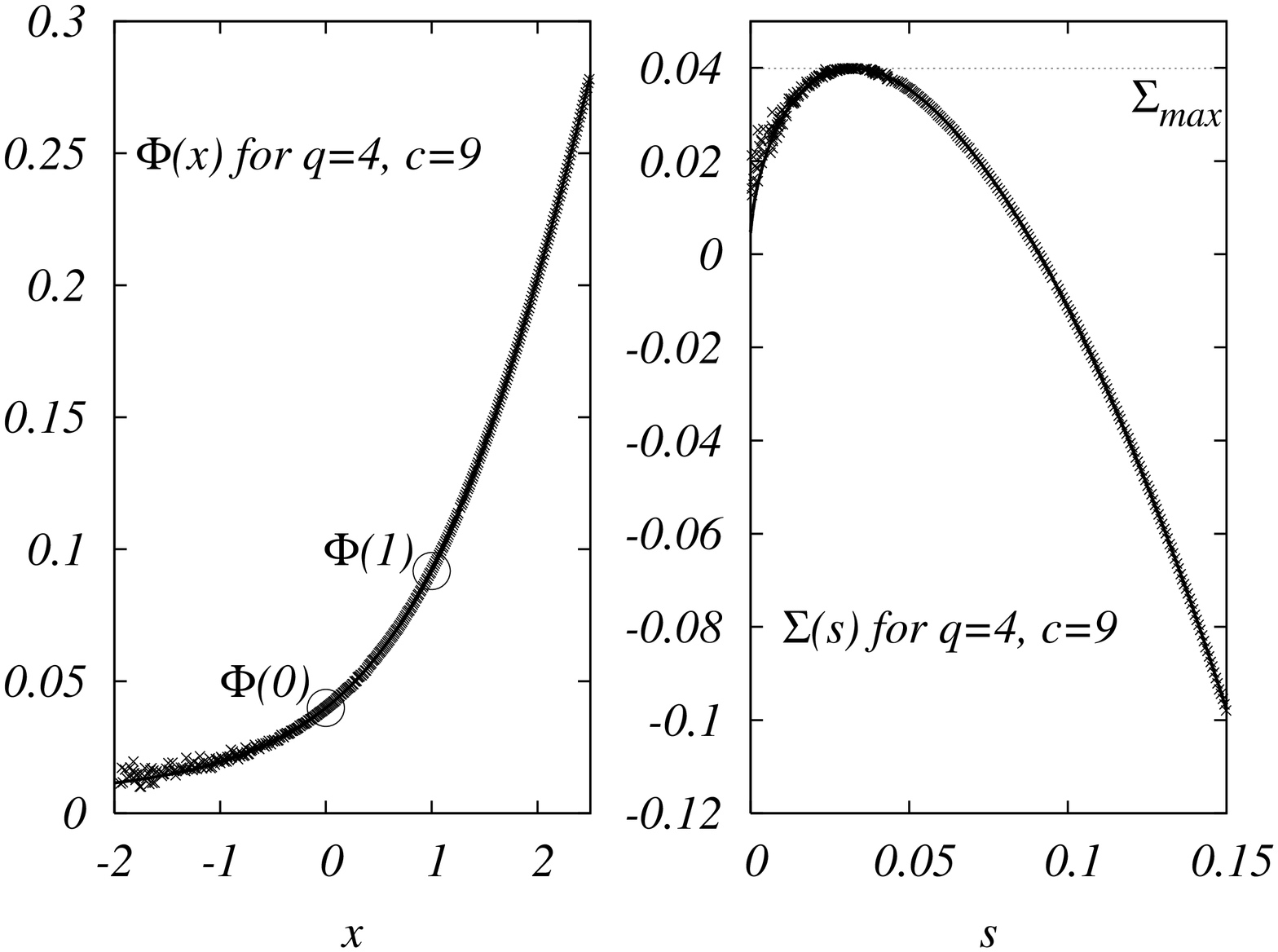}}
\end{center}
\end{minipage}
\label{fig:fit}
\caption{The numerical results for the free entropy (\ref{Sigma_ent}) and its fit with a
  function $a+b2^x+c3^x...$ for the $6$-coloring of 19-regular graphs and the
  $4$-coloring of 9-regular graphs. Circles give the analytical results at
  $m=0$ and $m=1$. On the right parts, we present the complexity versus internal
  entropy with the numerical points and the Legendre transform of the fit of
  the free entropy. The analytical result for $\Sigma_{max}$ is also
  shown.}
\end{figure}

Finally, we obtain the function $\Phi_s(m)$. We fit this function using an
ansatz $\Phi_s(m)=a+b 2^m + c 3^m \ldots$ and then perform the Legendre
transform to obtain the entropy and complexity. It is also possible to compute
directly the complexity from the population data using the expression of the
derivative of the potential directly in the code. Both methods lead to very
good results. We show an example of the raw data and their fit in
fig.~\ref{fig:fit}, where the data have been obtained with relatively small
population $(N=5000)$ but where the mixed strategy separating the hard and
soft fields have been used. For the purely soft-field branch, we used
$N=50000$. It took few hours up to few days to generate these curves on present
Intel PCs.

In the case of bi-regular random graphs, one needs two different populations:
one for the fields going from nodes with connectivity $c_1$ and one for 
the fields going from nodes with
connectivity $c_2$. Then each iteration for population $1$ ({\it resp.} $2$)
should be performed using as incoming messages the vectors of population $2$
({\it resp.} 1.). Again, one can separately perform the recursion for the
hard-field fractions in both population. 

The case of Erd\H{o}s-R\'enyi random graphs is more involved, as one needs a large number
$N_{\rm pop}$ of populations, each of them of size $N$. In this case, using the
separate hard/soft fields implementation and the formulae 
(\ref{hard_site_phi},\ref{hard_link_phi}) for complexity is crucial, as
it allows a good precision even for smaller population sizes. We used typically
$2N_{\rm pop}/c \approx (1-3)\cdot 10^3$ and $N \approx (1-3)\cdot 10^2$.
The error bars in table \ref{tab:results_complete} are computed from several 
independent runs of the population dynamics. In each case we were able to make 
the equilibration times and the population sizes large enough such that by 
doubling the time or the population size we did not observed any significant 
systematic changes in the average results. 

\section{High-$q$ asymptotics}
\label{app:largeq}

The quenched averages in the large $q$ limit are the same for the regular and
Erd\H{o}s-R\'enyi graphs and we thus consider directly the regular ensemble
of connectivity $c=k+1$. The appearance of a nontrivial 1RSB solution for $m=0$, which correspond to
$c_{\rm SP}$, was already computed in~\cite{ColoringFlo} and reads
\begin{eqnarray}
  c_r(m=0) &=& c_{\rm SP} = q\ \left[ \log q + \log \log q +1 -\log 2 + o(1)\right] \, ,\\
  \eta_d(m=0) &=& \frac 1q \left[ 1- \frac 1{\log q} + o\left(\frac{1}{\log q}\right)
  \right] \, , 
\end{eqnarray}
while the coloring threshold is~\cite{ColoringFlo}
\begin{equation}
  c_s = 2 q \log q - \log q - 1 + o(1).
\end{equation}
We now show how the connectivity where a solution with hard fields at $m=1$ first appears, 
and how the complete free entropy $\Phi_s(m)$ (\ref{Sigma_ent}) can be computed 
close to the COL/UNCOL transition

\subsection{The appearance of hard fields at $m=1$}

We first show that the correct scaling for the appearance of hard fields at
$m=1$ is
\begin{equation}
  k =  q [\log q + \log\log q + \alpha ].
\end{equation}
and compute the value of $\alpha$. In the order $O(q)$ we can write also
$k =  (q-1) [\log(q-1) + \log\log(q-1) + \alpha ]$. The starting point is the
equation (\ref{hard_m1_app}), with ${\cal Q}_1(x)=\delta(x-k)$. In the large 
$q$ limit the fraction of hard fields is $\mu(q,k)=q \eta(q,k) = 1-\theta(q,k)$, 
where $\theta(q,k)=o(1)$ is the fraction of soft fields. We check self-consistently 
at the end of the computation that only the two first 
terms of (\ref{hard_m1_app}) are important. Then we have
\begin{equation}
    \mu(q,k)= 1 -(q-1) \left( 1-\frac{1}{q-1}\mu(q,k) \right)^k 
    = 1 - (q-1) \e^{-\frac{k\mu(q,k)}{q-1}} \, .
\end{equation}
A self-consistent equation for $\theta(q,k)$ follows 
\begin{equation}
    \log(q-1) \theta(q,k)= (q-1)^{\theta(q,k)} \e^{-\alpha}\, ,
\end{equation}
which is solved by $\theta(q,k)=\gamma(\alpha)/\log(q-1)$ where 
\begin{equation}
        \gamma(\alpha) \e^{-\gamma(\alpha)} = \e^{-\alpha} \, . 
      \label{gamma}
\end{equation} 
The maximum of the left hand side is $1/e$ for $\gamma(\alpha)=1$. It means that a solution
of (\ref{gamma}) exists for $\alpha>1$. Finally the hard fields appear in the 1RSB 
solution for $m=1$ at connectivity
\begin{equation}
   c_r(m=1)= q [\log q + \log\log q + 1 +o(1)].
\end{equation}

The clustering transition $c_d$ should be between $c_{\rm SP}$ and $c_r(m=1)$
as this is what we observed for finite $q$. We see that $c_{\rm SP}$ and
$c_r(m=1)$ differs only in the third order and both are very far from the
coloring threshold and also from the condensation transition as we show in
section~\ref{q-cond}. 
It would be interesting to compute a large $q$ expansion of the connectivity at 
which the hard fields appears in all the clusters (for all $m$ such that $\Sigma(m)>0$).
Together with our conjecture about rigidity being responsible for the computational 
hardness that might give a hint about the answer on the long-standing 
question~\cite{FrMcDi}: ``Is 
there a polynomial algorithm and $\epsilon$ such that the algorithm would color random 
graphs of average connectivity $(1+\epsilon)q\log q$ for all large $q$?''

\subsection{The condensation transition}
\label{q-cond}

To compute the large-$q$ asymptotic of the condensation transition, we first need to derive the large-$q$ expansion of the free entropy (\ref{Sigma_ent}) in the connectivity regime
$c=2q\log{q}$. Let us show self-consistently that the following scaling is
relevant for the condensation transition in the large $q$ limit
\bea
  \label{eq:cd1_asymp}
  c_s &=& 2 q \log q  - \gamma \log q + \alpha \, , \\
  \label{eq:cd1_asympbis}
  \eta &=& \frac {1}{q} - \frac{B}{q^2} \, ,
\eea
and compute the constants $\gamma$, $\alpha$, $B$. Using the above scaling,
the function $w(\eta)$ (\ref{SP}) is dominated by the first two terms in
numerator and denominator, and reads in the first two leading orders 
$1 - qw(\eta) \approx \frac{q e^{-k\eta}}2$ so that
\be w(\eta) = \frac 1q - \frac 1 {2q^2} + O\left(\frac {\log{q}} {q^3} \right)  \ee
independently of $\gamma$, $\alpha$, and $B$. To take into account the reweighting we
expand eq. (\ref{hard_m}) in the two leading orders
\bea 
\eta &=& 
 \frac 1q - \frac 1{2q^2}  \frac{\overline{Z^m_s}}{\overline{Z^m_h}} + 
  {\cal{O}}\((\frac {\log {q}}{q^3}\)) \, .
\eea
Note that almost all the incoming fields are hard, {\it i.e.} have one
component of value $1$. Since there are on average only $2B\log{q}$ incoming
soft fields, the leading order of the hard-field reweighting (the
normalization in eq.~(\ref{update})) is different from $1$ with 
a probability only $O(\log{q}/q)$.  Similarly, almost all the soft fields
have two nonzero and equal components.  The normalization in eq.~(\ref{update})
is thus almost surely $2$, thus the average reweighting factor of the soft
fields is
\bea
\label{weight_soft}
\overline{Z^m_s} &=& 2^m + {\cal O}\((\frac{\log q}{q}\)) \, .
\eea
Finally,
\bea 
\eta &=& \frac 1q - \frac {2^m} {2q^2} + {\cal{O}}\((\frac {\log {q}}{q^3}\))\, .
\label{large_eta}
\eea
Therefore the constant $B$ in (\ref{eq:cd1_asympbis}) is $B=2^m/2$, independently of 
$\gamma$ and $\alpha$.

The computation of the complexity requires the next order in the hard-field
reweighting. Indeed the normalization in (\ref{update}) might not be $1$ but
$1/2$; and this happens when there is a soft field arriving of the color
corresponding to the hard field in consideration. The probability of this
event is $ \frac {2 c \((1-q\eta \))}q = O( \frac q{\log{q}} )$.  The hard-field reweighting is thus
\bea
\label{weight_hard}
\overline{Z^m_h} &=& 1 \left[ 1- \frac {2 c \((1-q\eta \))}q \right] + \frac {2 c
  \((1-q\eta \))}q  \frac{1}{2^m} + o\((\frac{\log q}{q}\)) \, .
\eea

We now expand the replicated free energy (\ref{Sigma_ent}) in the large $q$
limit and regime (\ref{eq:cd1_asymp}). Remind that from (\ref{free_i},
\ref{free_ij})
\be
\label{eq:phi}
\Phi_s(m)    =  \log {  \overline {(Z^i_{0})^m} } - \frac{c}{2} 
\log { \overline { \((Z^{ij}_0\))^m }}\, .
\ee
The averages are over the population in the sense of (\ref{Psi}). 

The site free energy is the logarithm of the average of the total field
normalization.  This average can be split into three parts when (i) the total
field is a hard field, (ii) the total field is a soft field and (iii) the
total field is contradictory (and its normalization zero).  The probability
that the total field is not contradictory is the denominator in eq. (\ref{SP})
\be g(\eta) = \sum_{l=0}^{q-1} (-1)^l {q \choose l+1} [1-(l+1)\eta]^c, \ee
where again only the first two terms are relevant in the expansion. 
The site free energy is then
\begin{eqnarray}
  \log{ \overline {(Z^i_{0})^m}} &=& 
 \log{ g(\eta)} + \log{ \left[ q\,  w(\eta)  \overline{Z^m_h} + (1-q\, w(\eta))\overline{Z^m_s} \right]} \nonumber \\
   &\simeq& \log{ \left[q (1 -\eta) ^c - \frac{q(q-1)}2  (1 -2\eta) ^c \right] }
+ \log {\left[1- \frac 1{2q} - \frac{2 c \((1-q\eta\))}q \left(1-\frac
 1{2^m}\right) + \frac{2^m}{2q}\right]} +o\left(\frac{1}{q}\right) \, . 
\label{site_phi}
\end{eqnarray}
where
\bea
\log{\[[q\[[1-\eta\]]^c+\frac {q\((q-1\))}2 \[[1-2\eta\]]^c\]]} &=& \log{q} + c\log{\[[1-\eta\]]}+\log{\((1-\frac{q-1}2\[[\frac{1-2\eta}{1-\eta}\]]^c\))} \\
&\approx& \log{q} + c\log{\[[1-\frac 1q + \frac{2^m}{2q^2}\]]} +
\log{\[[1-\frac 1{2q}+o(1/q)\]]}. \eea

To compute the link contribution in (\ref{eq:phi}) we need to consider two
fields $\psi_s^{i\to j}$ and $\psi_s^{j\to i}$ and to compute \be Z^{ij}_0=
1-\sum_{s=1}^q \psi_s^{i\to j} \psi_s^{j\to i}.  \ee There are three different
cases:
\begin{enumerate}
\item Two hard fields are chosen, then $Z^{ij}_0=0$ with probability $q \eta^2$
  (this is of order $1/q$) and $Z^{ij}_0=1$ with probability $q(q-1) \eta^2$ (this
  is of order $1$).
\item Two soft fields are chosen then $Z^{ij}_0=1$ with probability $(1-q\eta)^2$
  (this is of order $1/q^2$), all other situations being $O(1/q^3)$. Let 
   us remind that the dominant soft fields are two-component of type $(1/2,1/2,0,0,\dots)$.
\item One hard and one soft field is chosen, then $Z^{ij}_0=1$ with probability 
  $2 \eta (1-q\eta) (q-2)/q$, and
  $Z^{ij}_0=1/2$ with probability $4 \eta (1-q\eta)$ (this is of order
  $1/q^2$).
\end{enumerate}
On average, one thus obtain for the link contribution \be \log
\overline{\(({Z^{ij}_0}\))^m} =\log{\left[ \((1-q \eta^2-4 \eta\, (1-q\eta)\))
    1^m + 4 \eta \, (1-q\eta) \frac{1}{2^m} \right] }+
o\left(\frac{1}{q^2}\right)\, .
\label{link_phi}
\ee

Putting together the two pieces (\ref{site_phi}) and (\ref{link_phi}),
expanding $\eta$ according to (\ref{large_eta}) and considering only the
highest order in $c$, we can finally write the free energy as 
\bea \Phi_s(m) &=& 
\log{q} - \frac{c}{2q} + \frac {2^m-2}{2q} -\frac{c}{4q^2} +
o\left(\frac{1}{q}\right)\, .  \eea

The internal entropy $s(m)$ and the complexity $\Sigma = \Phi_s(m) - m s(m)$
are then \bea
s(m) &=& \frac{\partial \Phi_s(m)}{\partial m} = \frac {2^m \log{2}}{2q}\, , \\
\Sigma(m) &=& \log{q} - \frac{c}{2q} + \frac{2^m-2-m\, 2^m \log{2}}{2q} -
\frac{c}{4q^2} \, , \eea and the complexity is thus zero for $c_{\Sigma=0} = 2 q
\log {q} - \log{q} -2 + 2^m \[[1-m \log{2}\]]+o(1)$. In particular, one has
for the coloring and the condensation thresholds 
\bea
c_{\Sigma=0}(m=0) &=&  2q \log{q} -\log{q} -1 + o(1) \, , \\
c_{\Sigma=0}(m=1) &=& 2q \log{q} - \log{q} - 2 \log{2} + o(1)  \, .\eea
For connectivity $c=2q \log {q} - \log{q} + \alpha$, one gets \bea
2 q s(m) &\simeq& 2^m \log{2}\, , \\
2 q \Sigma(m) &\simeq& 2^m -2 -m2^m \log{2} - \alpha \, .  \eea


\end{document}